\documentclass[twocolumn]{aa}

\date{Received 23 December 2025 / Accepted 15 April 2026 }

\usepackage[varg]{txfonts}
\usepackage[utf8]{inputenc}
\usepackage[english]{babel}
\usepackage{tabularx}
\usepackage{aas_macros}

\usepackage{setspace}

\usepackage{indentfirst}

\usepackage{enumitem}
\setlist[itemize]{noitemsep, topsep=0pt}

\usepackage{graphicx}
\graphicspath{{./}}

\usepackage{wrapfig}
\usepackage{subfigure}

\usepackage[dvipsnames]{xcolor}

\usepackage{hyperref}
\hypersetup{
    colorlinks,
    urlcolor=blue,
    linkcolor=blue,
    citecolor=blue
}
\defcitealias{2025A&A...700A..53M}{Paper I}
\defcitealias{2026A&A...705A.165M}{Paper II}
\usepackage{csquotes}

\usepackage{listings}
\usepackage{courier}\definecolor{bluekeywords}{rgb}{0,0,1}
\definecolor{greencomments}{rgb}{0,0.5,0}
\definecolor{redstrings}{rgb}{0.64,0.08,0.08}
\definecolor{types}{rgb}{0.17,0.57,0.68}
\lstdefinestyle{py}
{
    language=Python,
    frame=l,
    framesep=5pt,
    captionpos=b,
    numbers=left,
    numberstyle=\tiny,
    showspaces=false,
    showtabs=false,
    breaklines=true,
    showstringspaces=false,
    breakatwhitespace=true,
    commentstyle=\color{greencomments},
    keywordstyle=\color{bluekeywords},
    stringstyle=\color{redstrings},
    basicstyle=\footnotesize\ttfamily,
}
         \usepackage{natbib}
\bibpunct{(}{)}{;}{a}{}{,} 
\begin{document}

\title{Stellar flare-driven evolution of primordial early exo-Earth atmospheres: Insights from a Young M Dwarf Flare model}
\author{E. Mamonova\inst{1}, K. Herbst \inst{1},  V. Kofman \inst{1}, O. Ozgurel \inst{1}, A. F. Kowalski\inst{2,3,4}, S. Wedemeyer\inst{5,6}, S. C. Werner\inst{1}}

\institute{Centre for Planetary Habitability (PHAB), University of Oslo, 0315 Oslo, Norway
\and National Solar Observatory, University of Colorado Boulder, 3665 Discovery Drive, Boulder, CO 80303, USA
\and Department of Astrophysical and Planetary Sciences, University of Colorado, Boulder, 2000 Colorado Ave, CO 80305, USA
\and Laboratory for Atmospheric and Space Physics, University of Colorado Boulder, 3665 Discovery Drive, Boulder, CO 80303, USA
\and Rosseland Centre for Solar Physics, University of Oslo, 0315 Oslo, Norway
\and Institute of Theoretical Astrophysics, University of Oslo, 0315 Oslo, Norway}

\abstract{\textit{Context.} M dwarfs are key targets for terrestrial exoplanet studies, with prospects for atmospheric spectroscopy. However, strong stellar magnetic activity and frequent flaring require modelling efforts to assess their impact on planetary atmospheres.\\
\textit{Aims.} We aim to investigate one year of atmospheric chemical evolution of a young exo-Earth orbiting an active M dwarf by coupling our Young M Dwarfs Flare (YMDF) model of stellar activity with the VULCAN chemistry kinetic code. \\
\textit{Methods.} The YMDF model provides time-resolved spectral energy distributions for high- and low-energy electron beam–driven flares, which are used as external radiative inputs to VULCAN to compute the time-dependent photochemistry and kinetics for different primordial atmospheric scenarios.\\
\textit{Results.} We present the impact of stellar flares on atmospheres with varying water vapour content, ranging from a plausible primordial atmosphere with solar abundances, representative of a planet-forming region in a dissipating protoplanetary disk, to an extreme water-steam atmosphere with minimal other species. This was explored across several configurations: variable flux in the YMDF model, the previous model representing an active but older M dwarf with added 10K or 400K bottom boundary heat flux, and a constant stellar flux model. \\
\textit{Conclusions.} Our study suggests that, compared to the previous model, the YMDF model produces synthetic flares that exert significantly greater stress on primordial atmospheres, regardless of the water-vapour content. Increased activity and prevalence of mid-size flares has the potential to induce permanent changes in atmospheric mixing ratios, especially in species with low abundances.}
\titlerunning{Exo-Earth atmospheres with the Young M Dwarf Flare model.}
\authorrunning{Mamonova et al.}
\maketitle

\section{Introduction}
\label{sec:introduction}

M dwarfs are the most abundant stars in the solar neighbourhood and harbour the greatest number of exo-Earths within the liquid-water habitable zone (HZ, \citealt{Owen1980,2013ApJ...762...41G,2015ApJ...807...45D}) compared to other stellar types. Therefore, they are prime targets in the search for terrestrial exoplanets. Their HZs lie at much smaller orbital separations than those around Sun-like stars, which enhances the geometric probability of detecting transits and increases the frequency of transit events. Crucially, these stars exhibit significantly higher magnetic activity levels, including frequent and energetic flares. During flare events, the M dwarf’s luminosity in ultraviolet (UV), X-ray and extreme-UV (XUV) and visible wavelength ranges can increase by up to three orders of magnitude \citep{2007AsBio...7...85S}. Our understanding of the activity of M stars has advanced considerably over the past two decades through high-precision, long-baseline photometric surveys such as Kepler \citep{2010Sci...327..977B}, K2 \citep{2014PASP..126..398H}, and TESS \citep{2015JATIS...1a4003R}, which have provided an extensive statistical characterization of flare rates and properties across different stellar ages and types \citep{2014ApJ...797..122D,2016ApJ...829...23D,2014ApJ...797..121H,2024AJ....168...60F}. Additionally, numerous UV and multi-wavelength studies using space-based observatories like the Hubble Space Telescope (HST) have complemented these surveys by capturing near-UV (NUV) and far-UV (FUV) flare emissions, revealing flare behaviour and energy distributions that are critical for assessing the impact of stellar activity on planetary atmospheres \citep{2019ApJ...871L..26F,2019ApJ...871..167K,2024MNRAS.533.1894J,2024ApJ...971...24P,2023ApJ...944....5B,2019ApJ...881....9H}. These energetic stellar events primarily impact exoplanets orbiting M dwarfs, inducing significant photochemical alterations in their upper atmospheres \citep{2010AsBio..10..751S,2016ApJ...820...89F,2019AsBio..19...64T}.

Studies of terrestrial planet atmospheres require a comprehensive understanding of their photochemistry. Unlike giant exoplanets, where composition is primarily perturbed by photolysis without surface emissions, terrestrial atmospheres arise from complex interplay among photolysis, reaction kinetics, vertical diffusion, escape, deposition, and condensation/sedimentation of condensible species, collectively controlling spectroscopically active trace gases whose lifetimes are governed by photochemical interactions \citep{2012ApJ...761..166H}. Whilst vertical diffusion, escape, and condensation also operate in gaseous
planet atmospheres, the coupling of these processes with surface interactions and the absence of a deep thermal reservoir make terrestrial atmospheres particularly sensitive to their combined effects. After a protoplanetary disk dissipation, terrestrial planets may retain primary H$_2$-rich envelopes \citep{2020SSRv..216..129O,2011Natur.470...53L}, yielding atmospheres dominated by volatile hydrogen compounds or even water vapour \citep{2012ApJ...745....3M}. The investigation of primordial atmospheres on terrestrial planets provides essential insights into the divergent evolutionary trajectories and habitability potential of rocky exoplanets.

In M dwarfs, stellar flares release energy in NUV and optical range \citep{2024ApJ...971...24P}, and in the UV and extreme UV (EUV) wavelengths \citep{2016ApJ...820...89F}. \citet{2024MNRAS.532.4436B} found that the time-integrated energy ratio of FUV to NUV during flares is on average three times greater than what would be expected from a constant 9000 K blackbody spectrum. Due to the strong absorption of these high-energy photons, as planetary atmospheres tend to be optically thick in FUV and NUV, it results in significant energy deposition in the upper atmospheric layers, such as the thermosphere and exosphere \citep{2006P&SS...54.1425K,2010AsBio..10..751S}. This absorption leads to heating at low pressures (high altitudes), while deeper layers receive minimal direct UV flux.

The NUV and FUV spectral regions are the primary drivers of photochemistry in planetary atmospheres, carrying sufficient photon energies to dissociate key molecules such as H$_2$O, CO$_2$, CH$_4$, and O$_3$. The exact role of NUV radiation in prebiotic chemistry on Earth remains uncertain \citep{2017ApJ...843..110R}, and M dwarfs may not emit sufficient quiescent NUV flux to drive such chemistry \citep{2019ApJ...871L..26F}. However, their high flare frequencies deliver intense bursts across the FUV and NUV bands that could compensate for the low steady emission \citep{2019ApJ...871L..26F, 2023AJ....165..195B}, making the time-averaged UV environment, particularly for young active stars such as AU~Mic, a critical factor for both atmospheric photochemistry and the potential for prebiotic synthesis on orbiting planets.

Atmospheric escape from terrestrial planets orbiting active M dwarfs is driven by thermal processes, such as hydrodynamic outflows induced by XUV heating of the exosphere \citep{2010AsBio..10..751S,2019AsBio..19...64T,2020AJ....160..237F,2023MNRAS.525.5168M,2015AsBio..15..119L}, alongside non-thermal mechanisms including enhanced polar winds, ion pickup, and steady or transient stellar wind interactions \citep{2016A&A...596A.111R,2017ApJ...836L...3A,2017ApJ...843L..33G,2022ApJ...941L...8G}. In hydrogen-rich primordial atmospheres, the escape of light species such as molecular hydrogen may become diffusion-limited near the homopause, where upward fluxes are constrained by diffusion through heavier background gases under strong stellar irradiation \citep{2013E&PSL.375..312K,2020PSJ.....1...11Z,2017ApJ...843..122Z,2020A&A...643A..81K}. Prior studies have quantified substantial losses, equivalent to several Earth oceans, for such atmospheres under young M dwarf irradiation, yet these calculations remain decoupled from the photochemical kinetics governing volatile replenishment and transformation \citep{2022ApJ...934..137Y,2015AsBio..15..119L}. Coupled photochemical-kinetic simulations of primordial H-H$_2$O atmospheres, incorporating variable stellar fluxes, are thus required to interpret volatile retention and habitability potential.

Several studies have employed three-dimensional coupled chemistry-climate models to investigate the impact of stellar flares on rocky planet Earth-like atmospheres. \citet{2023MNRAS.518.2472R} coupled a general circulation model with photochemical kinetics to find that flares increase atmospheric ozone by a factor of 20. \citet{2021NatAs...5..298C} used observations to show that recurring flares drive atmospheres of K and M dwarf planets into chemical equilibria substantially deviating from their pre-flare states, with flare-driven transmission features of species such as NO$_2$, N$_2$O, and HNO$_3$ offering promise for future detection.  Most recently, \citet{2025AJ....170...40C} extended their framework to TRAPPIST-1 e, finding that flare frequency and spectral shape govern cumulative chemical changes, and O$_3$ variability strongly mediates atmospheric temperature responses. These 3D studies reveal important hemispheric and spatio-temporal variability of relatively quiescent or old M dwarf stars.

Complementary to these 3D studies, several authors have additionally investigated the photochemical response of hydrogen-dominated atmospheres to stellar flares using 1D and pseudo-2D models \citep{2023MNRAS.521.3333L, 2025ApJ...993...41G, 2022A&A...667A..15K, 2023MNRAS.523.5681N}. This difference in dimensionality is physically motivated: Earth-like atmospheres benefit from decades of observational constraints and well-validated 3D frameworks, whereas atmospheres of exotic compositions, with no direct Solar system analogue, remain far less observationally constrained, making a systematic 1D parameter survey a natural and necessary foundation for more complex future modelling. Such primordial H$_2$O-atmosphere simulations, achievable with a 1D photochemical-kinetic model assuming full stellar flux redistribution, could reveal the chemical processes controlling volatile retention and the habitability potential of young worlds orbiting M dwarfs.

A first step toward such studies is taken by investigating the chemical evolution of the water vapour primordial envelope of a 1~M$_\oplus$, 1~R$_\oplus$ rocky planet (hereafter referred to as early exo-Earth) orbiting a young, active early M dwarf such as AU Mic, at a separation allowing for an equilibrium temperature of $\sim$500~K, representing an early, hot evolutionary stage rather than present-day surface habitability.
While many earlier studies (see e.q. \citealt{2010AsBio..10..751S,2023MNRAS.520.3867T}) have focused on short-term atmospheric responses or responses to individual flare events, and long-term simulations have to date addressed such compositions on tidally locked planets (see e.g. \citealt{2021NatAs...5..298C}), here we extend simulation timescales to capture the cumulative photochemical evolution of water-vapour-dominated atmospheres of terrestrial planets over approximately one year of repeated flare forcing. This duration balances the need to capture long-term chemical evolution with computational feasibility, allowing us to investigate both the direct effect and the long-term response of the atmosphere to the flares. We further evaluate how efficiently various species are transported to the upper layers of such primordial atmospheres by diffusion-limited escape. The manuscript is organized as follows: Section~\ref{sec:methods} describes the variable stellar flux model employed, the detailed parametrisation of the atmospheric grid and the chemical network of species along with vertical mixing parameters. In Section~\ref{sec:results}, we present a grid of atmospheric simulations evolved over approximately one year, comparing several configurations including constant quiescent flux. Section~\ref{sec:discussion} places our findings in the broader context of planetary atmospheric evolution and habitability.

\section{Methods}
\label{sec:methods}

\subsection{Simulations' setup}
\begin{figure}
  \centering
  \includegraphics[width=0.95\linewidth]{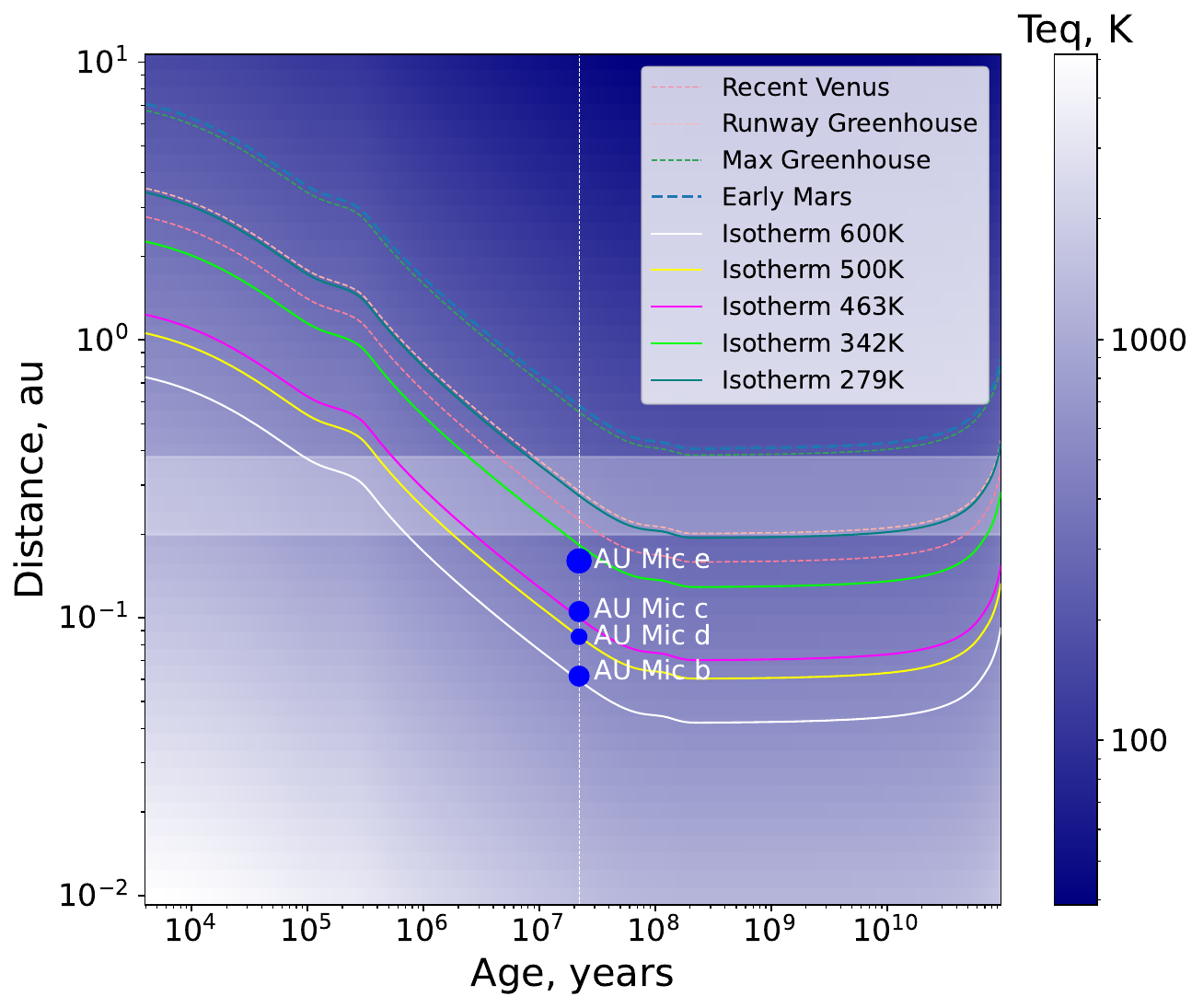}
  \caption{Habitable zone evolution for M dwarfs with 0.5~$M_\odot$, showing planetary equilibrium temperature ($T_\mathrm{eq}$) and atmospheric condition shifts with stellar age. Calculated using MESA Isochrones and Stellar Tracks (MIST; \citealt{2016ApJS..222....8D,2019ApJS..243...10P}). Dashed lines denote HZ boundaries (Recent Venus, Runaway Greenhouse, Maximum Greenhouse, Early Mars; \citealt{2014ApJ...787L..29K}); solid lines trace isotherms (600, 500, 460, 342, 279~K). Circles mark known exoplanets in the AU Mic system (22~Myr) across orbital separations.}
  \label{fig:101}
\end{figure}

Our modelling focuses on the AU Mic system \citep{2023AJ....166..232W} with 0.5 M$_\odot$, 0.02$\pm$0.002 Gyr M dwarf, and its planet AU Mic d with a rotation period of 12.736$\pm$0.008 days. This planet is non-transiting and characterised by a known Earth-like mass determined by transit time variations (TTVs). AU Mic d orbits at a separation consistent with an equilibrium temperature of $T_\mathrm{eq} \sim$ 500~K under the current stellar luminosity, a value that will decrease as the star settles onto the main sequence and its bolometric output declines. The habitable zone around a typical M dwarf star evolves substantially during its pre-main-sequence contraction, shifting inward as the star cools and stabilises. Figure~\ref{fig:101} illustrates this concept for the AU Mic system and HZ evolution around this 0.5~$M_\odot$ star, along with isotherms tracing changes in exoplanet equilibrium temperature. MIST isochrones \citep{2016ApJS..222....8D} predict more temperate planetary conditions for d planet as its hosting M dwarf will enter the main sequence. Although the planet is non-transiting, this system is studied due to its well-characterised nature, serving as a template for other terrestrial planets orbiting M dwarfs; moreover, at a distance of 9.7 pc \citep{2020yCat.1350....0G}, it ranks among the closer such systems and thus represents a promising target for future observations.

In a series of recent studies (\citealt{2025A&A...700A..53M}, hereafter \citetalias{2025A&A...700A..53M}, and \citealt{2026A&A...705A.165M}, hereafter \citetalias{2026A&A...705A.165M}), we introduced the Young M Dwarf Flare (YMDF) model, which characterises the spectral properties and temporal evolution of the flare activity in young M dwarfs. In this work, we use the synthetic spectra and temporal evolution data produced by the YMDF model as input to the atmospheric photochemistry. Considering the young age of the AU Mic system, we assume primordial atmospheres dominated by hydrogen with varying water vapour abundances.

To explore the influence of the atmospheric water content on flare-driven chemistry, we construct an atmospheric grid of 16 atmospheres spanning H$_2$O volume mixing ratios ranging from 0.1\% to 100\%, with the remaining composition scaled to solar elemental abundances, in two different regimes of internal heating: 10K and 400K. These two values of internal heating sample the edges of the physically plausible parameter space, from minimal interior heating to the hotter conditions theorised to allow ongoing silicate melting and magma ocean formation at the surface. Throughout this study, we compare the results of the chemical kinetic simulations within the grid of our atmospheres, that were subjected to variable stellar flux generated by our YMDF model, with those obtained using the framework described in \citet{2018ApJ...867...71L}, known as the `fiducial flare' model (hereafter FF). Additionally, we perform simulations using a constant quiescent flux (CF) based on the AU Mic panchromatic flux. The  simulations, incorporating interior heating of the atmosphere to 400 K at the lower boundary, are denoted as FF400K and YMDF400K, respectively. Different flare models enable a systematic assessment of varying UV fluxes on atmospheric composition and evolutionary trajectories, while probing the physically plausible parameter space extremes of interior heating.

\subsection{Constructing the synthetic flare populations}
\label{sec:methodcss}
We produced the population of flare events in form of an equivalent duration (ED, \citealt{1972Ap&SS..19...75G}) distribution based on the broken power law relation with $\alpha_1$ = 1.39 and $\alpha_2$ = 1.8, found for AU Mic in \citetalias{2025A&A...700A..53M}. The broken power law function is mathematically expressed as:

\begin{equation}\label{eq:17}
f(x) = \left \{
         \begin{array}{ll}
           A (x / x_{break}) ^ {-\alpha_1+1} & : x < x_{break} \\
           A (x / x_{break}) ^ {-\alpha_2+1} & :  x > x_{break}, \\
         \end{array}
       \right.
\end{equation}
where  $A$ is amplitude, and x$_{break}$ is the ED value at which the power law changes from coefficient $\alpha_1$ to $\alpha_2$.

Comparing to the observed flare frequency distribution (FFD), we added small flares to the population by choosing minimum ED smaller than observed and allow large flares by choosing maximum ED larger than observed. For producing a temporal evolution of stellar activity, we compare the obtained synthetic FFD to the observed AU Mic FFD (\citetalias{2025A&A...700A..53M}), and choose only populations passing a Kolmogorov–Smirnov (K–S) test, which assesses the null hypothesis that both samples are drawn from the same parent distribution. Using these conditions, we produced temporally resolved variable spectra in 36 chunks, each with a duration of 10 days, covering almost a year of stellar activity.

In \citetalias{2026A&A...705A.165M}, we simulated a synthetic spectrum of the stellar atmosphere during a flare event. This spectrum represents flaring on the entire projected surface area of the star.  For each flare, the synthetic spectrum is scaled according to the flare's ED and represents the flux density at the peak of the flare. In physical terms, this corresponds to the flux emerging from the portion of the stellar surface that is actively flaring at the end of the impulsive phase. The temporal evolution of individual flares (i.e., spectra at timestamps before and after the peak) was then modelled using the flare time profile from \citet{2022AJ....164...17T}.

To put this work in context of the previous studies, we also produce stochastic flaring utilising the FF framework \citep{2018ApJ...867...71L,2022ascl.soft02012L} with the AU Mic’s panchromatic spectrum \citep{2022AJ....164..110F} and its implementation of stellar activity. This enables us to evaluate the model-dependent effects of flaring on our atmospheric grid  in the same timescale of 360 days. The FF framework enables the generation of light curves containing a sequence of flare events, where the frequency of flares is determined from the FFD, assumed to follow a single power-law:
\begin{equation}\label{eq:18}
f(>E) = \frac{\beta}{\alpha - 1} E^{-\alpha + 1}.
\end{equation}
Here we assume the power law coefficient $\alpha$=1.8 and to be equal to the second coefficient $\alpha_2$, calculated previously for AU Mic. This value of $\alpha$ is consistent with \citet{2018ApJ...867...71L} results for active M dwarfs.

\subsection{Synthetic flares}
\label{sec:metodssf}

We produced the synthetic populations of stochastic flare events as described in Sect.~\ref{sec:methodcss}, with the largest synthetic flare produced by the YMDF model possessing a bolometric energy of 6.575 $\times$ 10$^{35}$ erg and a corresponding TESS band energy of 1.475 $\times$ 10$^{35}$ erg. In contrast, the largest flare produced by the FF model yields bolometric and TESS energies of 1.061 $\times$ 10$^{35}$ erg and 2.657 $\times$ 10$^{34}$ erg, respectively. For the AU Mic star analysed in \citetalias{2025A&A...700A..53M}, we established flare occurrence rates above a cut-off energy of 10$^{31}$ erg in the Kepler/TESS band, finding a daily rate of 14.3 flares at this threshold. Since more energetic flares follow the secondary power-law slope $\alpha_2$ in the broken power-law distribution we found for this star ($\alpha_1$ = 1.38, $\alpha_2$ = 1.8, see \citetalias{2025A&A...700A..53M} for details), we recalculated the expected occurrence rates for flares at energies comparable to those of our model flares. Over the 360-day span of our simulation, the predicted cumulative rates for such energetic flares are 4.8 events in the FF model and 2.4 events in the YMDF model. These results imply that such flares are plausible within the modelled time frame and that our estimates may be on the conservative side. The flux density spectra of these flares produced YMDF and FF models are presented and discussed in Appendix~\ref{sec:appendixsf}.

\begin{figure}\resizebox{\hsize}{!}{
   \centering
   \includegraphics{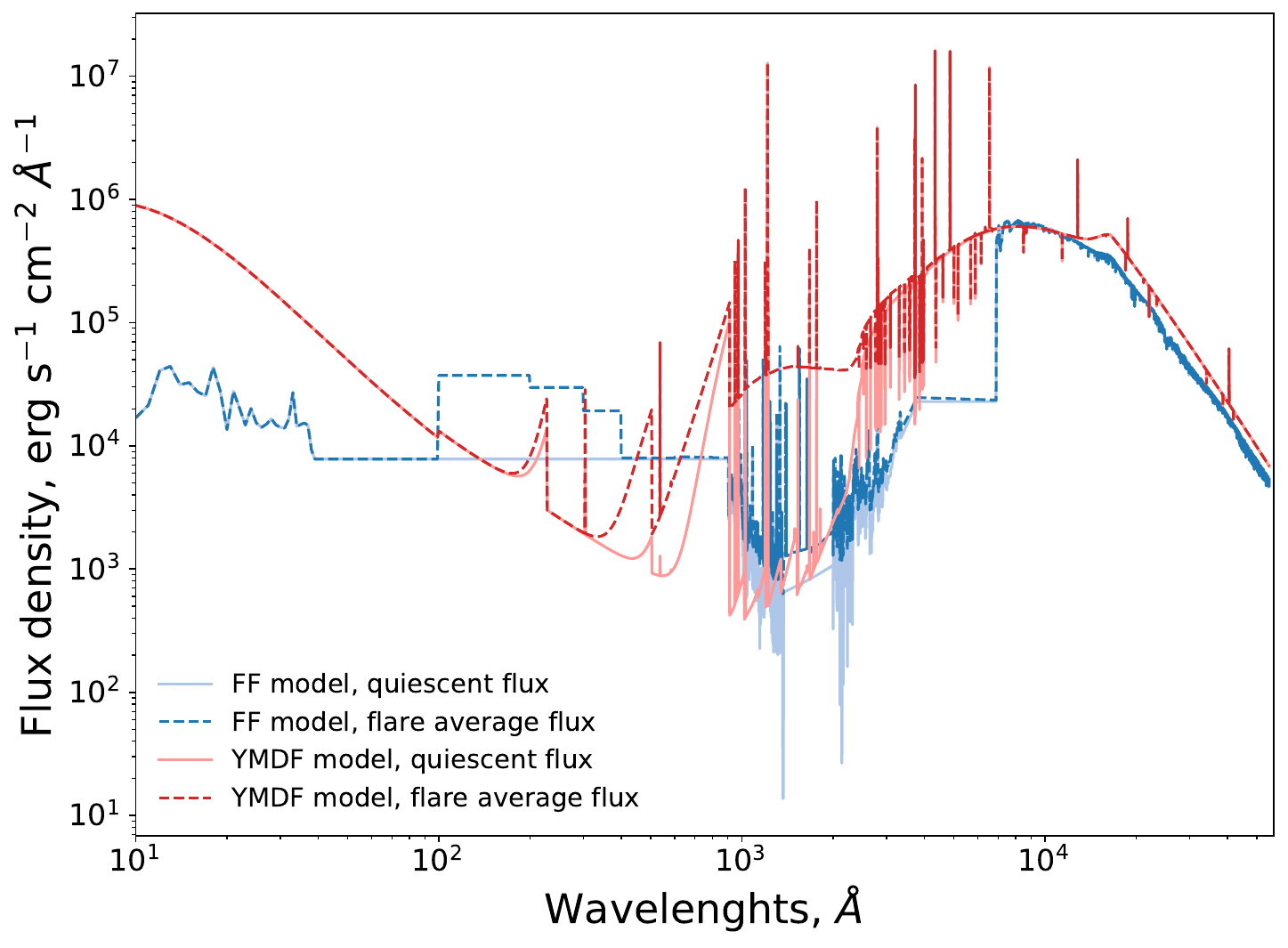}
    }
\caption{Quiescent and averaged flux densities for the FF and YMDF models. Fluxes are shown for the quiescent state (solid lines) and simulation-averaged state (dashed lines), averaged over all flaring events contributions within 360 days for two flare activity models: YMDF model demonstrates a clear UV and NUV flux excess during flares compared to FF model, while longer wavelengths show good agreement across models and small excess in flux density comparing to the quiescent stellar radiation.}
\label{fig:1}
\end{figure}

We computed the average flux densities over the 360-day period for two stellar flare activity models, FF and YMDF. In Fig.~\ref{fig:1}, we present the preflare (quiescent) flux alongside the average flux over this time span. Both models demonstrate enhanced emission in the UV and NUV regimes attributed to the flare events compared to quiescent flux, with the YMDF model producing a higher increase than the FF model. Deviations from the quiescent state are substantially reduced at longer wavelengths in both models. It is expected that M dwarfs flare activity predominantly increases energy input at the shortest wavelengths.

\subsection{Temperature and pressure profiles}
\label{sec:metodstp}
Following the modelling framework presented by \citet{2023ApJ...953...57K}, we adopt their approach to simulate hydrogen–helium–water atmospheres with variable water abundances. Specifically, we generate one-dimensional (1D) temperature-pressure (T-P) profiles in radiative-convective equilibrium using the open-source radiative-transfer code HELIOS\footnote{https://github.com/exoclime/HELIOS} \citep{2017AJ....153...56M,2019ApJ...886..142M}. Consistent with the aforementioned setup, we assume zero albedo and planet-wide heat redistribution, corresponding to a heat redistribution factor of 0.25 (heat is perfectly distributed over both hemispheres). In these calculation we used quiescent panchromatic flux of AU Mic \citep{2022AJ....164..110F} as before. We downloaded the opacity data used in this study from the Database for the Analysis of Complex Envelopes (DACE)\footnote{https://dace.unige.ch/opacityDatabase/}.
Collision-induced absorption (CIA): H$_2$-H$_2$ and H$_2$-He pairs and Rayleigh scattering of H$_2$O, H$_2$, and He are included using established datasets and formulations~\citep[e.g.,][]{2002A&A...390..779B, 2012JQSRT.113.1276R, 2017AJ....154..261C, 2005JQSRT..92..293S}. Including the scattering of water was done with consideration that the H$_2$O cross-section depends directly on the atmospheric abundance of water vapour.

The initial chemical compositions of the model atmospheres were computed under the assumption of thermochemical equilibrium using FastChem\footnote{https://github.com/NewStrangeWorlds/FastChem} \citep{2018MNRAS.479..865S}, which efficiently determines equilibrium abundances for a wide variety of atomic and molecular species across relevant pressure and temperature ranges. In this work, we used FastChem to model 16 atmosphere abundances according to the adopted grid: we sample atmospheric compositions at 100, 60, 40, 25, 15, and 10\% H$_2$O, repeating at one-tenth of these values successively down to 0.1\% H$_2$O. The atomic abundances supplied to FastChem are scaled to reproduce each target H$_2$O mixing ratio by adjusting O/H accordingly, with C, N, and He depleted proportionally, anchored to solar values at 0.1\% H$_2$O. These equilibrium abundances serve as the starting point for radiative-convective modelling with HELIOS.

The HELIOS code iteratively solves for 1D temperature-pressure profiles by balancing radiative transfer and convective adjustment, using a constant quiescent stellar flux. The code applies opacity data and allows for the self-consistent treatment of radiative-convective equilibrium (in this work, under cloud-free assumptions). The radiative transfer is implemented in a two-step process: first, the equilibrium T-P profile is determined using the k-distribution method with 20 Gaussian quadrature points in 384 wavelength bins spanning 0.244--500~$\mu m$, which ensures a sufficiently high resolution treatment of opacity. Then, the planetary emission spectrum is generated from the converged T-P profile using opacity sampling at spectral resolution $\lambda$/$\Delta \lambda$~=1000. For every atmospheric scenario, we employ "on-the-fly" opacity calculations weighted by the atmospheric abundances, calculated previously in FastChem.

Convection within the HELIOS framework is modelled using a convective adjustment scheme, which relies on the adiabatic coefficient governing the lapse rate in convective zones and the specific heat capacity. Therefore, we followed the approach by \citet{2019ApJ...886..142M}, which utilizes ideal gas approximations.

In this study, we assume all temperature-pressure profiles are for cloud-free atmospheres. Given that water constitutes a significant fraction of these atmospheres, we verify the resulting profiles against the thermodynamic conditions for water condensation, either as liquid or solid. Throughout our model grids, all atmospheric conditions remain within the gaseous phase regime.

\begin{figure*}
\sidecaption
\includegraphics[width=12cm]{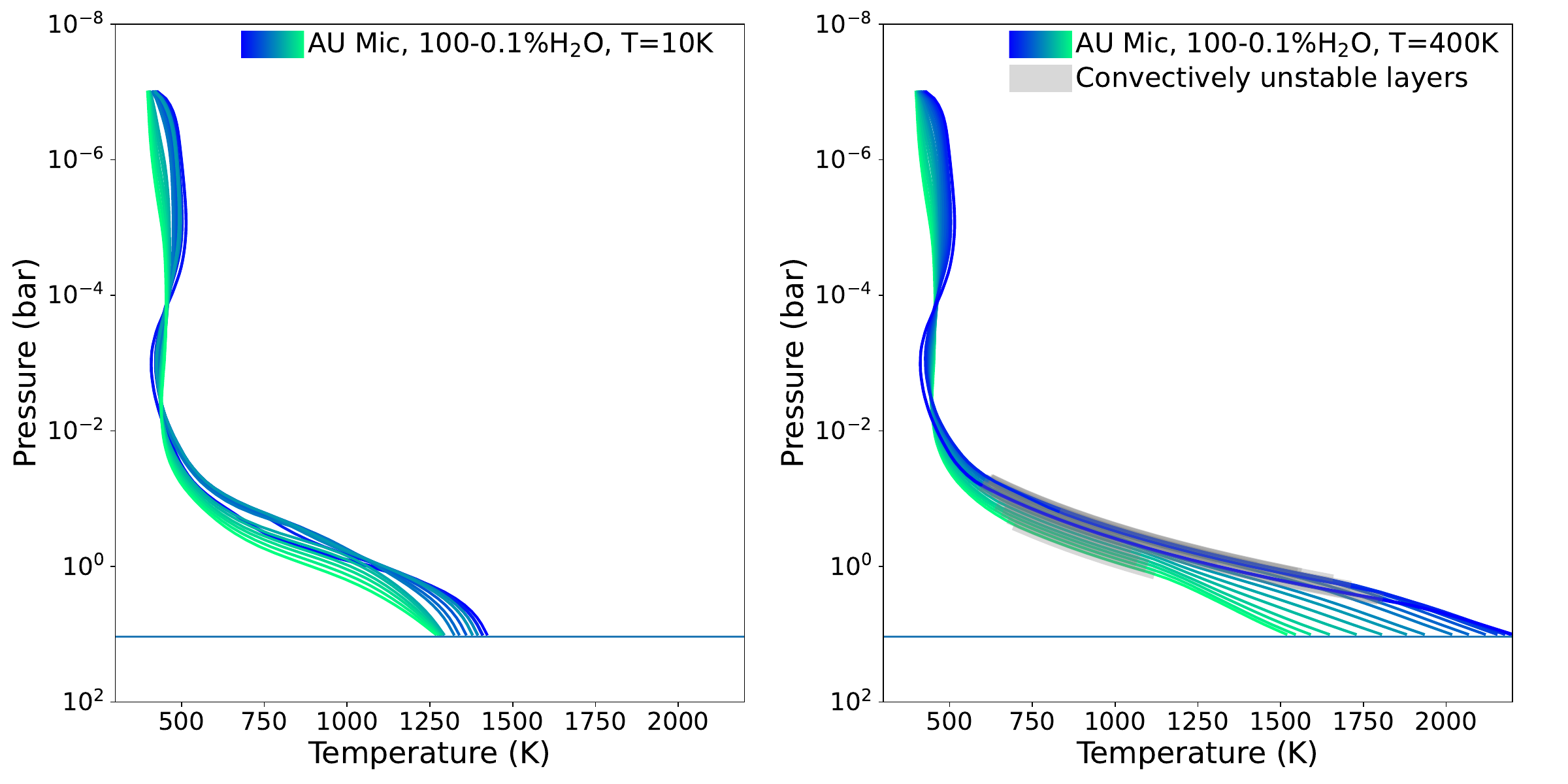}
\caption{Temperature and pressure profiles (x- and y-axis, respectively) for two interior heating scenarios: 10 K (left panel) and 400 K (right panel), computed for quiescent AU Mic irradiation. The profiles represent a hypothetical early exo-Earth with 1~M$_\oplus$, 1~R$_\oplus$ located at an orbital distance producing an equilibrium temperature of 500 K. The atmospheric grid spans primordial compositions with water vapour abundances ranging from 100\% down to 0.1\%, with the remaining fraction composed primarily of hydrogen and helium at solar abundances. Convective zones are displayed as broader light grey shadowed lines. All atmospheric structures were self-consistently calculated using the HELIOS code from equilibrium state reached in Fastchem.}
\label{fig:13}
\end{figure*}

\begin{figure*}\resizebox{\hsize}{!}{
    \centering
    \includegraphics{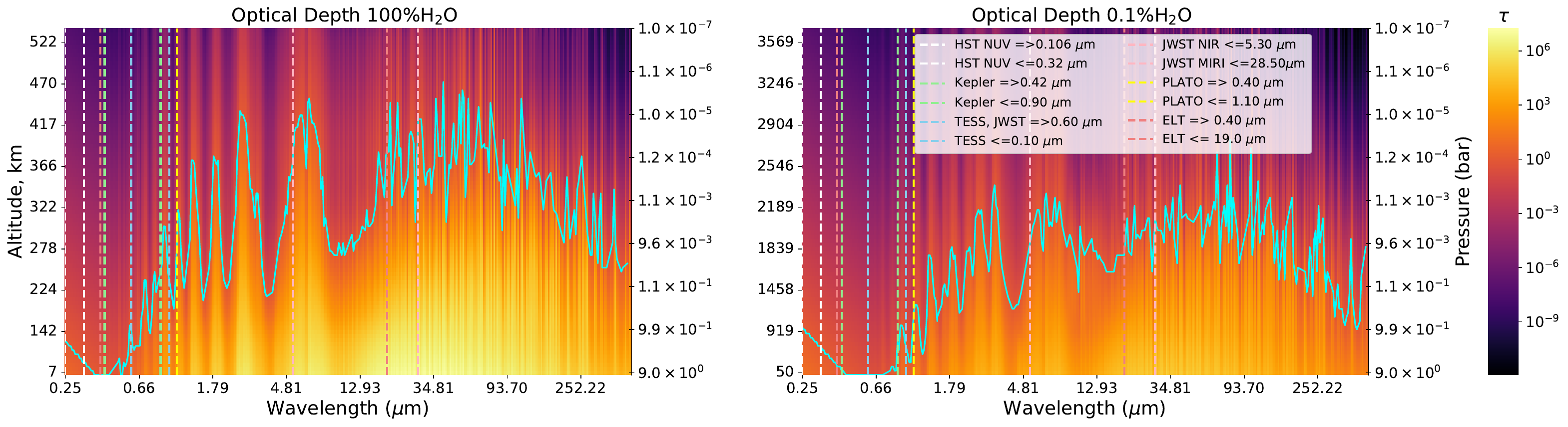}

    }
\caption{Optical depth profiles for the atmospheres with H$_2$O concentrations of 100\% and 0.1\%, calculated using HELIOS (left and right panels, respectively). Light blue lines represent the optical depths $\tau \sim$~1, while the colour map shows varying $\tau$ values across the atmospheres. The x-axis shows wavelength and the y-axis indicates altitude (left) and pressure (right).
Dashed vertical lines mark the observational ranges accessible by instruments onboard the HST (COS, <0.32~$\mu$m), Kepler (0.42--0.90~$\mu$m), TESS (0.60--1.05~$\mu$m), JWST [NIRCam (NIR): 0.60--5.00~$\mu$m, MIRI (MIR): 5.00--28.8~$\mu$m]. Future telescope ranges are stated for ELT (0.35--2.45~$\mu$m) and PLATO (0.40--1.10~$\mu$m).
}
\label{fig:23}
\end{figure*}

In Figure~\ref{fig:13}, we present T-P profiles computed with HELIOS for a hypothetical early exo-Earth planet orbiting our reference star AU Mic under quiescent stellar irradiation. Two interior heating scenarios are shown: low heating at 10 K (left panel) and high heating at 400 K (right panel). Convectively unstable layers are visually highlighted by the broader light grey shading, and indicate regions where the Schwarzschild criterion is violated; these layers are subject to convective adjustment within HELIOS calculations. Figure~\ref{fig:23} illustrates the optical depth profiles on the wavelength-altitude diagram for two atmospheres, that represent the edges of our grid,  100\% H$_2$O (left panel) and 0.1\% H$_2$O (right panel). Key wavelength ranges accessible to current observatories are marked by the dashed vertical lines, including currently operating HST-COS in FUV and NUV (<3200 $\AA$, \citealt{2012ApJ...744...60G}), TESS\footnote{http://www.nasa.gov/tess-transiting-exoplanet-survey-satellite} (6000–10500 $\AA$, \citealt{2016AGUFM.P13C..01R}), JWST instruments in near infra-red (NIR) and mid-IR (6000–288,000 $\AA$, \citealt{2018ConPh..59..251K}), decommissioned Kepler (4200–9000 $\AA$, \citealt{2010Sci...327..977B}), and future ground-based ELT (3500–24500 $\AA$, \citealt{2023ConPh..64...47P}) and space-based PLATO (4000-11000 $\AA$, \citealt{2025ExA....59...26R}) facilities.

The flare energy release occurring at wavelengths shorter than 2464~\AA\ is predominantly absorbed in the upper atmosphere at low pressures and high altitudes. Whether this partial UV deposition has a measurable effect on the temperature-pressure profile is examined below, in Sect.~\ref{sec:resultfa}.

\subsection{Averaged flare flux}
\label{sec:resultfa}

In order to determine whether the addition heating of the upper atmosphere layers is present, we further analysed the impact of the flux of flaring star averaged over the total simulation time compared to the quiescent flux in the solar abundance atmosphere containing 0.1\% H$_2$O. The T-P profile were computed using the k-distribution method as before, calculating over 9,213 wavelength bins spanning 0.02–200~$\mu$m, and we refer to it as the UV setup hereafter. The planetary emission spectrum was generated from the converged T-P profile using opacity sampling at the same spectral resolution, $\lambda / \Delta \lambda = 1000$. This approach allowed us to focus on a treatment of UV opacities while enabling to omit collision-induced absorption, which primarily occurs at greater atmospheric depths where pressures are sufficiently high. Our goal is to assess whether the additional heating will be present in the upper atmospheric layers, that could accelerate atmospheric escape processes and influence mixing dynamics. For opacities calculations in HELIOS, we used the averaged flux densities computed in the YMDF model. The choice was motivated by the fact that the YMDF model yielded an average integrated flux over the 360-day period of $\sim$1.0 $\times 10^{10}$ erg s$^{-1}$ cm$^{-2}$ $\AA^{-1}$, which exceeds the corresponding value of $\sim$0.7 $\times 10^{10}$ erg s$^{-1}$ cm$^{-2}$ $\AA^{-1}$ for the FF model. As our goal was to investigate whether the temperature-pressure profile changes significantly under an increased flux, we thereby selected the model with the higher energy input.

\begin{figure*}\resizebox{\hsize}{!}{
   \centering
   \raisebox{1ex}{\begin{subfigure}{}
      \includegraphics{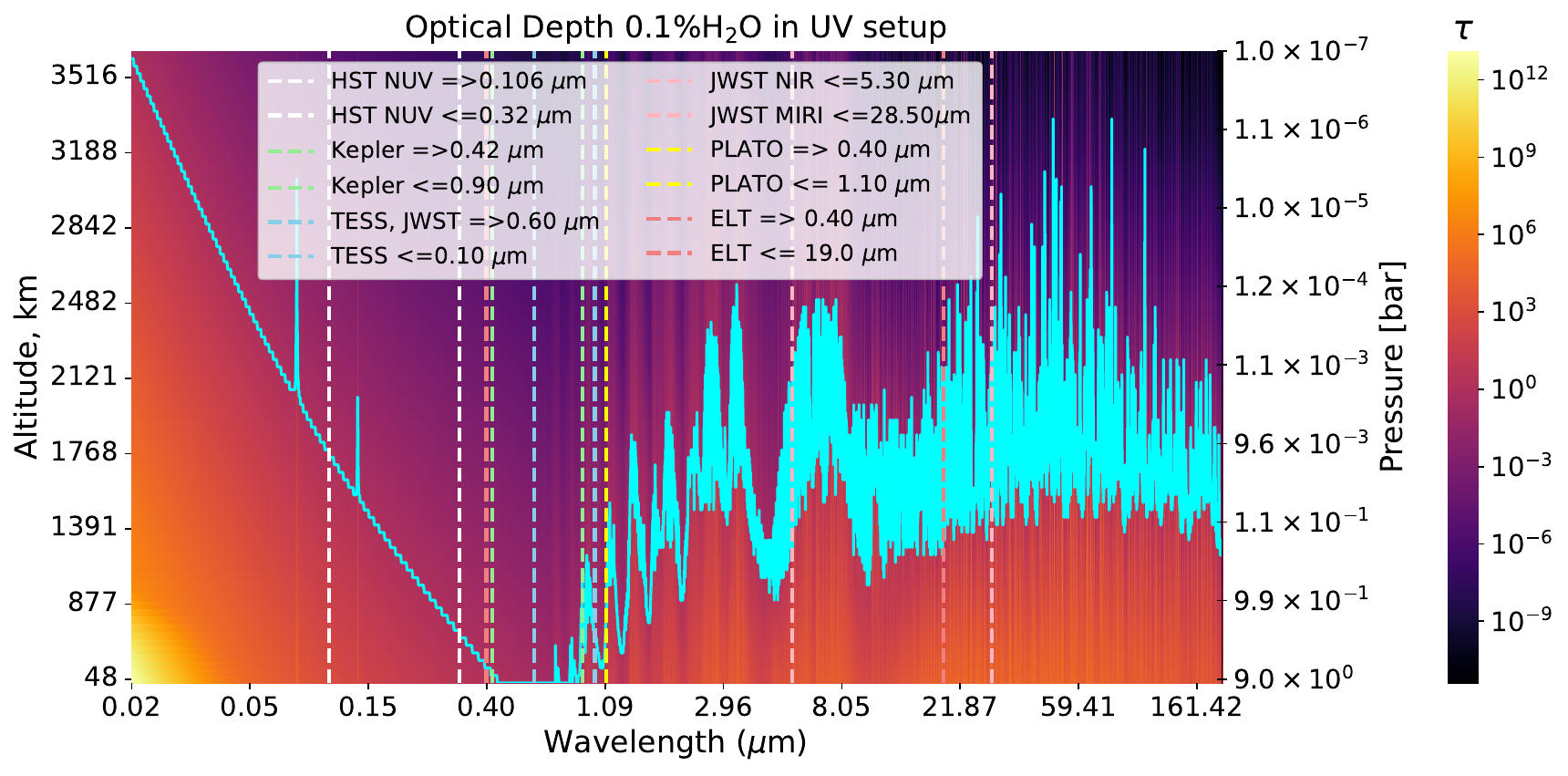}
   \end{subfigure}
      }
\begin{subfigure}
   \centering{}
      \includegraphics{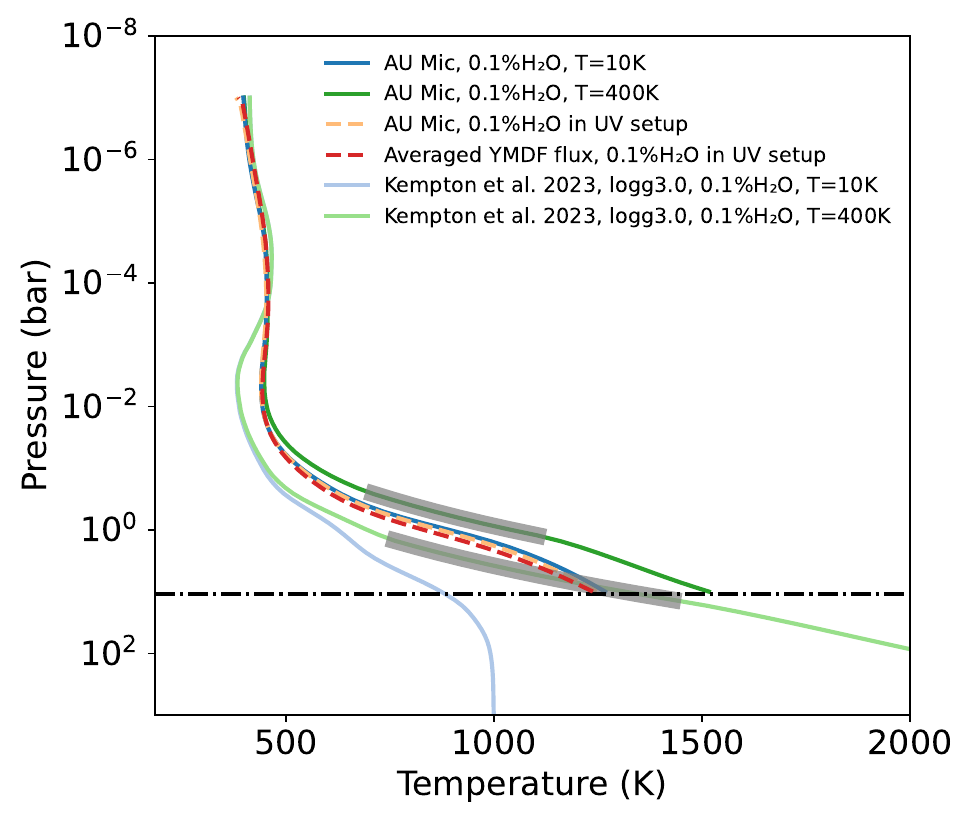}
   \end{subfigure}
   }

   \caption{UV setup implementation for optical depth and TP profile calculations. Left panel: optical depth profile for a 0.1\% H$_2$O atmosphere irradiated by the average flux from AU Mic, representing a flaring young M dwarf reference. As before, the colour map shows varying $\tau$ values across the atmospheres with light blue line as $\tau \sim$~1. Right panel: comparison of T-P profiles, including: the UV setup profiles for the 0.1\% H$_2$O atmosphere under quiescent and flare-averaged fluxes (dashed orange and red lines, respectively); the T-P profile based on a wavelength grid used throughout this study (solid blue line); the same atmosphere heated by an additional 400~K from the interior (solid green line, the light-green shadowed area denotes convectively unstable layers); and a mini-Neptune T-P profile with surface gravity comparable to Earth from \citealt{2023ApJ...953...57K} (solid light-blue line).}

\label{fig:18}
\end{figure*}

The left panel of Fig.~\ref{fig:18} illustrates the optical depth profile (in the UV setup) of the 0.1\% H$_2$O atmosphere irradiated by the flare-average flux from AU Mic, our reference for a flaring young M dwarf. The right panel compares several T-P profiles: the 0.1\% H$_2$O atmosphere under quiescent and flare-averaged fluxes in the UV setup; the wavelength grid T-P profile used throughout this study; a profile for the same atmosphere with an additional 400~K interior heating; and a mini-Neptune profile with a similar surface gravity to Earth ($\log g$= 3.0) and the equilibrium temperature of 500K from \citet{2023ApJ...953...57K}, consistent with the parameters adopted in our study. The T-P profiles computed in the UV setup do not significantly deviate from the T-P profile adopted in this study, particularly in the upper atmospheric layers where photochemical processes predominantly occur. This confirms that the flare energy deposited at wavelengths shorter than 2464~\AA\ affects predominantly the upper atmosphere at low pressures and high altitudes, leaving the deeper atmospheric layers largely unaffected. The discrepancies observed in the lower layers can be attributed to our exclusion of CIA effects from H$_2$-H$_2$ and H$_2$-He pairs, as these processes are not influenced by UV radiation. This simplification reduces computations and allows the layers to converge within the adequate simulation runtime.

\subsection{Chemistry kinetic code setup}

The atmospheric chemical kinetics are modelled using the VULCAN\footnote{https://github.com/exoclime/VULCAN} code \citep{2017ApJS..228...20T,2021ApJ...923..264T}, a 1D thermochemical solver that tracks atmospheric species evolution across diverse temperature and pressure conditions. VULCAN solves the coupled continuity equations for a network of chemical reactions, incorporating photochemistry, thermochemistry, vertical mixing, and molecular diffusion, thereby capturing the interplay between stellar irradiation and atmospheric composition.

We adjust the VULCAN code to work with variable stellar flux\footnote{https://github.com/cepylka/vulcan-flares}. Triggered by the flare event, the atmosphere is being re-built with the changing flare flux at every time step. At the start of the flare event, the integration time step is minimized to enable the atmospheric chemistry to respond rapidly to changes. The time step sizes are controlled adaptively based on an estimate of the truncation error, which typically decreases over time, allowing for progressively larger steps as the system evolves toward numerical steady state as defined in \citet{2017ApJS..228...20T}. To ensure that no flare spectra are missed during the integration, the maximum permissible time step is limited to the interval between the current simulation time and the onset of the next flare event.

In this study, we utilised the N-C-H-O full photochemical kinetics network, which includes nitrogen-, carbon-, and oxygen-bearing species for computing chemical reaction rates. The atomic abundances supplied to FastChem follow the atomic abundance scaling described in Sect.~\ref{sec:metodstp}, where O/H is adjusted to reproduce each target H$_2$O mixing ratio and the remaining species are depleted accordingly. The initial mixing ratios for selected atmospheric setups are presented in Fig.~\ref{fig:93} in Appendix~\ref{sec:appendixb}. Photochemical reaction rates are determined based on the local flux density at each layer, incorporating species-specific cross sections. Thermal reaction rates are computed simultaneously, reflecting the local temperature-dependent kinetics. Both types of reactions are evaluated continuously for every layer given stellar flux variability as we described above.

\subsection{Vertical mixing}
\label{sec:methodsvm}
Determining an accurate diffusion coefficient presents a significant challenge in modelling atmospheric mixing. In the adopted framework, atmospheric mixing in the vertical direction is represented by both eddy and molecular diffusion. Eddy diffusion generally dominates in the lower and intermediate atmospheric layers, where species are relatively well mixed.

In a typical 1D model, the eddy diffusion coefficient, $K_{zz}$ is often estimated using a mixing-length theory (MLT) approach, given by $K_{zz} = C l w_t$, where $C$ is a constant, $l$ is the mixing length, and $w_t$ is the turbulent velocity, under the assumption of free convection. For H$_2$-dominated atmospheres, a range of models \citet{2013A&A...558A..91P,2013ApJ...777...34M,2015MNRAS.446..345M} utilize $K{zz}$ values between $10^8$ and $10^{12}~\mathrm{cm}^2~\mathrm{s}^{-1}$. In this study, the $K_{zz}$ from \citet{2022ExA....53..279M} and \citet{2023MNRAS.521.3333L} is adopted:
\begin{equation}
K_{zz} = 5 \times 10^8 \left( \frac{P}{1~\mathrm{bar}} \right)^{0.5} \left( \frac{H_{1~\mathrm{mbar}}}{620~\mathrm{km}} \right) \left( \frac{T_{\mathrm{eff}}}{1450~\mathrm{K}} \right)^4
\end{equation}
where $P$ is the atmospheric pressure in bars, $H_{1~\mathrm{mbar}}$ is the scale height at 1 mbar, and $T_{\mathrm{eff}}$ is the effective temperature. For more detail, see Appendix \ref{sec:appmlt}.

In the uppermost atmospheric regions, mixing is primarily due to molecular diffusion. Layer-related vertical molecular diffusion coefficients $D_{zz}$ are adopted from \citet{2000Icar..143..244M}, and selected based on the dominant species. Water vapour is considered dominant in atmospheres with 40–100\% H$_2$O, while for the rest of the grid, H$_2$ is the main constituent.

Boundary conditions are specified as follows: the lower boundary is closed (no net flux), while the upper boundary allows for diffusion-limited escape of particles into the exosphere. The surface is treated as a solid boundary, and thermal coupling with the interior is included through the lower boundary temperature. A diffusion-limited flux defines the maximum rate at which particles escape, governed by their rate of diffusion upward into the exosphere, thereby providing an upper bound on potential atmospheric escape rather than a self-consistent escape rate. Further discussion of boundary effects and escape mechanisms appears in Sect.~\ref{sec:resultfa}.

\section{Results}
\label{sec:results}

\subsection{Cumulative changes in water vapour atmospheres of early exo-Earths}
\label{sec:resultfa}

The flare activity response was studied across our grid of water vapour atmospheres. By analysing these relatively long-term changes in comparison to the typical timescales of atmospheric processes, we characterise how sustained flare forcing reshapes their chemical composition and vertical structure. We first describe the end-state atmospheric compositions across the water vapour grid, and then examine the time-dependent behaviour of key species in the upper atmosphere.

\begin{figure*}\resizebox{\hsize}{!}{
   \centering
   \includegraphics{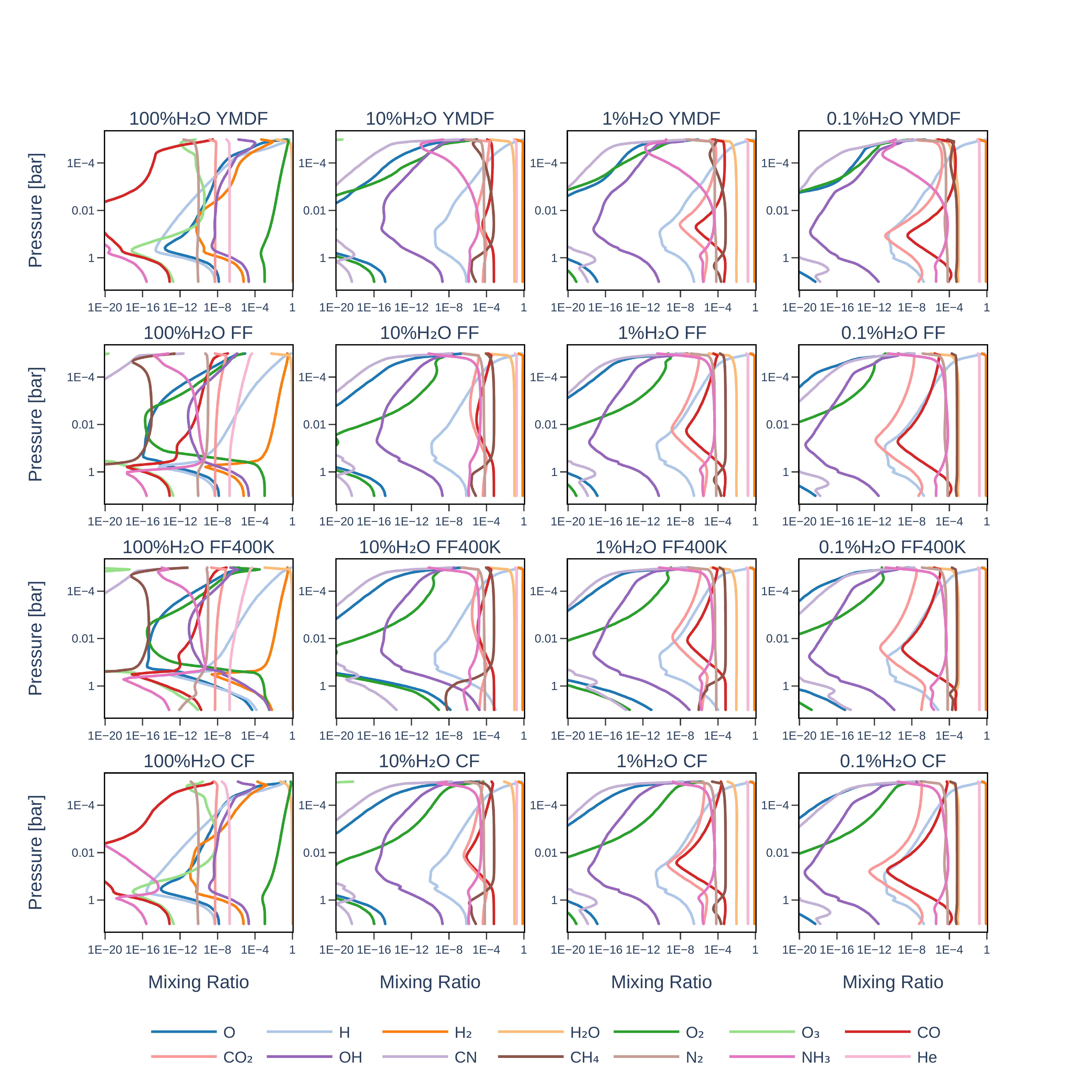}
    }
\caption{Atmospheric chemical compositions after one year of simulation time for four atmospheres with water fractions of 100\%, 10\%, 1\%, and 0.1\% arranged by columns, and four stellar forcing regimes arranged by rows: the YMDF, FF, FF400K, and the CF models. Each panel shows the mixing ratios of key species after one year of chemical kinetic integration. In the constant flux regime it was allowed to the system to achieve a numerical steady state, if occurs before one year period. See supplementary animation (DOI:10.5281/zenodo.18029710, in prep.) for 16 atmospheres under the YMDF model flaring.}
\label{fig:11}
\end{figure*}

In Figure~\ref{fig:11}, we present the chemical compositions of four atmospheres of different in water fractions in columns, while the stellar activity regimes are arranged by rows: the YMDF, FF, FF400K models, and the CF model characterised by a constant stellar flux. Each panel depicts the mixing ratios of key species after 360 days of chemical kinetic integration or upon reaching the numerical steady state, whichever occurs first. Notably, the steady state solution was achieved only for H$_2$-dominated atmospheres under the constant flux regime, requiring between 100 and 222 days (the shortest time observed for simulations with 10\%~H$_2$O and the longest for those with 0.25\%~H$_2$O).

At the high H$_2$O content, in the YMDF and CF models, photodissociation of water by intense radiation continuously produces H and O, with molecular and atomic oxygen accumulating in the upper atmosphere. In contrast, the FF model with frequent, low-energy flares maintain persistently perturbed upper layers of the atmosphere, dominated by hydrogen species. The continuous small flares prevents the atmosphere from recovering between events, suppressing the build-up of oxygen species. At lower H$_2$O content, all models exhibit high mixing ratios of hydrogen species throughout the atmosphere at all altitudes. In these atmospheres, the profiles of pressure and mixing ratios for species such as CO$_2$, and CH$_4$ appear similar across models, with the exception of the YMDF model, which predicts slightly higher concentrations of CO$_2$, and lower levels of methane in the upper atmosphere. CO number densities are generally slightly elevated in hydrogen-dominated atmospheric compositions under the YMDF model, whereas in the 100\% H$_2$O atmosphere the YMDF and CF model yield noticeably lower CO abundances compared to the FF models. The YMDF and CF models allow oxygen species to persist in the upper atmospheric layers, reflecting differences in the time-resolved flare forcing on the atmosphere. These results demonstrate that the FFDs of stellar flares play a critical role in governing the long-term redox balance of primordial atmospheres under sustained stellar irradiation, and underscore the necessity of incorporating stellar flare statistics into the interpretation of exoplanetary spectra.

The lower layers of the atmosphere influenced by the interior heating in the FF400K model show that species such as H, H$_2$, H$_2$O, and He remain largely unaffected, whereas atomic oxygen and O$_2$ are present in increased abundances, and methane is depleted at this depths throughout the grid of the water vapour concentrations, if compared to simulations without internal heating in the FF and CF models. It highlights that the heating conditions at the lower boundary primarily affect the deeper parts of the atmosphere, altering the abundances there. Nonetheless, in the investigated grid of atmospheres, the mixing ratios in the upper atmosphere remain largely unaffected by the interior heating, indicating limited coupling between temperature-dependent kinetics in the lower atmosphere and photolysis-driven chemistry in the upper atmosphere. We additionally examined the absolute number density profiles across all models and for two extreme compositions (100\% and 0.1\% of water content), finding agreement with the mixing ratio analysis presented above. The number density profiles, together with further discussion, are provided in Appendix~\ref{sec:appendixnd} and Fig.~\ref{fig:37}.

\begin{figure*}\resizebox{\hsize}{!}{
   \centering
   \includegraphics{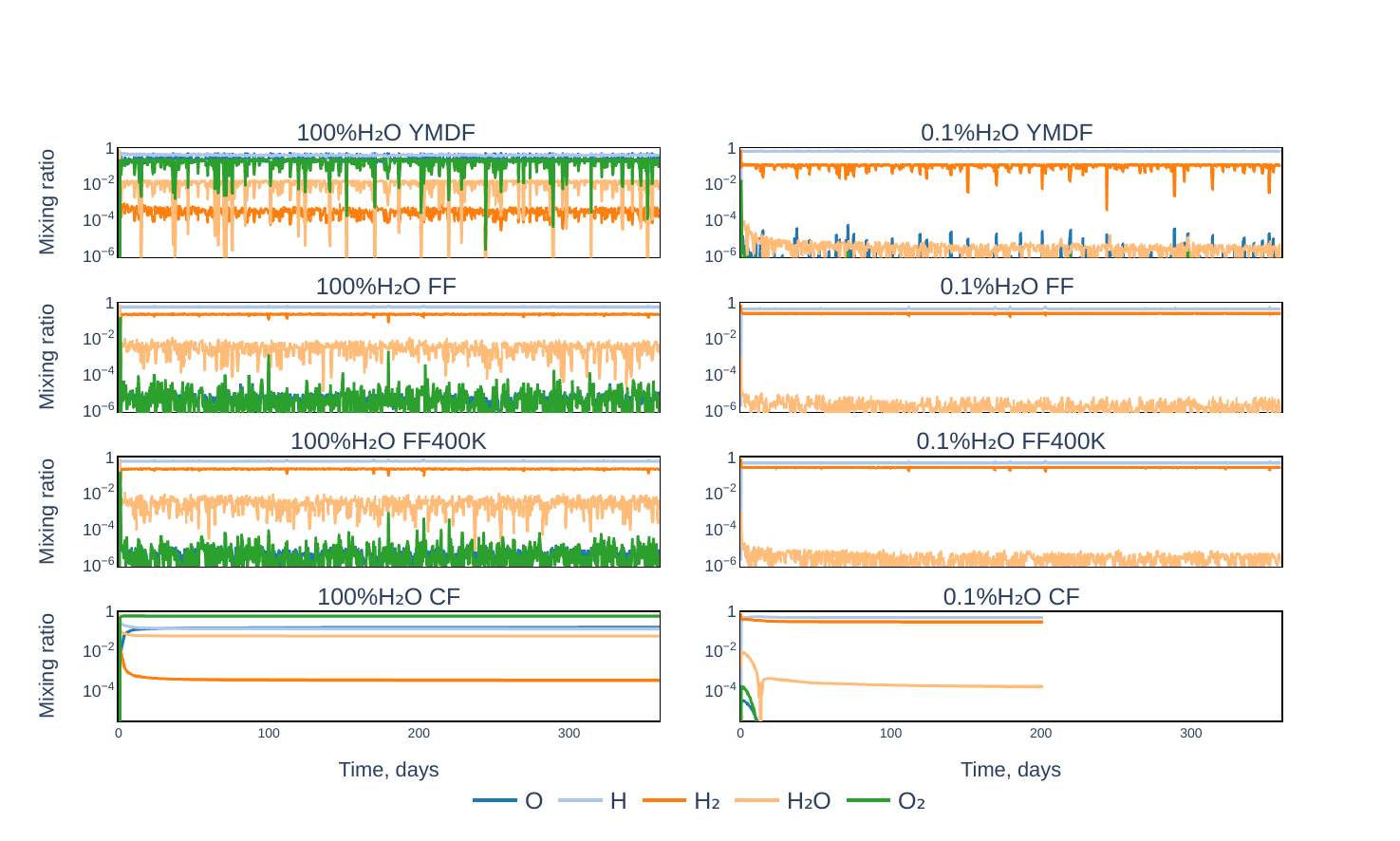}
    }
\caption{360-days evolution of the mixing ratios of photodissociation-driven species with concentrations above 10$^{-4}$  in the uppermost atmospheric layer are shown for four atmospheres with water fractions arranged by rows: 100\%, 10\%, 1\%, and 0.1\%. The left and right columns correspond to the YMDF and FF models, respectively. Each panel presents the temporal evolution of species mixing ratios over the span of one year, with the y-axis showing mixing ratio and the x-axis representing time. The CF model at 0.1\%H$_2$O shows the temporal evolution until the model reaches a steady state at day 201 of the simulation.}
\label{fig:2}
\end{figure*}

The temporal evolution of the mixing ratios of photodissociation-driven species in the uppermost atmospheric layer are presented in Fig.~\ref{fig:2} at the upper (100\%) and lower (0.1\%) extremes of water content, arranged by columns, while the rows correspond to the models. All models with variable stellar flux input exhibit rapid changes in species mixing ratios associated with flares, while the CF model, after some initial variation, stabilises to constant mixing ratios as expected. The YMDF and FF models effectively produce and transport H to the uppermost atmospheric layers, where it becomes available for diffusion-limited escape. The FF and FF400K models produce very similar results regardless of water content, suggesting that the photochemical composition of the upper atmosphere is not significantly influenced by interior heating.

In the uppermost layer of the 100\% H$_2$O atmosphere, H$_2$ reaches mixing ratios of 0.01--0.20 in the FF and FF400K models, whereas it remains at trace levels in the YMDF and CF models. The CF model sustains high O$_2$ with some atomic O, minimal H$_2$, and $\sim$20\% atomic H over one year, while the FF is dominated by H ($\sim$70\%) and H$_2$ ($\sim$30\%). The YMDF shows interchangeable O and O$_2$ abundances (20--60\%), reflecting dynamically stressed conditions from mid-sized flares. In the 0.1\% H$_2$O case, water mixing ratios hover around 10$^{-5}$ across variable flux models (slightly higher, $>10^{-4}$, in CF), with H and H$_2$ dominant; the YMDF uniquely exhibits sporadic atomic O ($\sim$10$^{-4}$) from flares, and H$_2$ lower than others. Across the grid, water vapour is efficiently reduced to $\sim$1\% of initial values there, driven by both flare activity in all variable flux models, and, to a somewhat lesser extent, under constant stellar flux.

Our simulations reveal that atmospheres comprised almost entirely of water vapour (we denoted it as 100\%H$_2$O) under the YMDF and CF models exhibit significant amount of molecular and atomic oxygen in the uppermost layer. Interestingly, the FF and FF400K models results indicate a markedly different atmospheric composition at higher altitudes. Here, atomic and molecular hydrogen rapidly dominate the upper layers of the atmosphere, along with a dramatic depletion of oxygen species to parts-per-billion (ppb) levels, leaving permille concentrations only in the lower part of the atmosphere. The atmospheres affected by the YMDF model output exhibit the presence of atomic O up to 10$^{-4}$ concentrations even in solar abundances atmosphere (0.1\%H$_2$O). This stark divergence suggests that the increased production of low-energy flares profoundly alters the upper atmospheric chemistry, which can be understood in terms of the balance between radical species production and recycling. Photodissociation of water (H$_2$O~+~$h\nu$~$\rightarrow$~H~+~OH) provides the primary source of hydrogen radicals, which rapidly recycle hydrogen through reactions forming H$_2$ and other reduced species. In the YMDF scenario, episodic high-energy flare events enhance photolysis rates and increase the production of atomic oxygen, facilitating the build-up of O and O$_2$ in the upper atmosphere. In contrast, the frequent but lower-energy events in the FF scenario maintain a more hydrogen-dominated radical environment, where oxygen radicals are more efficiently recycled back into water or other reduced species, limiting the accumulation of oxidised products. Therefore, the frequency and distribution of stellar flares can be crucial factors influencing the production or inhibition of oxygen species in primordial atmospheres.

To further explore the effect of additional interior heating on the atmospheres, we simulated 100\% and 0.1\% H$_2$O atmosphere scenarios with an interior heating of 400K in the YMDF model, shown in Fig.~\ref{fig:9} in Appendix~\ref{sec:appendixymdf}. The additional heating accelerates reaction kinetics in the lower atmosphere, with most changes compared to the model without interior heating occurring there. Hydrogen and oxygen species have slightly elevated mixing ratios there. In the upper layers, changes are negligible with the exception of the uppermost layers, where the lower right panel of the figure reveals a sporadic presence of both O and O$_2$ at mixing ratios around $\sim$10$^{-4}$, analogous to atmospheres without additional interior heating. For further discussion and additional analyses, see Appendix~\ref{sec:appendixymdf}.

\subsection{Spontaneous flare-induced changes in water vapour atmospheres of early exo-Earths}
\label{sec:resultsc}

\begin{figure*}
\resizebox{\hsize}{!}{
   \centering
   \includegraphics[width=0.95\textwidth]{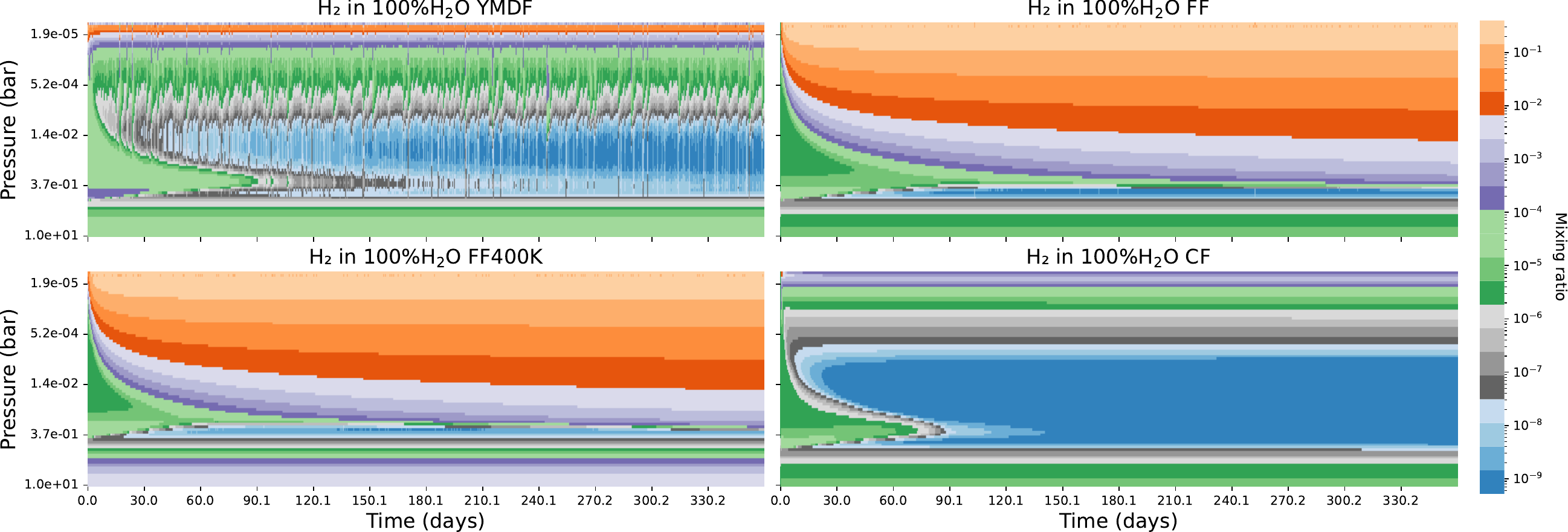}
    }
\caption{Time evolution of H$_2$ mixing ratio for an atmosphere with 100\% H$_2$O. The heatmaps display vertical profiles over the 360-day simulation duration, with pressure (in bars) on the y-axis and time (in days) on the x-axis. The upper row shows results from the YMDF and FF models, while the bottom row corresponds to the FF 400 K and the CF model, as indicated on the panels.
}
\label{fig:299}
\end{figure*}

In water-rich compositions, H$_2$O itself contributes significant opacity in the ultraviolet, acting as a self-shielding species that buffers photolysis rates at higher pressures even when the incident NUV/FUV flux varies appreciably. Additional absorbers, such as CO$_2$ or hydrocarbons, may further modulate the penetration of short-wavelength radiation into deeper atmospheric layers. In the 100\% H$_2$O atmospheres, water shows effectively no changes in the mixing ratios except for uppermost layers due to shielding; however, the initially low H$_2$ concentrations react remarkable different in this regime under the YMDF and FF activity models. In the former H$_2$ concentrations are visibly disturbed by flare activity, and in the latter, it slowly accumulates in the upper and middle atmosphere, as shown in Fig.~\ref{fig:299}. This behaviour can be understood as follows: species with low mixing ratios are particularly susceptible to flare-driven perturbations, an effect most pronounced under the YMDF model where mid-sized, energetic flares perturb trace species across a broad altitude range. Conversely, species gained higher mixing ratios by efficient background production show little or no detectable response under the FF model, where predominantly low-energy flares are insufficient to overcome background production rates.

\begin{figure*}\resizebox{\hsize}{!}{
   \centering
   \includegraphics{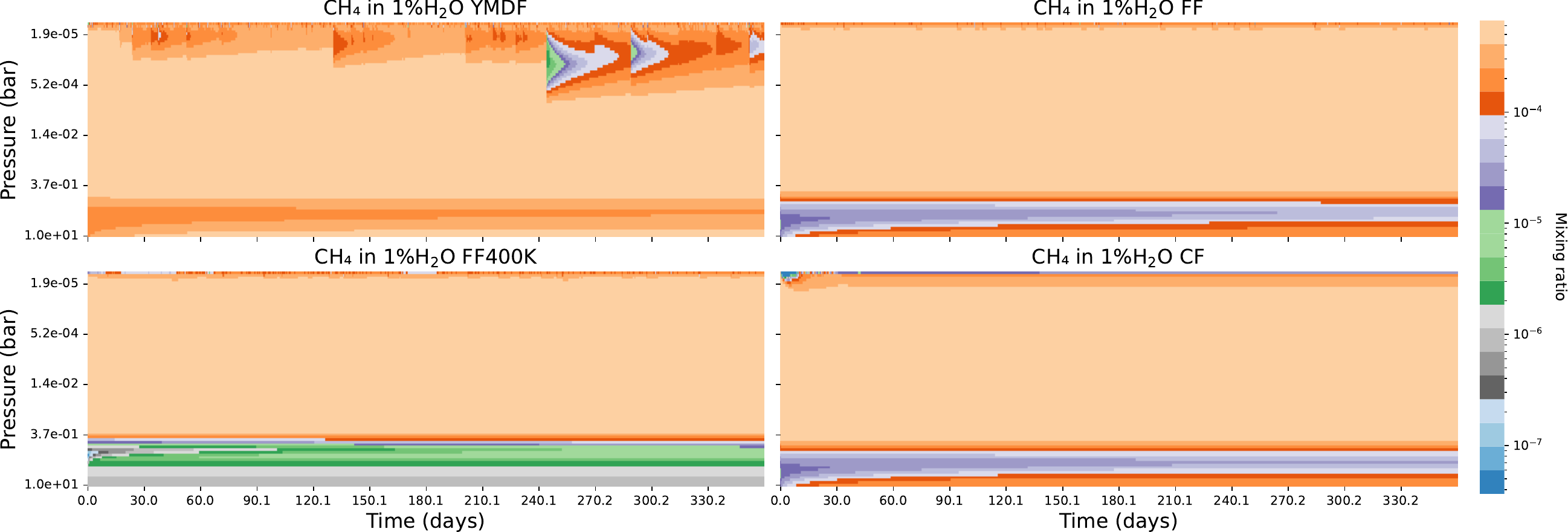}
    }
\caption{Heatmaps of CH$_4$ mixing ratio in the atmosphere with 1\% H$_2$O over the 360-day simulation period for an atmosphere containing 100\% water vapour analogous to the previous Fig.~\ref{fig:299}}
\label{fig:25}
\end{figure*}

The magnitude and vertical distribution of photolysis rates are governed by the wavelength-dependent actinic flux that directly reflect changes in the stellar flux due to flare activity, and the absorption cross-sections of the dominant atmospheric species (see Appendix \ref{sec:appendixrt} for details). In the upper atmosphere, variations in the stellar UV flux during flare events produce noticeable changes in photolysis rates for several species (see Fig.~\ref{fig:99}); however, deeper in the atmospheric column, strong UV shielding substantially suppresses any further enhancement of photolysis rates. The enhanced CH$_4$ abundances in the lower atmosphere under the YMDF model are consistent with flare-driven vertical redistribution from upper to lower layers, whereas in the FF model, persistent but shallow radical production depletes upper atmospheric CH$_4$ without significantly perturbing the deeper layers beyond the state established by the quiescent flux. The FF400K case shows even stronger CH$_4$ depletion at depth, which hints on thermal nature of this effect.

The four panels in Fig.~\ref{fig:25} represent heatmaps of the methane evolution in 1\% H$_2$O vapour atmospheres. The results reveal a pronounced response of the CH$_4$ mixing ratio to the flare input of the YMDF model corresponding to the most energetic flare event (at 243.671 day of the simulation), characterised by sharp temporal variations. Although a full return to preflare conditions is not observed within that the simulation timespan, it is inferred that the atmosphere would eventually re-equilibrate given the clear trend in the mixing ratio, if we assume the absence of subsequent large flares. However, this recovery is likely to be interrupted by the first successive very energetic flare. Longer simulation durations would be required to confirm this behaviour conclusively.

In contrast, the atmospheric response to the FF model stellar flaring, shows limited response to the most energetic flare in the model (at the 136.069 days of the simulation) under these conditions. However, as demonstrated in Fig.~\ref{fig:72}, the temporal evolution of O$_2$ in the 10\% H$_2$O atmosphere reveals that the response to flare activity remains pronounced in the FF model as well. When we consider the FF400K model, the atmospheric response to flares becomes more pronounced, suggesting that increased thermal conditions at the bottom of the atmosphere amplify the chemical perturbations induced by stellar activity by increasing reaction rates and vertical mixing. Both panels for the response on the FF and FF400K models show that slightly less methane accumulate in the lower atmosphere, and with introduced interior heating the effect is more pronounced. The response to the CF model displays the evolution of CH$_4$ with reaching a steady solution at $\sim$209 day under stable stellar forcing conditions. An analogous atmosphere response was observed in the temporal evolution of ammonia (NH$_3$) mixing ratios (see Fig.~\ref{fig:27} in Appendix~\ref{sec:appendix1}). For these two species, the long-term response to the largest flares in the corresponding simulations is pronounced throughout our atmospheric grid, with the exception of the 100\% H$_2$O atmosphere, where the high abundance of water vapour suppresses the other species to minimal concentrations, preventing their accumulation over the duration of the simulation. This behaviour emerges as larger portions of the atmosphere approaching solar abundances due to decreasing water content. The effect similar to the persistent changes in CH$_4$ and NH$_3$ across all water concentrations in hydrogen-dominated atmospheres (H$_2$O < 40\%), is also observed at somewhat lesser extent in other species including CO, CO$_2$, H, and OH under stellar forcing of the YMDF model (see Fig.~\ref{fig:288} in Appendix~\ref{sec:appendix1}). Reduced species such as CH$_4$ and NH$_3$ are particularly sensitive to flare-driven photochemistry, as they are efficiently destroyed through reactions with radical species produced during water photolysis, leading to their rapid conversion into the reactive intermediates CH$_3$ and NH$_2$.

Water abundance is exceptionally uniform throughout the atmospheric grid and models, showing a comparable decrease in mixing ratio at higher altitudes, with the sole exception of the 0.1\% H$_2$O atmosphere under the YMDF model, where self-shielding likely becomes ineffective under sufficient stellar flaring (see Fig.~\ref{fig:28} in Appendix~\ref{sec:appendix1}). Here, the strongest flare in the simulation induces atmospheric reactions observable possibly attributable to the minimal water content in this atmosphere. Consequently, across the grid, water vapour acts as the stable and primary source of opacity in the relevant atmospheric layers, emphasising its dominant role in controlling radiative transfer processes within these environments, while H$_2$ contributes through CIA, and He contributes through Rayleigh scattering and CIA in combination with H$_2$. We additionally find that the mixing ratio of H$_2$ remains largely stable from atmospheres with 60\% H$_2$O down to 0.1\% H$_2$O, except in the uppermost layers where dynamics are observed, even in the CF model.

\subsection{Diffusion limited escape}
\label{sec:resultfa}

The upper layers of the atmosphere are available for particle escape driven by both thermal and non-thermal mechanisms, including hydrodynamic outflow caused by X-UV and EUV heating, as well as non-thermal escape such as sputtering, charge exchange from stellar winds, and photochemical losses. In our modelling framework, the atmosphere is simulated in the pressure regime from bottom 10 bar to top 10$^{-5}$. This upper boundary lies near the exobase, the altitude where the atmosphere transitions from collisional to collisionless flow and where escape mechanisms become dominant. We, therefore, focus on the diffusion-limited escape regime where molecular and eddy diffusion control species transport. The escape of light species (e.g., H, H$_2$, He) can be limited by their availability at the top of the atmosphere; this supply is modulated by vertical diffusion from the homopause \citep{2021ApJ...923..264T}, leading to a diffusion-limited escape with the flux:

\begin{equation}
\Phi_{\mathrm{DL}} = D_t n_i \left( \frac{1}{H_t} - \frac{m_i g}{N_{\mathrm{Avo}} k_B T} \right)
\end{equation}

where $\Phi_{\mathrm{DL}}$ denotes the diffusion-limited flux, $D_t$ is vertical molecular diffusion coefficient D$_{zz}$ calculated for the uppermost layer, $n_i$ is the number density of species $i$ there, $H_t$ is the scale height and $T$ is the temperature, all evaluated at the top of the atmosphere, $m_i$ is molecular mass of species $i$, and the constants $N_{\mathrm{Avo}}$, and $k_B$ represent the Avogadro's constant and Boltzmann’s constant, respectively. We note that the direct proportional relationship between diffusion-limited flux and number density should be noted. Additional information is presented in Appendix~\ref{sec:app3} for completeness.

\begin{figure*}\resizebox{\hsize}{!}{
   \centering
   \begin{subfigure}{}
      \includegraphics{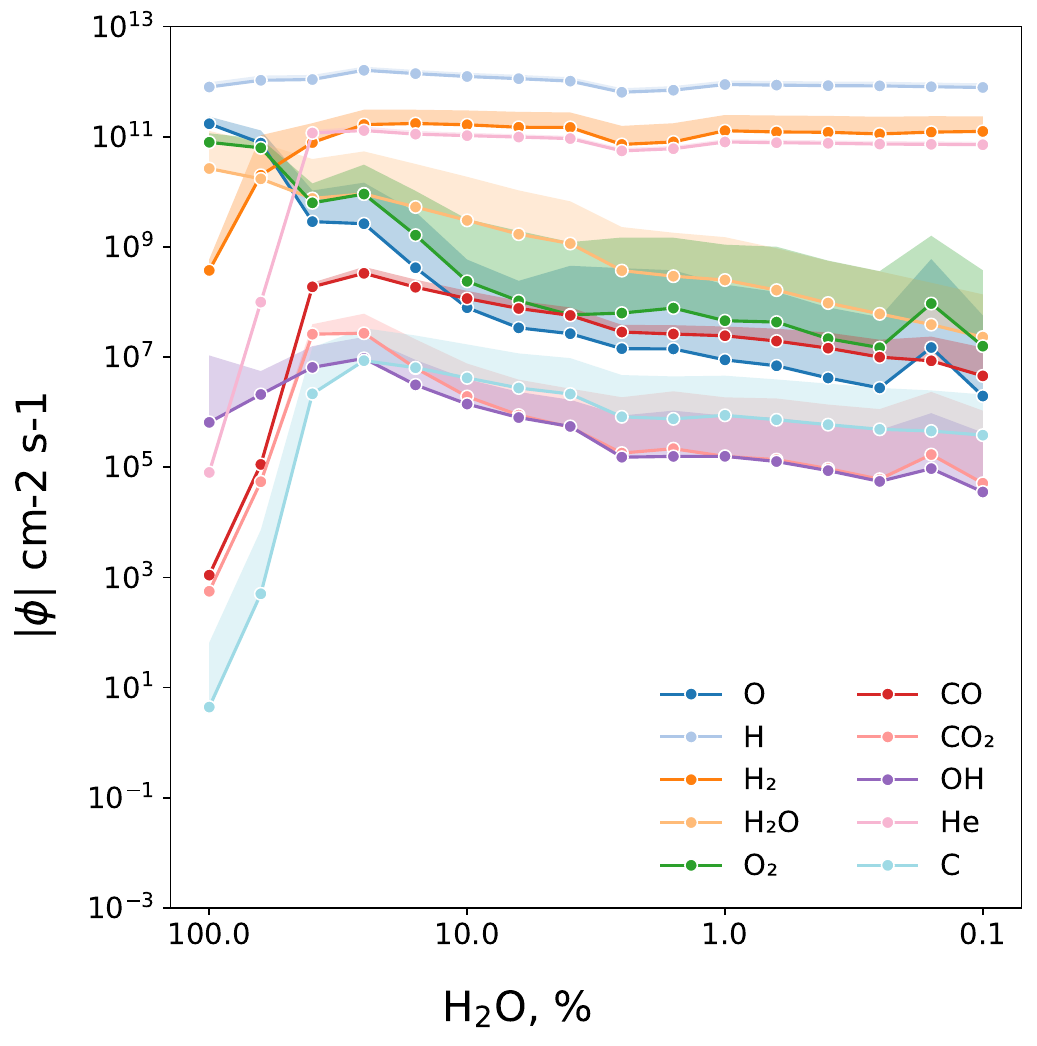}
   \end{subfigure}
   \begin{subfigure}
   \centering{}
      \includegraphics{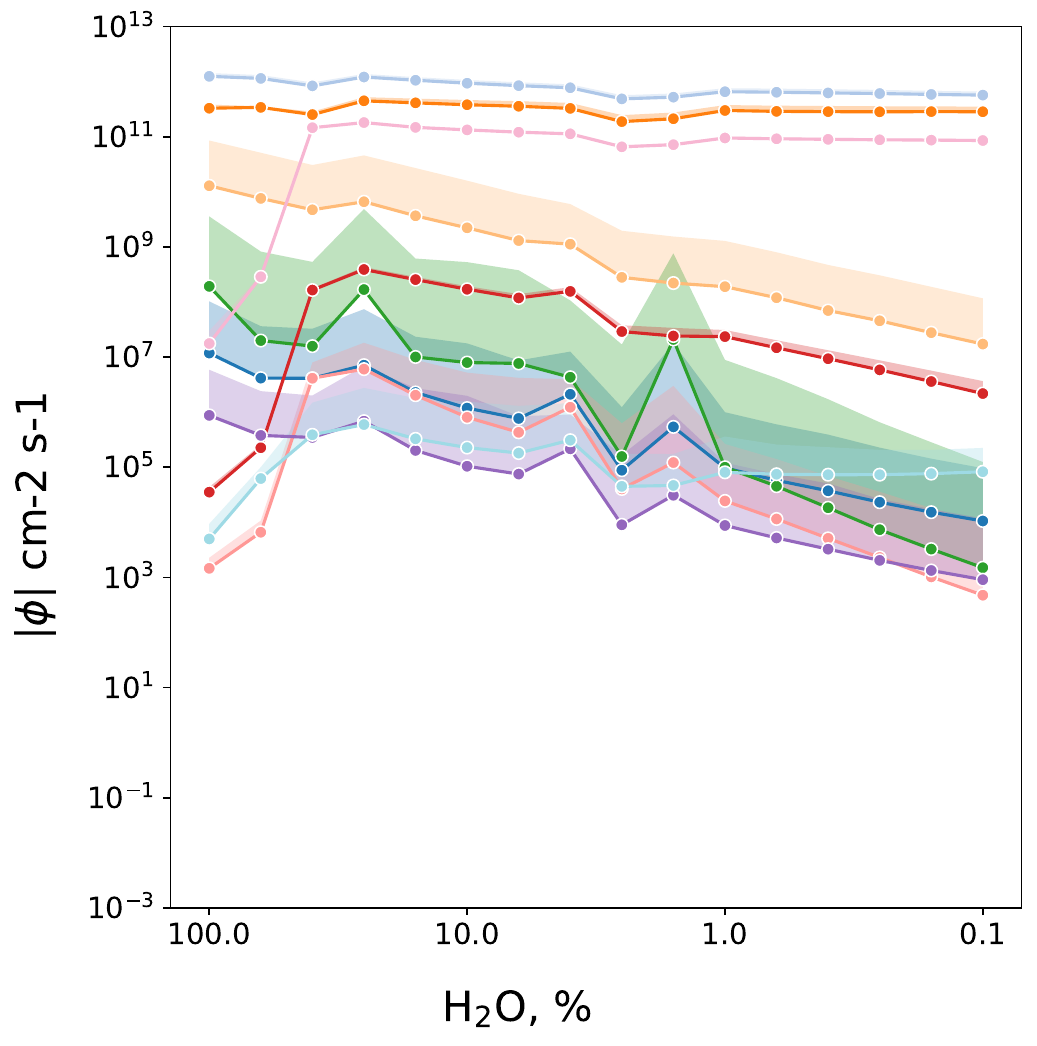}
   \end{subfigure}
   \begin{subfigure}
   \centering{}
      \includegraphics{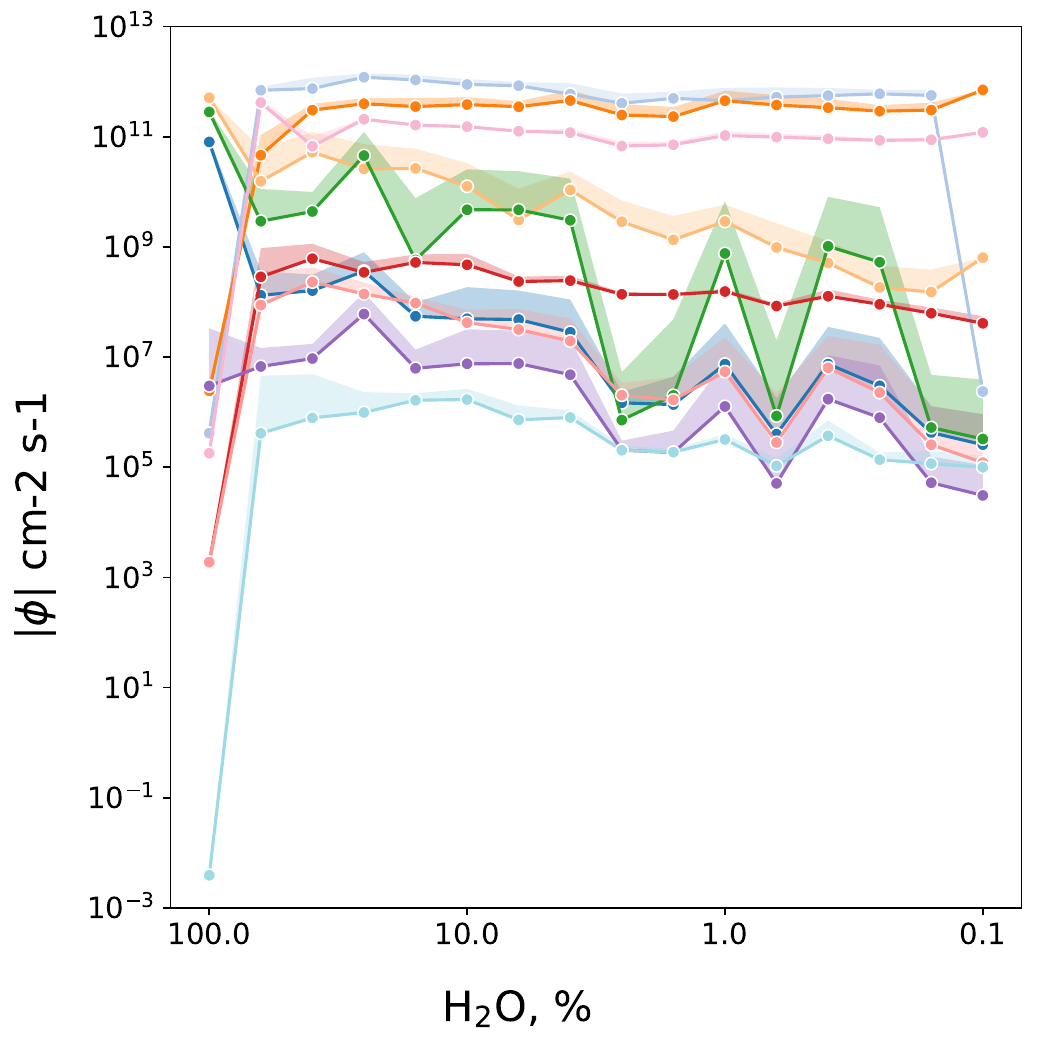}
   \end{subfigure}
   }
   \caption{Diffusion-limited escape averaged over one year for the principal atmospheric species across a range of atmospheres with water vapour concentrations varying from 100\%-0.1\%. The left panel shows results of the calculations using the YMDF variable flux, the centre panel presents the FF model application, and the right panel shows averaged rates for the simulation with the constant stellar flux. The coloured envelopes show the standard deviation of the upper bounds only; the lower bounds are omitted for clarity as they are symmetric.}

\label{fig:4}
\end{figure*}

Figure~\ref{fig:4} shows the diffusion-limited escape rates averaged over the 360-day period for key atmospheric species across a broad spectrum of atmospheres characterised by water vapour concentrations from 100\% down to 0.1\%,  for the YMDF, FF (left and centre panel, respectively) along with the CF model (right panel). This figure reflects the upward flux at the uppermost atmospheric layer, weighted by the species-specific thermal diffusion factor $\alpha$ (see Appendix~\ref{sec:app3}) and the local number density, and is intended to inform expectations for layers closer to the exobase rather than to quantify escape rates directly. Therefore, it characterises the reservoir of atmospheric species available for loss below the exobase.

Wider envelopes, representing the a standard deviation from the mean values, indicate enhanced variability resulting from species intermittently moving into and out of the upper layer. When species momentarily leave, the mixing ratio can approach zero, causing fluctuations. Conversely, narrow envelopes signify a stable, continuous presence of species, indicating uninterrupted availability for diffusion-limited escape.

The variability with H$_2$O content in all models arises from the ability of even the quiescent AU~Mic flux (CF model) to photodissociate water, driving an intermittent production of O, O$_2$, and OH in the upper layers; atomic C is also marginally affected but remains at low number densities due to rapid conversion into more complex carbon-bearing species. The stark transition between the 100\% and 60\% H$_2$O regimes reflects the shift from a water-dominated atmosphere, where all other species exist only at trace abundances, to a compositionally richer regime where H, He, C, N, and O are more substantially represented, though still following solar ratios within the non-water fraction. Under the YMDF model, energetic flares drive enhanced photodissociation of H$_2$O, producing greater variability in oxygen-bearing species, whilst H$_2$ shows the largest variability, reflecting the stronger overall atmospheric stress. The FF model, by contrast, shows greater relative variability in O and O$_2$, as its predominantly low-energy flares incrementally perturb oxygen radical abundances without imposing comparable stress on H$_2$. Towards the 0.1\% H$_2$O end of the grid, hydrogen-bearing species converge across models, however, in the CF model H$_2$ replaces atomic H as the dominant escaping species, whilst flaring models continue to dissociate H$_2$ and transport atomic H upward. The systematically lower upward flux of oxygen-bearing species in FF relative to YMDF reflects the less effective production of O, O$_2$, and OH under low-energy flaring.

\section{Discussion}
\label{sec:discussion}

\subsection{Primordial atmosphere of an early exo-Earth}

The population of known terrestrial exoplanets orbiting very young M dwarfs at ages of 5-10 Myr remains sparse, primarily due to the observational challenges posed by these super-young and highly active systems \citep{2016ApJ...821L..19N,2015MNRAS.448.3053A}. Such stars are often embedded in dusty environments and exhibit intense magnetic activity, which complicates detection and characterization efforts. Consequently, empirical constraints on the atmospheric compositions and evolutionary pathways of terrestrial planets in these nascent systems are limited.

The most important constituent in the explored grid of atmospheres is water, owing to its pivotal influence on habitability as well as atmospheric radiative properties and photochemical pathways. Our results therefore characterize the atmospheric conditions preceding potential habitability. Interestingly, as shown in Fig.~\ref{fig:101}, an Earth-like planet with a similar equilibrium temperature, as we used in our simulations, orbiting a star younger than our reference star AU Mic would enter the habitable zone as the star settles onto the main sequence. Therefore, if we hypothesize that these atmospheres adequately represent systems at an early evolutionary stage, our simulations with varying water content can be linked to the potential future presence of liquid water reservoirs on the planetary surface, once conditions become favourable for water condensation. For example, in \citetalias{2026A&A...705A.165M}, we found that flares in cooler M dwarfs with longer evolution timescales, such as Karmn J07446+035 (YZ CMi, spectral type M4.5), are well represented by the YMDF model. Further research is required to evaluate whether the YMDF model accurately captures the flare characteristics of young M dwarfs.

The chemical behaviour observed across the atmospheric grid can be understood in terms of several dominant photochemical processes. In all cases, photolysis of H$_2$O in the upper atmosphere provides the primary source of H, O, and OH radicals under both quiescent and flaring conditions. The subsequent fate of oxygen is governed by the local redox environment: in hydrogen-rich upper atmospheres, O and O$_2$ are efficiently recycled back into reduced species through reactions with H and H$_2$, suppressing long-term oxygen accumulation, whereas in compositions where hydrogen is less abundant or more readily lost, recombination pathways allow O and O$_2$ to persist. Carbon-bearing species respond primarily to enhanced UV-driven destruction during flare events, with CH$_4$ being particularly susceptible owing to its rapid photolysis, while CO and CO$_2$ are comparatively more stable and can be replenished from deeper layers through vertical mixing. The interplay between these competing processes, namely photolysis, radical recycling, and vertical transport, governs the time-dependent composition of the upper atmosphere and accounts for the contrasting chemical outcomes observed across the different flare scenarios.

Throughout the atmospheric grid, certain species, both inert, such as helium, and reactive, like water vapour, maintain consistent mixing ratios across all pressure layers, except in the uppermost layer, where even in the CF model, stellar radiation is sufficient to deplete these species. Conversely, other species become increasingly responsive to flare activity in the hydrogen-dominated atmospheres. This effect is most pronounced in the YMDF model, but nearly absent in the FF and FF400K model.

Numerous complex interactions occur in the lower atmosphere; for instance, redox reactions between a hot silicate surface and a steam-rich atmosphere can profoundly influence the oxidation states of both the surface and atmospheric species. Oxygen radicals generated from water vapour tend to oxidize surface minerals, altering their chemistry, while reducing agents such as H$_2$ and CO drive the reduction of metal oxides, fostering intricate chemical cycles at this interface. These processes critically modulate atmospheric composition and surface mineralogy, impacting volatile cycling and planetary habitability \citep{2020ChEG...80l5594F, 2020PSJ.....1...11Z,2012ApJ...761..166H}. Additionally, the present simulations employ a simplified N-C-H-O chemical network, excluding metal-bearing and sulphur-containing species (e.g., SiO, Na, K, FeO, sulphur compounds). These are vital for understanding processes in the secondary planetary atmospheres \citep{2023MNRAS.525.3703G}, which are theorized as the next stage in the evolution of terrestrial planet atmospheres after the primordial phase we focused on in this study. Our simulation setup does not fully incorporate processes related to surface–atmosphere exchange and lower atmospheric chemistry, which introduces inherent limitations to interpreting the results.

The simulations within the atmospheric grid suggest a distinct discontinuity between the upper and lower atmospheric layers, resulting in their chemical and physical responses being partially decoupled. Similar findings were reported in \citealt{2023MNRAS.521.3333L,2025ApJ...993...41G,2022A&A...667A..15K,2023MNRAS.523.5681N}. Stellar radiation, particularly the UV flux from flares, predominantly affects the upper layers, with species of low mixing ratios showing signs of this impact extending deeper. Although our simulations are parametrised for H$_2$--H$_2$O-dominated atmospheres, several of the controlling mechanisms are expected to be generic. The depth of perturbation is set by UV and self-shielding, while long-term evolution depends on the balance between radical production, chemical recycling, and vertical transport. Quantitative results will differ for other compositions, but the dependence on flare statistics and shielding efficiency should hold broadly.

\citet{2023MNRAS.521.3333L} similarly employed VULCAN with a stochastic flaring routine on M-dwarf exoplanet atmospheres, finding gradual accumulation of chemical perturbations in species such as H, OH, and CH$_4$ in hydrogen- or nitrogen-dominated atmospheres of similar T$_\mathrm{eq}$, consistent with our result that trace species at low mixing ratios are most susceptible to flare-driven changes under the YMDF model. \citet{2022A&A...667A..15K} found that high-energy flares permanently modify upper atmospheric composition in a tidally locked gas giant using a pseudo-2D framework; whilst their target differs substantially from our water-rich primordial grid of terrestrial planets, the persistent upper-atmospheric perturbations are likely analogous, and a similar upper and lower atmosphere decoupling was recovered. \citet{2023MNRAS.523.5681N} also demonstrated that temperature--chemistry feedback plays an important role in determining long-term atmospheric evolution under flares, particularly for cooler planets, with the UV radiation is absorbed at lower pressures in planets with similar T$_\mathrm{eq}$ to our setup compared to the hotter planets.

Comparing models with and without interior heating within our grid, such as FF and FF400K, the layering becomes more evident: the additional heating primarily alters chemistry in the lower layers without significant upward propagation. Similarly, whilst the YMDF model produces long-term changes in the upper atmosphere, the YMDF400K setup confines the effect on mixing ratios predominantly to the lower atmosphere. This could indicate that processes not explicitly accounted for in the model, such as surface redox chemistry or outgassing, exert minimal influence on the upper atmospheric composition, particularly in hydrogen-dominated primordial atmospheres with solar elemental abundances. Consequently, the apparent decoupling between upper and lower atmospheric chemical regimes remains a robust feature within the framework of the current model assumptions.

Our results further indicate that flare activity does not contribute significantly to heating the upper atmospheric layers; instead, photochemistry plays the dominant role. This is consistent with \citet{2025ApJ...993...41G}, who demonstrate, using a brown dwarf as a hydrogen-rich analogue, that a clear discontinuity between upper and lower atmospheric regions persists and that despite significant photochemical effects the thermal influence of even the most energetic flares remains minimal, hereby hinting on the applicability of our approach with our T-P profile is held fixed, following the negligible thermal response to UV irradiation shown in Section~\ref{sec:resultfa}. However, self-consistent feedback across our diverse composition grid remains an avenue for future work.

Finally, \citet{2021NatAs...5..298C} used a 3D coupled chemistry--climate model to show that recurring flares drive rocky planet atmospheres in older M dwarf systems into persistent chemical disequilibrium, with flare-driven signatures of bio-indicating species potentially detectable with future instruments;  \citet{2025AJ....170...40C} further showed, using a 3D general circulation model with interactive photochemistry, that cumulative flare effects are largely determined by flare frequency whilst instantaneous effects depend on spectral shape and energy, consistent with the contrasting behaviours we observe between the YMDF and FF models. Our results extend this picture to a broader primordial composition space, though their 3D framework captures dynamical effects beyond our current 1D treatment, motivating future extensions of this work.

\subsection{FFDs may shape primordial upper atmospheres of early exo-Earths}

The largest flare simulated in the YMDF model exhibits a bolometric energy approximately 6.2 times greater and a TESS band energy about 5.5 times greater than the corresponding flare in the FF model. The larger flare amplitude simulated in the YMDF compared to the FF model can be attributed to limitations in the model flare spectrum, as detailed in \citetalias{2026A&A...705A.165M} (see, e.g., Figure 3, right inset, specifically for 6000-7000 $\AA$ spectral range). These limitations stem from the FF model's approach to handling an incomplete observed panchromatic spectrum combined with a simplified box-car temporal flare evolution. It should be noted that such deficiencies are inherent wherever observational coverage of the full spectral range is lacking. The FF model fundamentally relies on observed panchromatic fluxes as input, and any gaps or limitations in this observational dataset can introduce uncertainness in the simulated flux density at specific temporal points during the flare event, thereby affecting the accuracy of the integrated flare energy in the simulation. Throughout the atmospheric grid in this study, the FF model effectively induces flare-related changes confined to much shallower layers in the upper atmosphere compared to the YMDF model. The deficiency in flare energies simulated in the FF model limits its capacity to produce significant atmospheric changes, whereas the variable flare energies in the YMDF model enable changes that may persist over longer timescales. It remains plausible that if the FF model adequately captured these higher-energy flares, it could similarly drive such persistent atmospheric modifications.

In this work, we studied the atmospheric scenarios and found that the FFDs used in atmospheric modelling affect the mixing ratios of species in primordial early exo-Earth atmospheres around M dwarfs. The YMDF model applies a broken power-law FFD, representing flare behaviour typical of younger M dwarfs as established in \citetalias{2025A&A...700A..53M}, while the FF model uses a single power-law FFD based on the observations of older M dwarf population \citep{2018ApJ...867...71L}. Under similar assumption about planet-star geometry, planetary parameters and atmospheric conditions, the YMDF and FF stellar flare forcing lead to distinct atmospheric responses. The flare energy distribution critically governs the photochemical evolution of the upper atmosphere. Frequent low-energy flares efficiently deplete the limited H$_2$O reservoir in the tenuous, radiation-dominated upper layers, driving the local chemistry toward an H-rich, O-poor state; once this shallow layer is cleared of water, subsequent small flares contribute little additional photodissociation, while oxygen-bearing photoproducts are redistributed to deeper layers where recombination is more effective. By contrast, less frequent but more energetic flares penetrate deeper into the atmosphere, probe larger water reservoirs, and sustain photodissociation over a broader altitude range, maintaining a non-negligible population of oxygen-bearing species and preserving a more balanced local H/O chemistry. These results imply that it is the shape of the flare energy distribution, rather than the time-averaged high-energy flux alone, that determines whether water-rich upper atmospheres evolve toward transient H-dominated layers or retain observable oxygen signatures at altitudes accessible to future transmission spectroscopy.

The accumulation of O$_2$ to substantial levels in some of our simulations (up to 10\% in the 100\% water-vapour-dominated configuration) raises the question of biosignature false positives \citep{2018AsBio..18..630M}. However, the planetary configurations considered in this study are not amenable to atmospheric characterisation with James Webb Space Telescope (JWST, \citealt{2006SSRv..123..485G}) and ESA's Ariel \citep{2018ExA....46..135T} missions, which are both targeted to detect spectral features in larger planets at these orbital separations. Future facilities will, however, be capable of addressing this directly. The Habitable World Observatory (HWO)\footnote{https://habitableworldsobservatory.org/home} will achieve $R\sim150$ visible resolution at 10-pc distances, enabling simultaneous O$_2$/O$_4$ collision-induced absorption observations at 0.34--0.7~$\mu$m. LIFE\footnote{https://life-space-mission.com/}, a proposed mid-infrared nulling interferometer array, will characterise thermal emission ($R\sim50$, 4--18.5~$\mu$m) from nearby temperate terrestrial planets, with co-detection of O$_3$ (9.6~$\mu$m) and CH$_4$ (7.7~$\mu$m) as biosignatures.

The FF and YMDF flare frequency distributions, therefore,  produce markedly divergent upper atmospheric compositions with potential implication for the future observations.
The divergence originates in how each FFD shapes the time-dependent energy input, governing both the production rates of reactive species (see Appendix~\ref{sec:appendixrt}) and their vertical transport throughout the atmospheric column. These differences propagate into substantially distinct chemical compositions and structural profiles in the upper atmosphere, with potential consequences for long-term atmospheric evolution. For photochemical studies of young M-dwarf systems, the YMDF model is the more physically motivated choice: the contrast with the FF model is sufficiently pronounced that adopting the latter would risk substantially underestimating flare-driven perturbations in the primordial regime most relevant to early atmospheric evolution. The YMDF framework therefore provides a stronger basis for interpreting future observational data and for exploring how stellar flaring shapes the evolutionary pathways of terrestrial exoplanet atmospheres.

\subsection{Atmospheric escape and volatile retention}

Using our modelling approach, we can probe young M dwarf planets by simulating the evolution of their atmospheric compositions and potentially predicting the fate of water within explored parameter space. In Fig.~\ref{fig:4}, we presented the diffusion-limited escape flux averaged over one year for the main atmospheric species. It is important to emphasise that these values represent not the actual escape rates but rather quantify the diffusion-limited supply of species below the exobase, which marks the altitude where escape processes may operate.

Atmospheric escape processes can be categorised into thermal escape, controlled by stellar radiation including X-UV, EUV, and IR fluxes (accounted for in this study), and non-thermal escape mechanisms, driven by the complex interplay between the stellar wind and the planetary atmosphere, including magnetic field interactions (for a comprehensive review on atmospheric escape, see \citealt{2025RvMPP...9...18H}). These processes are inherently a three-dimensional phenomenon influenced by complex processes such as day-night temperature asymmetries, planetary magnetic fields, and spatial variations in stellar irradiation. 1D modelling, while simplify many of these features, have proven to be extremely valuable for long-term planetary evolution studies where computational efficiency and parameter exploration are essential and capture the fundamental physics governing mass-loss rates and upper atmospheric structure while enabling systematic investigations across a broad parameter space \citep{2018ApJ...866L..18K,2018A&A...617A.107J,2019MNRAS.490.3760A}. Therefore, utilising 1D kinetic chemistry models can be beneficial, but it is essential to couple them with advanced atmospheric escape modelling approaches to comprehensively account for multiple loss processes.

In this study, we calculated the diffusion-limited escape rates at the exobase of the similar atmospheres to \citet{2022ApJ...934..137Y}. They showed that H$_2$O vapour and photochemical products enhance radiative cooling in hydrodynamic outflows around M-dwarf HZ terrestrial exobase, reducing H escape by ten times at basal H$_2$O/H$_2$=0.1, potentially preserving H$_2$/H$_2$O beyond the pre-main-sequence runaway greenhouse phase.
Coupling diffusion-limited escape calculations that incorporate that incorporate modelled stellar variability in young stars with hydrodynamic escape simulations above the exobase would be a natural next step toward a more comprehensive model of atmospheric loss.

Our results suggest that water is remarkably persistent: mixing ratios remain stable throughout the atmospheric column below the exobase over the full one-year span of our simulations, with the exception of the uppermost layers. Water self-shields against photodissociation, allowing volatile retention over long timescales even under active flaring conditions. This has direct implications for volatile retention and long-term habitability potential on planets orbiting M dwarfs, hinting at a potentially available water reservoir should conditions on the planet become habitable, as illustrated in Figure~\ref{fig:101}.

By understanding the evolutionary pathways in these primordial atmospheres with varying water content, we can predict the observational signatures expected in systems evolving toward habitable planets.  However, to achieve this, it is necessary to improve our model by incorporating additional atmospheric escape processes beyond the current state-of-the-art, thereby providing a more comprehensive account of water loss mechanisms and planetary evolution.

\section{Conclusion}

We explored the parameter space of atmospheric chemical kinetics in primordial (0.1-100\%H$_2$O) atmospheres of early exo-Earth, orbiting active M dwarfs. Due to observational challenges, empirical constraints on stellar activity remain limited, which motivated the development of the Young M dwarf flare model capable of producing temporally resolved flare fluxes. Given that M dwarf activity predominantly occurs in the UV range, our simulations provide predictions for the potential development of liquid water reservoirs on planets as their host stars evolve toward the main sequence. Flare activity primarily affects atmospheric chemistry in the upper layers, while the lower atmosphere appear decoupled from the photochemistry driven upper envelopes. These findings therefore provide a framework applicable beyond the specific atmospheric regime modelled, with flare statistics and shielding efficiency acting as key controls on chemical evolution in terrestrial atmospheres more generally.

Flare frequency distributions play a role in shaping upper atmospheric chemistry, with the YMDF model capturing high-energy and mid-size flares that induce deeper and potentially persistent atmospheric changes compared to the `fiducial flare' model. Our diffusion-limited escape calculations at the exobase provide important constraints on an atmospheric species transport, establishing a lower bound on atmospheric loss under hydrodynamic escape conditions. Future work should couple diffusion-limited escape calculations that incorporate stellar variability with hydrodynamic simulations including radiative cooling, thereby developing integrated atmospheric escape models.

It is essential to understand water loss and atmospheric evolution during the prolonged runaway greenhouse phase. The long-term retention of water below the exobase, sustained by photochemical self-shielding and robust against active flaring, implies that a significant volatile reservoir may persist on such worlds, with direct consequences for habitability assessments of M-dwarf exoplanet systems.
Exploring the observational detectability of the predicted atmospheric signatures presents a promising future direction and warrants dedicated investigation in subsequent analyses. Future facilities such as HWO and LIFE will enable discrimination of the abiotic O$_2$ accumulation identified here from biologically sustained disequilibria, through co-detection of complementary biosignature tracers. Addressing the gaps in current modelling approaches will enhance our predictive capabilities regarding atmospheric composition evolution, planetary habitability, and observational signatures in young, active M dwarf exoplanet systems.

\begin{acknowledgements}
We acknowledge financial support from the Research Council of Norway (RCN), through its Centres of Excellence funding scheme, projects number 332523 (PHAB, Centre for Planetary Habitability) and number 262622 (RoCS, Rosseland Centre for Solar Physics). We thank an anonymous referee for helpful comments and critiques, which have significantly improved the quality of this manuscript.
\end{acknowledgements}

\bibliography{bibliography}

@article{2013ApJ...762...41G,
    title        = {{Objects in Kepler's Mirror May be Larger Than They Appear: Bias and Selection Effects in Transiting Planet Surveys}},
    author       = {{Gaidos}, Eric and {Mann}, Andrew W.},
    year         = 2013,
    month        = jan,
    journal      = {\apj},
    volume       = 762,
    number       = 1,
    pages        = 41,
    doi          = {10.1088/0004-637X/762/1/41},
    eid          = 41,
    archiveprefix = {arXiv},
    eprint       = {1211.2279},
    primaryclass = {astro-ph.EP},
    adsurl       = {https://ui.adsabs.harvard.edu/abs/2013ApJ...762...41G}
}

@article{2010Sci...327..977B,
    title        = {{Kepler Planet-Detection Mission: Introduction and First Results}},
    author       = {{Borucki}, William J. and {Koch}, David and {Basri}, Gibor and {Batalha}, Natalie and {Brown}, Timothy and {Caldwell}, Douglas and {Caldwell}, John and {Christensen-Dalsgaard}, J{\o}rgen and {Cochran}, William D. and {DeVore}, Edna and {Dunham}, Edward W. and {Dupree}, Andrea K. and {Gautier}, Thomas N. and {Geary}, John C. and {Gilliland}, Ronald and {Gould}, Alan and {Howell}, Steve B. and {Jenkins}, Jon M. and {Kondo}, Yoji and {Latham}, David W. and {Marcy}, Geoffrey W. and {Meibom}, S{\o}ren and {Kjeldsen}, Hans and {Lissauer}, Jack J. and {Monet}, David G. and {Morrison}, David and {Sasselov}, Dimitar and {Tarter}, Jill and {Boss}, Alan and {Brownlee}, Don and {Owen}, Toby and {Buzasi}, Derek and {Charbonneau}, David and {Doyle}, Laurance and {Fortney}, Jonathan and {Ford}, Eric B. and {Holman}, Matthew J. and {Seager}, Sara and {Steffen}, Jason H. and {Welsh}, William F. and {Rowe}, Jason and {Anderson}, Howard and {Buchhave}, Lars and {Ciardi}, David and {Walkowicz}, Lucianne and {Sherry}, William and {Horch}, Elliott and {Isaacson}, Howard and {Everett}, Mark E. and {Fischer}, Debra and {Torres}, Guillermo and {Johnson}, John Asher and {Endl}, Michael and {MacQueen}, Phillip and {Bryson}, Stephen T. and {Dotson}, Jessie and {Haas}, Michael and {Kolodziejczak}, Jeffrey and {Van Cleve}, Jeffrey and {Chandrasekaran}, Hema and {Twicken}, Joseph D. and {Quintana}, Elisa V. and {Clarke}, Bruce D. and {Allen}, Christopher and {Li}, Jie and {Wu}, Haley and {Tenenbaum}, Peter and {Verner}, Ekaterina and {Bruhweiler}, Frederick and {Barnes}, Jason and {Prsa}, Andrej},
    year         = 2010,
    month        = feb,
    journal      = {Science},
    volume       = 327,
    number       = 5968,
    pages        = 977,
    doi          = {10.1126/science.1185402},
    adsurl       = {https://ui.adsabs.harvard.edu/abs/2010Sci...327..977B}
}

@article{2014PASP..126..398H,
    title        = {{The K2 Mission: Characterization and Early Results}},
    author       = {{Howell}, Steve B. and {Sobeck}, Charlie and {Haas}, Michael and {Still}, Martin and {Barclay}, Thomas and {Mullally}, Fergal and {Troeltzsch}, John and {Aigrain}, Suzanne and {Bryson}, Stephen T. and {Caldwell}, Doug and {Chaplin}, William J. and {Cochran}, William D. and {Huber}, Daniel and {Marcy}, Geoffrey W. and {Miglio}, Andrea and {Najita}, Joan R. and {Smith}, Marcie and {Twicken}, J.~D. and {Fortney}, Jonathan J.},
    year         = 2014,
    month        = apr,
    journal      = {\pasp},
    volume       = 126,
    number       = 938,
    pages        = 398,
    doi          = {10.1086/676406},
    archiveprefix = {arXiv},
    eprint       = {1402.5163},
    primaryclass = {astro-ph.IM},
    adsurl       = {https://ui.adsabs.harvard.edu/abs/2014PASP..126..398H}
}

@inproceedings{2016AGUFM.P13C..01R,
    title        = {{The Transiting Exoplanet Survey Satellite (TESS): Discovering Exoplanets in the Solar Neighborhood}},
    author       = {{Ricker}, G.~R.},
    year         = 2016,
    month        = dec,
    booktitle    = {AGU Fall Meeting Abstracts},
    pages        = {P13C-01},
    eid          = {P13C-01},
    adsurl       = {https://ui.adsabs.harvard.edu/abs/2016AGUFM.P13C..01R}
}

@article{2014ApJ...797..122D,
    title        = {{Kepler Flares. II. The Temporal Morphology of White-light Flares on GJ 1243}},
    author       = {{Davenport}, James R.~A. and {Hawley}, Suzanne L. and {Hebb}, Leslie and {Wisniewski}, John P. and {Kowalski}, Adam F. and {Johnson}, Emily C. and {Malatesta}, Michael and {Peraza}, Jesus and {Keil}, Marcus and {Silverberg}, Steven M. and {Jansen}, Tiffany C. and {Scheffler}, Matthew S. and {Berdis}, Jodi R. and {Larsen}, Daniel M. and {Hilton}, Eric J.},
    year         = 2014,
    month        = dec,
    journal      = {\apj},
    volume       = 797,
    number       = 2,
    pages        = 122,
    doi          = {10.1088/0004-637X/797/2/122},
    eid          = 122,
    archiveprefix = {arXiv},
    eprint       = {1411.3723},
    primaryclass = {astro-ph.SR},
    adsurl       = {https://ui.adsabs.harvard.edu/abs/2014ApJ...797..122D}
}

@article{2010AsBio..10..751S,
    title        = {{The Effect of a Strong Stellar Flare on the Atmospheric Chemistry of an Earth-like Planet Orbiting an M Dwarf}},
    author       = {{Segura}, Ant{\'\i}gona and {Walkowicz}, Lucianne M. and {Meadows}, Victoria and {Kasting}, James and {Hawley}, Suzanne},
    year         = 2010,
    month        = sep,
    journal      = {Astrobiology},
    volume       = 10,
    number       = 7,
    pages        = {751--771},
    doi          = {10.1089/ast.2009.0376},
    archiveprefix = {arXiv},
    eprint       = {1006.0022},
    primaryclass = {astro-ph.EP},
    adsurl       = {https://ui.adsabs.harvard.edu/abs/2010AsBio..10..751S}
}

@article{2019AsBio..19...64T,
    title        = {{Modeling Repeated M Dwarf Flaring at an Earth-like Planet in the Habitable Zone: Atmospheric Effects for an Unmagnetized Planet}},
    author       = {{Tilley}, Matt A. and {Segura}, Ant{\'\i}gona and {Meadows}, Victoria and {Hawley}, Suzanne and {Davenport}, James},
    year         = 2019,
    month        = jan,
    journal      = {Astrobiology},
    volume       = 19,
    number       = 1,
    pages        = {64--86},
    doi          = {10.1089/ast.2017.1794},
    adsurl       = {https://ui.adsabs.harvard.edu/abs/2019AsBio..19...64T}
}

@article{2017ApJ...843..110R,
    title        = {{The Surface UV Environment on Planets Orbiting M Dwarfs: Implications for Prebiotic Chemistry and the Need for Experimental Follow-up}},
    author       = {{Ranjan}, Sukrit and {Wordsworth}, Robin and {Sasselov}, Dimitar D.},
    year         = 2017,
    month        = jul,
    journal      = {\apj},
    volume       = 843,
    number       = 2,
    pages        = 110,
    doi          = {10.3847/1538-4357/aa773e},
    eid          = 110,
    archiveprefix = {arXiv},
    eprint       = {1705.02350},
    primaryclass = {astro-ph.EP},
    adsurl       = {https://ui.adsabs.harvard.edu/abs/2017ApJ...843..110R}
}

@article{2014ApJ...797..121H,
    title        = {{Kepler Flares. I. Active and Inactive M Dwarfs}},
    author       = {{Hawley}, Suzanne L. and {Davenport}, James R.~A. and {Kowalski}, Adam F. and {Wisniewski}, John P. and {Hebb}, Leslie and {Deitrick}, Russell and {Hilton}, Eric J.},
    year         = 2014,
    month        = dec,
    journal      = {\apj},
    volume       = 797,
    number       = 2,
    pages        = 121,
    doi          = {10.1088/0004-637X/797/2/121},
    eid          = 121,
    archiveprefix = {arXiv},
    eprint       = {1410.7779},
    primaryclass = {astro-ph.SR},
    adsurl       = {https://ui.adsabs.harvard.edu/abs/2014ApJ...797..121H}
}

@article{2018ApJ...867...71L,
    title        = {{The MUSCLES Treasury Survey. V. FUV Flares on Active and Inactive M Dwarfs}},
    author       = {{Loyd}, R.~O. Parke and {France}, Kevin and {Youngblood}, Allison and {Schneider}, Christian and {Brown}, Alexander and {Hu}, Renyu and {Segura}, Ant{\'\i}gona and {Linsky}, Jeffrey and {Redfield}, Seth and {Tian}, Feng and {Rugheimer}, Sarah and {Miguel}, Yamila and {Froning}, Cynthia S.},
    year         = 2018,
    month        = nov,
    journal      = {\apj},
    volume       = 867,
    number       = 1,
    pages        = 71,
    doi          = {10.3847/1538-4357/aae2bd},
    eid          = 71,
    archiveprefix = {arXiv},
    eprint       = {1809.07322},
    primaryclass = {astro-ph.SR},
    adsurl       = {https://ui.adsabs.harvard.edu/abs/2018ApJ...867...71L}
}

@article{2024MNRAS.533.1894J,
    title        = {{Optically quiet, but FUV loud: results from comparing the far-ultraviolet predictions of flare models with TESS and HST}},
    author       = {{Jackman}, James A.~G. and {Shkolnik}, Evgenya L. and {Loyd}, R.~O. Parke and {Richey-Yowell}, Tyler},
    year         = 2024,
    month        = sep,
    journal      = {\mnras},
    volume       = 533,
    number       = 2,
    pages        = {1894--1906},
    doi          = {10.1093/mnras/stae1570},
    archiveprefix = {arXiv},
    eprint       = {2406.15308},
    primaryclass = {astro-ph.SR},
    adsurl       = {https://ui.adsabs.harvard.edu/abs/2024MNRAS.533.1894J}
}

@article{2016ApJS..222....8D,
    title        = {{MESA Isochrones and Stellar Tracks (MIST) 0: Methods for the Construction of Stellar Isochrones}},
    author       = {{Dotter}, Aaron},
    year         = 2016,
    month        = jan,
    journal      = {\apjs},
    volume       = 222,
    number       = 1,
    pages        = 8,
    doi          = {10.3847/0067-0049/222/1/8},
    eid          = 8,
    archiveprefix = {arXiv},
    eprint       = {1601.05144},
    primaryclass = {astro-ph.SR},
    adsurl       = {https://ui.adsabs.harvard.edu/abs/2016ApJS..222....8D}
}

@article{2019ApJS..243...10P,
    title        = {{Modules for Experiments in Stellar Astrophysics (MESA): Pulsating Variable Stars, Rotation, Convective Boundaries, and Energy Conservation}},
    author       = {{Paxton}, Bill and {Smolec}, R. and {Schwab}, Josiah and {Gautschy}, A. and {Bildsten}, Lars and {Cantiello}, Matteo and {Dotter}, Aaron and {Farmer}, R. and {Goldberg}, Jared A. and {Jermyn}, Adam S. and {Kanbur}, S.~M. and {Marchant}, Pablo and {Thoul}, Anne and {Townsend}, Richard H.~D. and {Wolf}, William M. and {Zhang}, Michael and {Timmes}, F.~X.},
    year         = 2019,
    month        = jul,
    journal      = {\apjs},
    volume       = 243,
    number       = 1,
    pages        = 10,
    doi          = {10.3847/1538-4365/ab2241},
    eid          = 10,
    archiveprefix = {arXiv},
    eprint       = {1903.01426},
    primaryclass = {astro-ph.SR},
    adsurl       = {https://ui.adsabs.harvard.edu/abs/2019ApJS..243...10P}
}

@article{2014ApJ...787L..29K,
    title        = {{Habitable Zones around Main-sequence Stars: Dependence on Planetary Mass}},
    author       = {{Kopparapu}, Ravi Kumar and {Ramirez}, Ramses M. and {SchottelKotte}, James and {Kasting}, James F. and {Domagal-Goldman}, Shawn and {Eymet}, Vincent},
    year         = 2014,
    month        = jun,
    journal      = {\apjl},
    volume       = 787,
    number       = 2,
    pages        = {L29},
    doi          = {10.1088/2041-8205/787/2/L29},
    eid          = {L29},
    archiveprefix = {arXiv},
    eprint       = {1404.5292},
    primaryclass = {astro-ph.EP},
    adsurl       = {https://ui.adsabs.harvard.edu/abs/2014ApJ...787L..29K}
}

@incollection{Owen1980,
    title        = {The Search for Early Forms of Life in Other Planetary Systems: Future Possibilities Afforded by Spectroscopic Techniques},
    author       = {Owen, Tobias},
    year         = 1980,
    booktitle    = {Strategies for the Search for Life in the Universe: A Joint Session of Commissions 16, 40, and 44, Held in Montreal, Canada, During the IAU General Assembly, 15 and 16 August, 1979},
    publisher    = {Springer Netherlands},
    address      = {Dordrecht},
    pages        = {177--185},
    doi          = {10.1007/978-94-009-9115-6_17},
    isbn         = {978-94-009-9115-6},
    url          = {https://doi.org/10.1007/978-94-009-9115-6_17},
    editor       = {Papagiannis, Michael D.}
}

@article{1972Ap&SS..19...75G,
    title        = {{Some results of the cooperative photometric observations of the UV Cet-type flare stars in the years 1967 71}},
    author       = {{Gershberg}, R.~E.},
    year         = 1972,
    month        = nov,
    journal      = {\apss},
    volume       = 19,
    number       = 1,
    pages        = {75--92},
    doi          = {10.1007/BF00643168},
    adsurl       = {https://ui.adsabs.harvard.edu/abs/1972Ap&SS..19...75G}
}

@article{2015ApJ...807...45D,
    title        = {{The Occurrence of Potentially Habitable Planets Orbiting M Dwarfs Estimated from the Full Kepler Dataset and an Empirical Measurement of the Detection Sensitivity}},
    author       = {{Dressing}, Courtney D. and {Charbonneau}, David},
    year         = 2015,
    month        = jul,
    journal      = {\apj},
    volume       = 807,
    number       = 1,
    pages        = 45,
    doi          = {10.1088/0004-637X/807/1/45},
    eid          = 45,
    archiveprefix = {arXiv},
    eprint       = {1501.01623},
    primaryclass = {astro-ph.EP},
    adsurl       = {https://ui.adsabs.harvard.edu/abs/2015ApJ...807...45D}
}

@article{2022AJ....164..110F,
    title        = {{AU Microscopii in the Far-UV: Observations in Quiescence, during Flares, and Implications for AU Mic b and c}},
    author       = {{Feinstein}, Adina D. and {France}, Kevin and {Youngblood}, Allison and {Duvvuri}, Girish M. and {Teal}, D.~J. and {Cauley}, P. Wilson and {Seligman}, Darryl Z. and {Gaidos}, Eric and {Kempton}, Eliza M. -R. and {Bean}, Jacob L. and {Diamond-Lowe}, Hannah and {Newton}, Elisabeth and {Ginzburg}, Sivan and {Plavchan}, Peter and {Gao}, Peter and {Schlichting}, Hilke},
    year         = 2022,
    month        = sep,
    journal      = {\aj},
    volume       = 164,
    number       = 3,
    pages        = 110,
    doi          = {10.3847/1538-3881/ac8107},
    eid          = 110,
    archiveprefix = {arXiv},
    eprint       = {2205.09606},
    primaryclass = {astro-ph.SR},
    adsurl       = {https://ui.adsabs.harvard.edu/abs/2022AJ....164..110F}
}

@article{2024AJ....168...60F,
    title        = {{Evolution of Flare Activity in GKM Stars Younger Than 300 Myr over Five Years of TESS Observations}},
    author       = {{Feinstein}, Adina D. and {Seligman}, Darryl Z. and {France}, Kevin and {Gagn{\'e}}, Jonathan and {Kowalski}, Adam},
    year         = 2024,
    month        = aug,
    journal      = {\aj},
    volume       = 168,
    number       = 2,
    pages        = 60,
    doi          = {10.3847/1538-3881/ad4edf},
    eid          = 60,
    archiveprefix = {arXiv},
    eprint       = {2405.00850},
    primaryclass = {astro-ph.SR},
    adsurl       = {https://ui.adsabs.harvard.edu/abs/2024AJ....168...60F}
}

@article{2019ApJ...881....9H,
    title        = {{EvryFlare. I. Long-term Evryscope Monitoring of Flares from the Cool Stars across Half the Southern Sky}},
    author       = {{Howard}, Ward S. and {Corbett}, Hank and {Law}, Nicholas M. and {Ratzloff}, Jeffrey K. and {Glazier}, Amy and {Fors}, Octavi and {del Ser}, Daniel and {Haislip}, Joshua},
    year         = 2019,
    month        = aug,
    journal      = {\apj},
    volume       = 881,
    number       = 1,
    pages        = 9,
    doi          = {10.3847/1538-4357/ab2767},
    eid          = 9,
    archiveprefix = {arXiv},
    eprint       = {1904.10421},
    primaryclass = {astro-ph.SR},
    adsurl       = {https://ui.adsabs.harvard.edu/abs/2019ApJ...881....9H}
}

@article{2023ApJ...944....5B,
    title        = {{Constraints on Stellar Flare Energy Ratios in the NUV and Optical from a Multiwavelength Study of GALEX and Kepler Flare Stars}},
    author       = {{Brasseur}, C.~E. and {Osten}, Rachel A. and {Tristan}, Isaiah I. and {Kowalski}, Adam F.},
    year         = 2023,
    month        = feb,
    journal      = {\apj},
    volume       = 944,
    number       = 1,
    pages        = 5,
    doi          = {10.3847/1538-4357/acab59},
    eid          = 5,
    archiveprefix = {arXiv},
    eprint       = {2212.08696},
    primaryclass = {astro-ph.SR},
    adsurl       = {https://ui.adsabs.harvard.edu/abs/2023ApJ...944....5B}
}

@article{2019ApJ...871..167K,
    title        = {{The Near-ultraviolet Continuum Radiation in the Impulsive Phase of HF/GF-type dMe Flares. I. Data}},
    author       = {{Kowalski}, Adam F. and {Wisniewski}, John P. and {Hawley}, Suzanne L. and {Osten}, Rachel A. and {Brown}, Alexander and {Fari{\~n}a}, Cecilia and {Valenti}, Jeff A. and {Brown}, Stephen and {Xilouris}, Manolis and {Schmidt}, Sarah J. and {Johns-Krull}, Christopher},
    year         = 2019,
    month        = feb,
    journal      = {\apj},
    volume       = 871,
    number       = 2,
    pages        = 167,
    doi          = {10.3847/1538-4357/aaf058},
    eid          = 167,
    archiveprefix = {arXiv},
    eprint       = {1811.04021},
    primaryclass = {astro-ph.SR},
    adsurl       = {https://ui.adsabs.harvard.edu/abs/2019ApJ...871..167K}
}

@article{2016ApJ...829...23D,
    title        = {{The Kepler Catalog of Stellar Flares}},
    author       = {{Davenport}, James R.~A.},
    year         = 2016,
    month        = sep,
    journal      = {\apj},
    volume       = 829,
    number       = 1,
    pages        = 23,
    doi          = {10.3847/0004-637X/829/1/23},
    eid          = 23,
    archiveprefix = {arXiv},
    eprint       = {1607.03494},
    primaryclass = {astro-ph.SR},
    adsurl       = {https://ui.adsabs.harvard.edu/abs/2016ApJ...829...23D}
}

@article{2023ApJ...953...57K,
    title        = {{Where are the Water Worlds?: Self-consistent Models of Water-rich Exoplanet Atmospheres}},
    author       = {{Kempton}, Eliza M. -R. and {Lessard}, Madeline and {Malik}, Matej and {Rogers}, Leslie A. and {Futrowsky}, Kate E. and {Ih}, Jegug and {Marounina}, Nadejda and {Romero-Mirza}, Carlos E.},
    year         = 2023,
    month        = aug,
    journal      = {\apj},
    volume       = 953,
    number       = 1,
    pages        = 57,
    doi          = {10.3847/1538-4357/ace10d},
    eid          = 57,
    archiveprefix = {arXiv},
    eprint       = {2307.06508},
    primaryclass = {astro-ph.EP},
    adsurl       = {https://ui.adsabs.harvard.edu/abs/2023ApJ...953...57K}
}

@article{2019ApJ...886..142M,
    title        = {{Analyzing Atmospheric Temperature Profiles and Spectra of M Dwarf Rocky Planets}},
    author       = {{Malik}, Matej and {Kempton}, Eliza M. -R. and {Koll}, Daniel D.~B. and {Mansfield}, Megan and {Bean}, Jacob L. and {Kite}, Edwin},
    year         = 2019,
    month        = dec,
    journal      = {\apj},
    volume       = 886,
    number       = 2,
    pages        = 142,
    doi          = {10.3847/1538-4357/ab4a05},
    eid          = 142,
    archiveprefix = {arXiv},
    eprint       = {1907.13135},
    primaryclass = {astro-ph.EP},
    adsurl       = {https://ui.adsabs.harvard.edu/abs/2019ApJ...886..142M}
}

@software{2022ascl.soft02012L,
    title        = {{fiducial\_flare: Spectra and lightcurves of a standardized far ultraviolet flare}},
    author       = {{Loyd}, R.~O. Parke},
    year         = 2022,
    month        = feb,
    howpublished = {Astrophysics Source Code Library, record ascl:2202.012},
    eid          = {ascl:2202.012},
    adsurl       = {https://ui.adsabs.harvard.edu/abs/2022ascl.soft02012L}
}

@article{2023AJ....166..232W,
    title        = {{Validating AU Microscopii d with Transit Timing Variations}},
    author       = {{Wittrock}, Justin M. and {Plavchan}, Peter P. and {Cale}, Bryson L. and {Barclay}, Thomas and {Ludwig}, Mathis R. and {Schwarz}, Richard P. and {M{\'e}karnia}, Djamel and {Triaud}, Amaury H.~M.~J. and {Abe}, Lyu and {Suarez}, Olga and {Guillot}, Tristan and {Conti}, Dennis M. and {Collins}, Karen A. and {Waite}, Ian A. and {Kielkopf}, John F. and {Collins}, Kevin I. and {Dreizler}, Stefan and {El Mufti}, Mohammed and {Feliz}, Dax L. and {Gaidos}, Eric and {Geneser}, Claire S. and {Horne}, Keith D. and {Kane}, Stephen R. and {Lowrance}, Patrick J. and {Martioli}, Eder and {Radford}, Don J. and {Reefe}, Michael A. and {Roccatagliata}, Veronica and {Shporer}, Avi and {Stassun}, Keivan G. and {Stockdale}, Christopher and {Tan}, Thiam-Guan and {Tanner}, Angelle M. and {Vega}, Laura D.},
    year         = 2023,
    month        = dec,
    journal      = {\aj},
    volume       = 166,
    number       = 6,
    pages        = 232,
    doi          = {10.3847/1538-3881/acfda8},
    eid          = 232,
    archiveprefix = {arXiv},
    eprint       = {2302.04922},
    primaryclass = {astro-ph.EP},
    adsurl       = {https://ui.adsabs.harvard.edu/abs/2023AJ....166..232W}
}

@article{2000Icar..143..244M,
    title        = {{Photochemistry of Saturn's Atmosphere. I. Hydrocarbon Chemistry and Comparisons with ISO Observations}},
    author       = {{Moses}, Julianne I. and {B{\'e}zard}, Bruno and {Lellouch}, Emmanuel and {Gladstone}, G. Randall and {Feuchtgruber}, Helmut and {Allen}, Mark},
    year         = 2000,
    month        = feb,
    journal      = {\icarus},
    volume       = 143,
    number       = 2,
    pages        = {244--298},
    doi          = {10.1006/icar.1999.6270},
    adsurl       = {https://ui.adsabs.harvard.edu/abs/2000Icar..143..244M}
}

@article{2025A&A...700A..53M,
    title        = {{Flare frequency in M dwarfs belonging to young moving groups}},
    author       = {{Mamonova}, E. and {Shan}, Y. and {Kowalski}, A.~F. and {Wedemeyer}, S. and {Werner}, S.~C.},
    year         = 2025,
    month        = aug,
    journal      = {\aap},
    volume       = 700,
    pages        = {A53},
    doi          = {10.1051/0004-6361/202554614},
    eid          = {A53},
    archiveprefix = {arXiv},
    eprint       = {2506.04465},
    primaryclass = {astro-ph.SR},
    adsurl       = {https://ui.adsabs.harvard.edu/abs/2025A&A...700A..53M}
}

@article{2018MNRAS.479..865S,
    title        = {{FastChem: A computer program for efficient complex chemical equilibrium calculations in the neutral/ionized gas phase with applications to stellar and planetary atmospheres}},
    author       = {{Stock}, Joachim W. and {Kitzmann}, Daniel and {Patzer}, A. Beate C. and {Sedlmayr}, Erwin},
    year         = 2018,
    month        = sep,
    journal      = {\mnras},
    volume       = 479,
    number       = 1,
    pages        = {865--874},
    doi          = {10.1093/mnras/sty1531},
    archiveprefix = {arXiv},
    eprint       = {1804.05010},
    primaryclass = {astro-ph.EP},
    adsurl       = {https://ui.adsabs.harvard.edu/abs/2018MNRAS.479..865S}
}

@article{2017AJ....153...56M,
    title        = {{HELIOS: An Open-source, GPU-accelerated Radiative Transfer Code for Self-consistent Exoplanetary Atmospheres}},
    author       = {{Malik}, Matej and {Grosheintz}, Luc and {Mendon{\c{c}}a}, Jo{\~a}o M. and {Grimm}, Simon L. and {Lavie}, Baptiste and {Kitzmann}, Daniel and {Tsai}, Shang-Min and {Burrows}, Adam and {Kreidberg}, Laura and {Bedell}, Megan and {Bean}, Jacob L. and {Stevenson}, Kevin B. and {Heng}, Kevin},
    year         = 2017,
    month        = feb,
    journal      = {\aj},
    volume       = 153,
    number       = 2,
    pages        = 56,
    doi          = {10.3847/1538-3881/153/2/56},
    eid          = 56,
    archiveprefix = {arXiv},
    eprint       = {1606.05474},
    primaryclass = {astro-ph.EP},
    adsurl       = {https://ui.adsabs.harvard.edu/abs/2017AJ....153...56M}
}

@article{2002A&A...390..779B,
    title        = {{Collision-induced absorption coefficients of H$_{2}$ pairs at temperatures from 60 K to 1000 K}},
    author       = {{Borysow}, A.},
    year         = 2002,
    month        = aug,
    journal      = {\aap},
    volume       = 390,
    pages        = {779--782},
    doi          = {10.1051/0004-6361:20020555},
    adsurl       = {https://ui.adsabs.harvard.edu/abs/2002A&A...390..779B}
}

@article{2012JQSRT.113.1276R,
    title        = {{New section of the HITRAN database: Collision-induced absorption (CIA)}},
    author       = {{Richard}, C. and {Gordon}, I.~E. and {Rothman}, L.~S. and {Abel}, M. and {Frommhold}, L. and {Gustafsson}, M. and {Hartmann}, J. -M. and {Hermans}, C. and {Lafferty}, W.~J. and {Orton}, G.~S. and {Smith}, K.~M. and {Tran}, H.},
    year         = 2012,
    month        = jul,
    journal      = {\jqsrt},
    volume       = 113,
    number       = 11,
    pages        = {1276--1285},
    doi          = {10.1016/j.jqsrt.2011.11.004},
    adsurl       = {https://ui.adsabs.harvard.edu/abs/2012JQSRT.113.1276R}
}

@article{2017AJ....154..261C,
    title        = {{Trends in Atmospheric Properties of Neptune-size Exoplanets}},
    author       = {{Crossfield}, Ian J.~M. and {Kreidberg}, Laura},
    year         = 2017,
    month        = dec,
    journal      = {\aj},
    volume       = 154,
    number       = 6,
    pages        = 261,
    doi          = {10.3847/1538-3881/aa9279},
    eid          = 261,
    archiveprefix = {arXiv},
    eprint       = {1708.00016},
    primaryclass = {astro-ph.EP},
    adsurl       = {https://ui.adsabs.harvard.edu/abs/2017AJ....154..261C}
}

@article{2005JQSRT..92..293S,
    title        = {{Direct measurement of the Rayleigh scattering cross section in various gases}},
    author       = {{Sneep}, Maarten and {Ubachs}, Wim},
    year         = 2005,
    month        = may,
    journal      = {\jqsrt},
    volume       = 92,
    number       = 3,
    pages        = {293--310},
    doi          = {10.1016/j.jqsrt.2004.07.025},
    adsurl       = {https://ui.adsabs.harvard.edu/abs/2005JQSRT..92..293S}
}

@article{2017ApJS..228...20T,
    title        = {{VULCAN: An Open-source, Validated Chemical Kinetics Python Code for Exoplanetary Atmospheres}},
    author       = {{Tsai}, Shang-Min and {Lyons}, James R. and {Grosheintz}, Luc and {Rimmer}, Paul B. and {Kitzmann}, Daniel and {Heng}, Kevin},
    year         = 2017,
    month        = feb,
    journal      = {\apjs},
    volume       = 228,
    number       = 2,
    pages        = 20,
    doi          = {10.3847/1538-4365/228/2/20},
    eid          = 20,
    archiveprefix = {arXiv},
    eprint       = {1607.00409},
    primaryclass = {astro-ph.EP},
    adsurl       = {https://ui.adsabs.harvard.edu/abs/2017ApJS..228...20T}
}

@article{2021ApJ...923..264T,
    title        = {{A Comparative Study of Atmospheric Chemistry with VULCAN}},
    author       = {{Tsai}, Shang-Min and {Malik}, Matej and {Kitzmann}, Daniel and {Lyons}, James R. and {Fateev}, Alexander and {Lee}, Elspeth and {Heng}, Kevin},
    year         = 2021,
    month        = dec,
    journal      = {\apj},
    volume       = 923,
    number       = 2,
    pages        = 264,
    doi          = {10.3847/1538-4357/ac29bc},
    eid          = 264,
    archiveprefix = {arXiv},
    eprint       = {2108.01790},
    primaryclass = {astro-ph.EP},
    adsurl       = {https://ui.adsabs.harvard.edu/abs/2021ApJ...923..264T}
}

@article{2013A&A...558A..91P,
    title        = {{3D mixing in hot Jupiters atmospheres. I. Application to the day/night cold trap in HD 209458b}},
    author       = {{Parmentier}, Vivien and {Showman}, Adam P. and {Lian}, Yuan},
    year         = 2013,
    month        = oct,
    journal      = {\aap},
    volume       = 558,
    pages        = {A91},
    doi          = {10.1051/0004-6361/201321132},
    eid          = {A91},
    archiveprefix = {arXiv},
    eprint       = {1301.4522},
    primaryclass = {astro-ph.EP},
    adsurl       = {https://ui.adsabs.harvard.edu/abs/2013A&A...558A..91P}
}

@article{2013ApJ...777...34M,
    title        = {{Compositional Diversity in the Atmospheres of Hot Neptunes, with Application to GJ 436b}},
    author       = {{Moses}, J.~I. and {Line}, M.~R. and {Visscher}, C. and {Richardson}, M.~R. and {Nettelmann}, N. and {Fortney}, J.~J. and {Barman}, T.~S. and {Stevenson}, K.~B. and {Madhusudhan}, N.},
    year         = 2013,
    month        = nov,
    journal      = {\apj},
    volume       = 777,
    number       = 1,
    pages        = 34,
    doi          = {10.1088/0004-637X/777/1/34},
    eid          = 34,
    archiveprefix = {arXiv},
    eprint       = {1306.5178},
    primaryclass = {astro-ph.EP},
    adsurl       = {https://ui.adsabs.harvard.edu/abs/2013ApJ...777...34M}
}

@article{2023MNRAS.521.3333L,
    title        = {{The impact of time-dependent stellar activity on exoplanet atmospheres}},
    author       = {{Louca}, Amy J. and {Miguel}, Yamila and {Tsai}, Shang-Min and {Froning}, Cynthia S. and {Loyd}, R.~O. Parke and {France}, Kevin},
    year         = 2023,
    month        = may,
    journal      = {\mnras},
    volume       = 521,
    number       = 3,
    pages        = {3333--3347},
    doi          = {10.1093/mnras/stac1220},
    archiveprefix = {arXiv},
    eprint       = {2204.10835},
    primaryclass = {astro-ph.EP},
    adsurl       = {https://ui.adsabs.harvard.edu/abs/2023MNRAS.521.3333L}
}

@article{2022ExA....53..279M,
    title        = {{Chemical variation with altitude and longitude on exo-Neptunes: Predictions for Ariel phase-curve observations}},
    author       = {{Moses}, Julianne I. and {Tremblin}, Pascal and {Venot}, Olivia and {Miguel}, Yamila},
    year         = 2022,
    month        = apr,
    journal      = {Experimental Astronomy},
    volume       = 53,
    number       = 2,
    pages        = {279--322},
    doi          = {10.1007/s10686-021-09749-1},
    archiveprefix = {arXiv},
    eprint       = {2103.07023},
    primaryclass = {astro-ph.EP},
    adsurl       = {https://ui.adsabs.harvard.edu/abs/2022ExA....53..279M}
}

@article{2015MNRAS.446..345M,
    title        = {{The effect of Lyman {\ensuremath{\alpha}} radiation on mini-Neptune atmospheres around M stars: application to GJ 436b}},
    author       = {{Miguel}, Yamila and {Kaltenegger}, Lisa and {Linsky}, Jeffrey L. and {Rugheimer}, Sarah},
    year         = 2015,
    month        = jan,
    journal      = {\mnras},
    volume       = 446,
    number       = 1,
    pages        = {345--353},
    doi          = {10.1093/mnras/stu2107},
    archiveprefix = {arXiv},
    eprint       = {1410.2112},
    primaryclass = {astro-ph.EP},
    adsurl       = {https://ui.adsabs.harvard.edu/abs/2015MNRAS.446..345M}
}

@article{2006P&SS...54.1425K,
    title        = {{Atmospheric and water loss from early Venus}},
    author       = {{Kulikov}, Yu. N. and {Lammer}, H. and {Lichtenegger}, H.~I.~M. and {Terada}, N. and {Ribas}, I. and {Kolb}, C. and {Langmayr}, D. and {Lundin}, R. and {Guinan}, E.~F. and {Barabash}, S. and {Biernat}, H.~K.},
    year         = 2006,
    month        = nov,
    journal      = {\planss},
    volume       = 54,
    number       = {13-14},
    pages        = {1425--1444},
    doi          = {10.1016/j.pss.2006.04.021},
    adsurl       = {https://ui.adsabs.harvard.edu/abs/2006P&SS...54.1425K}
}

@article{2023MNRAS.520.3867T,
    title        = {{The climate and compositional variation of the highly eccentric planet HD 80606 b - the rise and fall of carbon monoxide and elemental sulfur}},
    author       = {{Tsai}, Shang-Min and {Steinrueck}, Maria and {Parmentier}, Vivien and {Lewis}, Nikole and {Pierrehumbert}, Raymond},
    year         = 2023,
    month        = apr,
    journal      = {\mnras},
    volume       = 520,
    number       = 3,
    pages        = {3867--3886},
    doi          = {10.1093/mnras/stad214},
    adsurl       = {https://ui.adsabs.harvard.edu/abs/2023MNRAS.520.3867T}
}

@article{2023MNRAS.525.5168M,
    title        = {{Impact of M-dwarf stellar wind and photoevaporation on the atmospheric evolution of small planets}},
    author       = {{Modi}, Ashini and {Estrela}, Raissa and {Valio}, Adriana},
    year         = 2023,
    month        = nov,
    journal      = {\mnras},
    volume       = 525,
    number       = 4,
    pages        = {5168--5179},
    doi          = {10.1093/mnras/stad2557},
    archiveprefix = {arXiv},
    eprint       = {2309.10942},
    primaryclass = {astro-ph.EP},
    adsurl       = {https://ui.adsabs.harvard.edu/abs/2023MNRAS.525.5168M}
}

@article{2020AJ....160..237F,
    title        = {{The High-energy Radiation Environment around a 10 Gyr M Dwarf: Habitable at Last?}},
    author       = {{France}, Kevin and {Duvvuri}, Girish and {Egan}, Hilary and {Koskinen}, Tommi and {Wilson}, David J. and {Youngblood}, Allison and {Froning}, Cynthia S. and {Brown}, Alexander and {Alvarado-G{\'o}mez}, Juli{\'a}n D. and {Berta-Thompson}, Zachory K. and {Drake}, Jeremy J. and {Garraffo}, Cecilia and {Kaltenegger}, Lisa and {Kowalski}, Adam F. and {Linsky}, Jeffrey L. and {Loyd}, R.~O. Parke and {Mauas}, Pablo J.~D. and {Miguel}, Yamila and {Pineda}, J. Sebastian and {Rugheimer}, Sarah and {Schneider}, P. Christian and {Tian}, Feng and {Vieytes}, Mariela},
    year         = 2020,
    month        = nov,
    journal      = {\aj},
    volume       = 160,
    number       = 5,
    pages        = 237,
    doi          = {10.3847/1538-3881/abb465},
    eid          = 237,
    archiveprefix = {arXiv},
    eprint       = {2009.01259},
    primaryclass = {astro-ph.EP},
    adsurl       = {https://ui.adsabs.harvard.edu/abs/2020AJ....160..237F}
}

@article{2007AsBio...7...85S,
    title        = {{M Stars as Targets for Terrestrial Exoplanet Searches And Biosignature Detection}},
    author       = {{Scalo}, John and {Kaltenegger}, Lisa and {Segura}, Ant{\'\i}gona and {Fridlund}, Malcolm and {Ribas}, Ignasi and {Kulikov}, Yu. N. and {Grenfell}, John L. and {Rauer}, Heike and {Odert}, Petra and {Leitzinger}, Martin and {Selsis}, F. and {Khodachenko}, Maxim L. and {Eiroa}, Carlos and {Kasting}, Jim and {Lammer}, Helmut},
    year         = 2007,
    month        = feb,
    journal      = {Astrobiology},
    volume       = 7,
    number       = 1,
    pages        = {85--166},
    doi          = {10.1089/ast.2006.0125},
    adsurl       = {https://ui.adsabs.harvard.edu/abs/2007AsBio...7...85S}
}

@article{2015AsBio..15..119L,
    title        = {{Extreme Water Loss and Abiotic O2Buildup on Planets Throughout the Habitable Zones of M Dwarfs}},
    author       = {{Luger}, R. and {Barnes}, R.},
    year         = 2015,
    month        = feb,
    journal      = {Astrobiology},
    volume       = 15,
    number       = 2,
    pages        = {119--143},
    doi          = {10.1089/ast.2014.1231},
    archiveprefix = {arXiv},
    eprint       = {1411.7412},
    primaryclass = {astro-ph.EP},
    adsurl       = {https://ui.adsabs.harvard.edu/abs/2015AsBio..15..119L}
}

@article{2016A&A...596A.111R,
    title        = {{The habitability of Proxima Centauri b. I. Irradiation, rotation and volatile inventory from formation to the present}},
    author       = {{Ribas}, Ignasi and {Bolmont}, Emeline and {Selsis}, Franck and {Reiners}, Ansgar and {Leconte}, J{\'e}r{\'e}my and {Raymond}, Sean N. and {Engle}, Scott G. and {Guinan}, Edward F. and {Morin}, Julien and {Turbet}, Martin and {Forget}, Fran{\c{c}}ois and {Anglada-Escud{\'e}}, Guillem},
    year         = 2016,
    month        = dec,
    journal      = {\aap},
    volume       = 596,
    pages        = {A111},
    doi          = {10.1051/0004-6361/201629576},
    eid          = {A111},
    archiveprefix = {arXiv},
    eprint       = {1608.06813},
    primaryclass = {astro-ph.EP},
    adsurl       = {https://ui.adsabs.harvard.edu/abs/2016A&A...596A.111R}
}

@article{2017ApJ...836L...3A,
    title        = {{How Hospitable Are Space Weather Affected Habitable Zones? The Role of Ion Escape}},
    author       = {{Airapetian}, Vladimir S. and {Glocer}, Alex and {Khazanov}, George V. and {Loyd}, R.~O.~P. and {France}, Kevin and {Sojka}, Jan and {Danchi}, William C. and {Liemohn}, Michael W.},
    year         = 2017,
    month        = feb,
    journal      = {\apjl},
    volume       = 836,
    number       = 1,
    pages        = {L3},
    doi          = {10.3847/2041-8213/836/1/L3},
    eid          = {L3},
    adsurl       = {https://ui.adsabs.harvard.edu/abs/2017ApJ...836L...3A}
}

@article{2022ApJ...941L...8G,
    title        = {{Revisiting the Space Weather Environment of Proxima Centauri b}},
    author       = {{Garraffo}, Cecilia and {Alvarado-G{\'o}mez}, Juli{\'a}n D. and {Cohen}, Ofer and {Drake}, Jeremy J.},
    year         = 2022,
    month        = dec,
    journal      = {\apjl},
    volume       = 941,
    number       = 1,
    pages        = {L8},
    doi          = {10.3847/2041-8213/aca487},
    eid          = {L8},
    archiveprefix = {arXiv},
    eprint       = {2211.15697},
    primaryclass = {astro-ph.SR},
    adsurl       = {https://ui.adsabs.harvard.edu/abs/2022ApJ...941L...8G}
}

@article{2017ApJ...843L..33G,
    title        = {{The Threatening Magnetic and Plasma Environment of the TRAPPIST-1 Planets}},
    author       = {{Garraffo}, Cecilia and {Drake}, Jeremy J. and {Cohen}, Ofer and {Alvarado-G{\'o}mez}, Julian D. and {Moschou}, Sofia P.},
    year         = 2017,
    month        = jul,
    journal      = {\apjl},
    volume       = 843,
    number       = 2,
    pages        = {L33},
    doi          = {10.3847/2041-8213/aa79ed},
    eid          = {L33},
    archiveprefix = {arXiv},
    eprint       = {1706.04617},
    primaryclass = {astro-ph.SR},
    adsurl       = {https://ui.adsabs.harvard.edu/abs/2017ApJ...843L..33G}
}

@article{2022AJ....164...17T,
    title        = {{Llamaradas Estelares: Modeling the Morphology of White-light Flares}},
    author       = {{Tovar Mendoza}, Guadalupe and {Davenport}, James R.~A. and {Agol}, Eric and {Jackman}, James A.~G. and {Hawley}, Suzanne L.},
    year         = 2022,
    month        = jul,
    journal      = {\aj},
    volume       = 164,
    number       = 1,
    pages        = 17,
    doi          = {10.3847/1538-3881/ac6fe6},
    eid          = 17,
    archiveprefix = {arXiv},
    eprint       = {2205.05706},
    primaryclass = {astro-ph.SR},
    adsurl       = {https://ui.adsabs.harvard.edu/abs/2022AJ....164...17T}
}

@article{2025ExA....59...26R,
    title        = {{The PLATO mission}},
    author       = {{Rauer}, Heike and {Aerts}, Conny and {Cabrera}, Juan and {Deleuil}, Magali and {Erikson}, Anders and {Gizon}, Laurent and {Goupil}, Mariejo and {Heras}, Ana and {Walloschek}, Thomas and {Lorenzo-Alvarez}, Jose and {Marliani}, Filippo and {Martin-Garcia}, C{\'e}sar and {Mas-Hesse}, J. Miguel and {O'Rourke}, Laurence and {Osborn}, Hugh and {Pagano}, Isabella and {Piotto}, Giampaolo and {Pollacco}, Don and {Ragazzoni}, Roberto and {Ramsay}, Gavin and {Udry}, St{\'e}phane and {Appourchaux}, Thierry and {Benz}, Willy and {Brandeker}, Alexis and {G{\"u}del}, Manuel and {Janot-Pacheco}, Eduardo and {Kabath}, Petr and {Kjeldsen}, Hans and {Min}, Michiel and {Santos}, Nuno and {Smith}, Alan and {Suarez}, Juan-Carlos and {Werner}, Stephanie C. and {Aboudan}, Alessio and {Abreu}, Manuel and {Acu{\~n}a}, Lorena and {Adams}, Moritz and {Adibekyan}, Vardan and {Affer}, Laura and {Agneray}, Fran{\c{c}}ois and {Agnor}, Craig and {Aguirre B{\o}rsen-Koch}, Victor and {Ahmed}, Saad and {Aigrain}, Suzanne and {Al-Bahlawan}, Ashraf and {Alcacera Gil}, Ma de los Angeles and {Alei}, Eleonora and {Alencar}, Silvia and {Alexander}, Richard and {Alfonso-Garz{\'o}n}, Julia and {Alibert}, Yann and {Allende Prieto}, Carlos and {Almeida}, Leonardo and {Alonso Sobrino}, Roi and {Altavilla}, Giuseppe and {Althaus}, Christian and {Alvarez Trujillo}, Luis Alonso and {Amarsi}, Anish and {Ammler-von Eiff}, Matthias and {Am{\^o}res}, Eduardo and {Andrade}, Laerte and {Antoniadis-Karnavas}, Alexandros and {Ant{\'o}nio}, Carlos and {Aparicio del Moral}, Beatriz and {Appolloni}, Matteo and {Arena}, Claudio and {Armstrong}, David and {Aroca Aliaga}, Jose and {Asplund}, Martin and {Audenaert}, Jeroen and {Auricchio}, Natalia and {Avelino}, Pedro and {Baeke}, Ann and {Bailli{\'e}}, Kevin and {Balado}, Ana and {Ballber Balaguer{\'o}}, Pau and {Balestra}, Andrea and {Ball}, Warrick and {Ballans}, Herve and {Ballot}, Jerome and {Barban}, Caroline and {Barbary}, Ga{\"e}le and {Barbieri}, Mauro and {Barcel{\'o} Forteza}, Sebasti{\`a} and {Barker}, Adrian and {Barklem}, Paul and {Barnes}, Sydney and {Barrado Navascues}, David and {Barragan}, Oscar and {Baruteau}, Cl{\'e}ment and {Basu}, Sarbani and {Baudin}, Frederic and {Baumeister}, Philipp and {Bayliss}, Daniel and {Bazot}, Michael and {Beck}, Paul G. and {Belkacem}, Kevin and {Bellinger}, Earl and {Benatti}, Serena and {Benomar}, Othman and {B{\'e}rard}, Diane and {Bergemann}, Maria and {Bergomi}, Maria and {Bernardo}, Pierre and {Biazzo}, Katia and {Bignamini}, Andrea and {Bigot}, Lionel and {Billot}, Nicolas and {Binet}, Martin and {Biondi}, David and {Biondi}, Federico and {Birch}, Aaron C. and {Bitsch}, Bertram and {Bluhm Ceballos}, Paz Victoria and {B{\'o}di}, Attila and {Bogn{\'a}r}, Zs{\'o}fia and {Boisse}, Isabelle and {Bolmont}, Emeline and {Bonanno}, Alfio and {Bonavita}, Mariangela and {Bonfanti}, Andrea and {Bonfils}, Xavier and {Bonito}, Rosaria and {Bonomo}, Aldo Stefano and {B{\"o}rner}, Anko and {Boro Saikia}, Sudeshna and {Borreguero Mart{\'\i}n}, Elisa and {Borsa}, Francesco and {Borsato}, Luca and {Bossini}, Diego and {Bouchy}, Francois and {Bou{\'e}}, Gwena{\"e}l and {Boufleur}, Rodrigo and {Boumier}, Patrick and {Bourrier}, Vincent and {Bowman}, Dominic M. and {Bozzo}, Enrico and {Bradley}, Louisa and {Bray}, John and {Bressan}, Alessandro and {Breton}, Sylvain and {Brienza}, Daniele and {Brito}, Ana and {Brogi}, Matteo and {Brown}, Beverly and {Brown}, David J.~A. and {Brun}, Allan Sacha and {Bruno}, Giovanni and {Bruns}, Michael and {Buchhave}, Lars A. and {Bugnet}, Lisa and {Buldgen}, Ga{\"e}l and {Burgess}, Patrick and {Busatta}, Andrea and {Busso}, Giorgia and {Buzasi}, Derek and {Caballero}, Jos{\'e} A. and {Cabral}, Alexandre and {Cabrero Gomez}, Juan-Francisco and {Calderone}, Flavia and {Cameron}, Robert and {Cameron}, Andrew and {Campante}, Tiago and {Campos Gestal}, N{\'e}stor and {Canto Martins}, Bruno Leonardo and {Cara}, Christophe and {Carone}, Ludmila and {Carrasco}, Josep Manel and {Casagrande}, Luca and {Casewell}, Sarah L. and {Cassisi}, Santi and {Castellani}, Marco and {Castro}, Matthieu and {Catala}, Claude and {Catal{\'a}n Fern{\'a}ndez}, Irene and {Catelan}, M{\'a}rcio and {Cegla}, Heather and {Cerruti}, Chiara and {Cessa}, Virginie and {Chadid}, Merieme and {Chaplin}, William and {Charpinet}, Stephane and {Chiappini}, Cristina and {Chiarucci}, Simone and {Chiavassa}, Andrea and {Chinellato}, Simonetta and {Chirulli}, Giovanni and {Christensen-Dalsgaard}, J{\o}rgen and {Church}, Ross and {Claret}, Antonio and {Clarke}, Cathie and {Claudi}, Riccardo and {Clermont}, Lionel and {Coelho}, Hugo and {Coelho}, Joao and {Cogato}, Fabrizio and {Colom{\'e}}, Josep and {Condamin}, Mathieu and {Conde Garc{\'\i}a}, Fernando and {Conseil}, Simon},
    year         = 2025,
    month        = jun,
    journal      = {Experimental Astronomy},
    volume       = 59,
    number       = 3,
    pages        = 26,
    doi          = {10.1007/s10686-025-09985-9},
    eid          = 26,
    archiveprefix = {arXiv},
    eprint       = {2406.05447},
    primaryclass = {astro-ph.IM},
    adsurl       = {https://ui.adsabs.harvard.edu/abs/2025ExA....59...26R}
}

@article{2023ConPh..64...47P,
    title        = {{The Extremely Large Telescope}},
    author       = {{Padovani}, Paolo and {Cirasuolo}, Michele},
    year         = 2023,
    month        = jan,
    journal      = {Contemporary Physics},
    volume       = 64,
    number       = 1,
    pages        = {47--64},
    doi          = {10.1080/00107514.2023.2266921},
    archiveprefix = {arXiv},
    eprint       = {2312.04299},
    primaryclass = {astro-ph.IM},
    adsurl       = {https://ui.adsabs.harvard.edu/abs/2023ConPh..64...47P}
}

@article{2012ApJ...744...60G,
    title        = {{The Cosmic Origins Spectrograph}},
    author       = {{Green}, James C. and {Froning}, Cynthia S. and {Osterman}, Steve and {Ebbets}, Dennis and {Heap}, Sara H. and {Leitherer}, Claus and {Linsky}, Jeffrey L. and {Savage}, Blair D. and {Sembach}, Kenneth and {Shull}, J. Michael and {Siegmund}, Oswald H.~W. and {Snow}, Theodore P. and {Spencer}, John and {Stern}, S. Alan and {Stocke}, John and {Welsh}, Barry and {B{\'e}land}, St{\'e}phane and {Burgh}, Eric B. and {Danforth}, Charles and {France}, Kevin and {Keeney}, Brian and {McPhate}, Jason and {Penton}, Steven V. and {Andrews}, John and {Brownsberger}, Kenneth and {Morse}, Jon and {Wilkinson}, Erik},
    year         = 2012,
    month        = jan,
    journal      = {\apj},
    volume       = 744,
    number       = 1,
    pages        = 60,
    doi          = {10.1088/0004-637X/744/1/6010.1086/141956},
    eid          = 60,
    archiveprefix = {arXiv},
    eprint       = {1110.0462},
    primaryclass = {astro-ph.IM},
    adsurl       = {https://ui.adsabs.harvard.edu/abs/2012ApJ...744...60G}
}

@article{2018ConPh..59..251K,
    title        = {{Scientific discovery with the James Webb Space Telescope}},
    author       = {{Kalirai}, Jason},
    year         = 2018,
    month        = jul,
    journal      = {Contemporary Physics},
    volume       = 59,
    number       = 3,
    pages        = {251--290},
    doi          = {10.1080/00107514.2018.1467648},
    archiveprefix = {arXiv},
    eprint       = {1805.06941},
    primaryclass = {astro-ph.IM},
    adsurl       = {https://ui.adsabs.harvard.edu/abs/2018ConPh..59..251K}
}

@article{2024MNRAS.532.4436B,
    title        = {{Stellar flares are far-ultraviolet luminous}},
    author       = {{Berger}, Vera L. and {Hinkle}, Jason T. and {Tucker}, Michael A. and {Shappee}, Benjamin J. and {van Saders}, Jennifer L. and {Huber}, Daniel and {Reep}, Jeffrey W. and {Sun}, Xudong and {Yang}, Kai E.},
    year         = 2024,
    month        = aug,
    journal      = {\mnras},
    volume       = 532,
    number       = 4,
    pages        = {4436--4445},
    doi          = {10.1093/mnras/stae1648},
    archiveprefix = {arXiv},
    eprint       = {2312.12511},
    primaryclass = {astro-ph.SR},
    adsurl       = {https://ui.adsabs.harvard.edu/abs/2024MNRAS.532.4436B}
}

@article{2016ApJ...820...89F,
    title        = {{The MUSCLES Treasury Survey. I. Motivation and Overview}},
    author       = {{France}, Kevin and {Loyd}, R.~O. Parke and {Youngblood}, Allison and {Brown}, Alexander and {Schneider}, P. Christian and {Hawley}, Suzanne L. and {Froning}, Cynthia S. and {Linsky}, Jeffrey L. and {Roberge}, Aki and {Buccino}, Andrea P. and {Davenport}, James R.~A. and {Fontenla}, Juan M. and {Kaltenegger}, Lisa and {Kowalski}, Adam F. and {Mauas}, Pablo J.~D. and {Miguel}, Yamila and {Redfield}, Seth and {Rugheimer}, Sarah and {Tian}, Feng and {Vieytes}, Mariela C. and {Walkowicz}, Lucianne M. and {Weisenburger}, Kolby L.},
    year         = 2016,
    month        = apr,
    journal      = {\apj},
    volume       = 820,
    number       = 2,
    pages        = 89,
    doi          = {10.3847/0004-637X/820/2/89},
    eid          = 89,
    archiveprefix = {arXiv},
    eprint       = {1602.09142},
    primaryclass = {astro-ph.SR},
    adsurl       = {https://ui.adsabs.harvard.edu/abs/2016ApJ...820...89F}
}

@article{2020ChEG...80l5594F,
    title        = {{Volatile element chemistry during accretion of the earth}},
    author       = {{Fegley}, Jr., Bruce and {Lodders}, Katharina and {Jacobson}, Nathan S.},
    year         = 2020,
    month        = apr,
    journal      = {Chemie der Erde / Geochemistry},
    volume       = 80,
    number       = 1,
    pages        = 125594,
    doi          = {10.1016/j.chemer.2019.125594},
    adsurl       = {https://ui.adsabs.harvard.edu/abs/2020ChEG...80l5594F}
}

@article{2020PSJ.....1...11Z,
    title        = {{Creation and Evolution of Impact-generated Reduced Atmospheres of Early Earth}},
    author       = {{Zahnle}, Kevin J. and {Lupu}, Roxana and {Catling}, David C. and {Wogan}, Nick},
    year         = 2020,
    month        = jun,
    journal      = {Planet. Sci. J.},
    volume       = 1,
    number       = 1,
    pages        = 11,
    doi          = {10.3847/PSJ/ab7e2c},
    eid          = 11,
    archiveprefix = {arXiv},
    eprint       = {2001.00095},
    primaryclass = {astro-ph.EP},
    adsurl       = {https://ui.adsabs.harvard.edu/abs/2020PSJ.....1...11Z}
}

@article{2023MNRAS.525.3703G,
    title        = {{A mineralogical reason why all exoplanets cannot be equally oxidizing}},
    author       = {{Guimond}, Claire Marie and {Shorttle}, Oliver and {Jordan}, Sean and {Rudge}, John F.},
    year         = 2023,
    month        = nov,
    journal      = {\mnras},
    volume       = 525,
    number       = 3,
    pages        = {3703--3717},
    doi          = {10.1093/mnras/stad2486},
    archiveprefix = {arXiv},
    eprint       = {2308.09505},
    primaryclass = {astro-ph.EP},
    adsurl       = {https://ui.adsabs.harvard.edu/abs/2023MNRAS.525.3703G}
}

@article{2012ApJ...745....3M,
    title        = {{The Atmospheric Chemistry of GJ 1214b: Photochemistry and Clouds}},
    author       = {{Miller-Ricci Kempton}, Eliza and {Zahnle}, Kevin and {Fortney}, Jonathan J.},
    year         = 2012,
    month        = jan,
    journal      = {\apj},
    volume       = 745,
    number       = 1,
    pages        = 3,
    doi          = {10.1088/0004-637X/745/1/3},
    eid          = 3,
    archiveprefix = {arXiv},
    eprint       = {1104.5477},
    primaryclass = {astro-ph.EP},
    adsurl       = {https://ui.adsabs.harvard.edu/abs/2012ApJ...745....3M}
}

@article{2016ApJ...821L..19N,
    title        = {{The Impact of Stellar Rotation on the Detectability of Habitable Planets around M Dwarfs}},
    author       = {{Newton}, Elisabeth R. and {Irwin}, Jonathan and {Charbonneau}, David and {Berta-Thompson}, Zachory K. and {Dittmann}, Jason A.},
    year         = 2016,
    month        = apr,
    journal      = {\apjl},
    volume       = 821,
    number       = 1,
    pages        = {L19},
    doi          = {10.3847/2041-8205/821/1/L19},
    eid          = {L19},
    archiveprefix = {arXiv},
    eprint       = {1604.03135},
    primaryclass = {astro-ph.EP},
    adsurl       = {https://ui.adsabs.harvard.edu/abs/2016ApJ...821L..19N}
}

@article{2015MNRAS.448.3053A,
    title        = {{Stellar activity as noise in exoplanet detection - II. Application to M dwarfs}},
    author       = {{Andersen}, J.~M. and {Korhonen}, H.},
    year         = 2015,
    month        = apr,
    journal      = {\mnras},
    volume       = 448,
    number       = 4,
    pages        = {3053--3069},
    doi          = {10.1093/mnras/stu2731},
    archiveprefix = {arXiv},
    eprint       = {1501.01302},
    primaryclass = {astro-ph.SR},
    adsurl       = {https://ui.adsabs.harvard.edu/abs/2015MNRAS.448.3053A}
}

@article{2025RvMPP...9...18H,
    title        = {{Atmospheric escape from exoplanets: recent observations and theoretical models}},
    author       = {{Hazra}, Gopal},
    year         = 2025,
    month        = may,
    journal      = {Reviews of Modern Plasma Physics},
    volume       = 9,
    number       = 1,
    pages        = 18,
    doi          = {10.1007/s41614-025-00195-6},
    eid          = 18,
    archiveprefix = {arXiv},
    eprint       = {2502.18124},
    primaryclass = {astro-ph.EP},
    adsurl       = {https://ui.adsabs.harvard.edu/abs/2025RvMPP...9...18H}
}

@article{2018ApJ...866L..18K,
    title        = {{Overcoming the Limitations of the Energy-limited Approximation for Planet Atmospheric Escape}},
    author       = {{Kubyshkina}, D. and {Fossati}, L. and {Erkaev}, N.~V. and {Cubillos}, P.~E. and {Johnstone}, C.~P. and {Kislyakova}, K.~G. and {Lammer}, H. and {Lendl}, M. and {Odert}, P.},
    year         = 2018,
    month        = oct,
    journal      = {\apjl},
    volume       = 866,
    number       = 2,
    pages        = {L18},
    doi          = {10.3847/2041-8213/aae586},
    eid          = {L18},
    archiveprefix = {arXiv},
    eprint       = {1810.06920},
    primaryclass = {astro-ph.EP},
    adsurl       = {https://ui.adsabs.harvard.edu/abs/2018ApJ...866L..18K}
}

@article{2018A&A...617A.107J,
    title        = {{Upper atmospheres of terrestrial planets: Carbon dioxide cooling and the Earth's thermospheric evolution}},
    author       = {{Johnstone}, C.~P. and {G{\"u}del}, M. and {Lammer}, H. and {Kislyakova}, K.~G.},
    year         = 2018,
    month        = sep,
    journal      = {\aap},
    volume       = 617,
    pages        = {A107},
    doi          = {10.1051/0004-6361/201832776},
    eid          = {A107},
    archiveprefix = {arXiv},
    eprint       = {1806.06897},
    primaryclass = {astro-ph.EP},
    adsurl       = {https://ui.adsabs.harvard.edu/abs/2018A&A...617A.107J}
}

@article{2019MNRAS.490.3760A,
    title        = {{Evolution of atmospheric escape in close-in giant planets and their associated Ly {\ensuremath{\alpha}} and H {\ensuremath{\alpha}} transit predictions}},
    author       = {{Allan}, A. and {Vidotto}, A.~A.},
    year         = 2019,
    month        = dec,
    journal      = {\mnras},
    volume       = 490,
    number       = 3,
    pages        = {3760--3771},
    doi          = {10.1093/mnras/stz2842},
    archiveprefix = {arXiv},
    eprint       = {1908.03510},
    primaryclass = {astro-ph.EP},
    adsurl       = {https://ui.adsabs.harvard.edu/abs/2019MNRAS.490.3760A}
}

@article{2020SSRv..216..129O,
    title        = {{Hydrogen Dominated Atmospheres on Terrestrial Mass Planets: Evidence, Origin and Evolution}},
    author       = {{Owen}, J.~E. and {Shaikhislamov}, I.~F. and {Lammer}, H. and {Fossati}, L. and {Khodachenko}, M.~L.},
    year         = 2020,
    month        = nov,
    journal      = {\ssr},
    volume       = 216,
    number       = 8,
    pages        = 129,
    doi          = {10.1007/s11214-020-00756-w},
    eid          = 129,
    archiveprefix = {arXiv},
    eprint       = {2010.15091},
    primaryclass = {astro-ph.EP},
    adsurl       = {https://ui.adsabs.harvard.edu/abs/2020SSRv..216..129O}
}

@article{2011Natur.470...53L,
    title        = {{A closely packed system of low-mass, low-density planets transiting Kepler-11}},
    author       = {{Lissauer}, Jack J. and {Fabrycky}, Daniel C. and {Ford}, Eric B. and {Borucki}, William J. and {Fressin}, Francois and {Marcy}, Geoffrey W. and {Orosz}, Jerome A. and {Rowe}, Jason F. and {Torres}, Guillermo and {Welsh}, William F. and {Batalha}, Natalie M. and {Bryson}, Stephen T. and {Buchhave}, Lars A. and {Caldwell}, Douglas A. and {Carter}, Joshua A. and {Charbonneau}, David and {Christiansen}, Jessie L. and {Cochran}, William D. and {Desert}, Jean-Michel and {Dunham}, Edward W. and {Fanelli}, Michael N. and {Fortney}, Jonathan J. and {Gautier}, III, Thomas N. and {Geary}, John C. and {Gilliland}, Ronald L. and {Haas}, Michael R. and {Hall}, Jennifer R. and {Holman}, Matthew J. and {Koch}, David G. and {Latham}, David W. and {Lopez}, Eric and {McCauliff}, Sean and {Miller}, Neil and {Morehead}, Robert C. and {Quintana}, Elisa V. and {Ragozzine}, Darin and {Sasselov}, Dimitar and {Short}, Donald R. and {Steffen}, Jason H.},
    year         = 2011,
    month        = feb,
    journal      = {\nat},
    volume       = 470,
    number       = 7332,
    pages        = {53--58},
    doi          = {10.1038/nature09760},
    archiveprefix = {arXiv},
    eprint       = {1102.0291},
    primaryclass = {astro-ph.EP},
    adsurl       = {https://ui.adsabs.harvard.edu/abs/2011Natur.470...53L}
}

@article{2020A&A...643A..81K,
    title        = {{Effect of mantle oxidation state and escape upon the evolution of Earth's magma ocean atmosphere}},
    author       = {{Katyal}, Nisha and {Ortenzi}, Gianluigi and {Lee Grenfell}, John and {Noack}, Lena and {Sohl}, Frank and {Godolt}, Mareike and {Garc{\'\i}a Mu{\~n}oz}, Antonio and {Schreier}, Franz and {Wunderlich}, Fabian and {Rauer}, Heike},
    year         = 2020,
    month        = nov,
    journal      = {\aap},
    volume       = 643,
    pages        = {A81},
    doi          = {10.1051/0004-6361/202038779},
    eid          = {A81},
    archiveprefix = {arXiv},
    eprint       = {2009.14599},
    primaryclass = {astro-ph.EP},
    adsurl       = {https://ui.adsabs.harvard.edu/abs/2020A&A...643A..81K}
}

@article{2013E&PSL.375..312K,
    title        = {{Effective hydrodynamic hydrogen escape from an early Earth atmosphere inferred from high-accuracy numerical simulation}},
    author       = {{Kuramoto}, Kiyoshi and {Umemoto}, Takafumi and {Ishiwatari}, Masaki},
    year         = 2013,
    month        = aug,
    journal      = {Earth and Planetary Science Letters},
    volume       = 375,
    pages        = {312--318},
    doi          = {10.1016/j.epsl.2013.05.050},
    adsurl       = {https://ui.adsabs.harvard.edu/abs/2013E&PSL.375..312K}
}

@article{2017ApJ...843..122Z,
    title        = {{The Cosmic Shoreline: The Evidence that Escape Determines which Planets Have Atmospheres, and what this May Mean for Proxima Centauri B}},
    author       = {{Zahnle}, Kevin J. and {Catling}, David C.},
    year         = 2017,
    month        = jul,
    journal      = {\apj},
    volume       = 843,
    number       = 2,
    pages        = 122,
    doi          = {10.3847/1538-4357/aa7846},
    eid          = 122,
    archiveprefix = {arXiv},
    eprint       = {1702.03386},
    primaryclass = {astro-ph.EP},
    adsurl       = {https://ui.adsabs.harvard.edu/abs/2017ApJ...843..122Z}
}

@article{2022ApJ...934..137Y,
    title        = {{Less Effective Hydrodynamic Escape of H$_{2}$-H$_{2}$O Atmospheres on Terrestrial Planets Orbiting Pre-main-sequence M Dwarfs}},
    author       = {{Yoshida}, Tatsuya and {Terada}, Naoki and {Ikoma}, Masahiro and {Kuramoto}, Kiyoshi},
    year         = 2022,
    month        = aug,
    journal      = {\apj},
    volume       = 934,
    number       = 2,
    pages        = 137,
    doi          = {10.3847/1538-4357/ac7be7},
    eid          = 137,
    archiveprefix = {arXiv},
    eprint       = {2207.06570},
    primaryclass = {astro-ph.EP},
    adsurl       = {https://ui.adsabs.harvard.edu/abs/2022ApJ...934..137Y}
}

@article{2015JATIS...1a4003R,
    title        = {{Transiting Exoplanet Survey Satellite (TESS)}},
    author       = {{Ricker}, George R. and {Winn}, Joshua N. and {Vanderspek}, Roland and {Latham}, David W. and {Bakos}, G{\'a}sp{\'a}r {\'A}. and {Bean}, Jacob L. and {Berta-Thompson}, Zachory K. and {Brown}, Timothy M. and {Buchhave}, Lars and {Butler}, Nathaniel R. and {Butler}, R. Paul and {Chaplin}, William J. and {Charbonneau}, David and {Christensen-Dalsgaard}, J{\o}rgen and {Clampin}, Mark and {Deming}, Drake and {Doty}, John and {De Lee}, Nathan and {Dressing}, Courtney and {Dunham}, Edward W. and {Endl}, Michael and {Fressin}, Francois and {Ge}, Jian and {Henning}, Thomas and {Holman}, Matthew J. and {Howard}, Andrew W. and {Ida}, Shigeru and {Jenkins}, Jon M. and {Jernigan}, Garrett and {Johnson}, John Asher and {Kaltenegger}, Lisa and {Kawai}, Nobuyuki and {Kjeldsen}, Hans and {Laughlin}, Gregory and {Levine}, Alan M. and {Lin}, Douglas and {Lissauer}, Jack J. and {MacQueen}, Phillip and {Marcy}, Geoffrey and {McCullough}, Peter R. and {Morton}, Timothy D. and {Narita}, Norio and {Paegert}, Martin and {Palle}, Enric and {Pepe}, Francesco and {Pepper}, Joshua and {Quirrenbach}, Andreas and {Rinehart}, Stephen A. and {Sasselov}, Dimitar and {Sato}, Bun'ei and {Seager}, Sara and {Sozzetti}, Alessandro and {Stassun}, Keivan G. and {Sullivan}, Peter and {Szentgyorgyi}, Andrew and {Torres}, Guillermo and {Udry}, Stephane and {Villasenor}, Joel},
    year         = 2015,
    month        = jan,
    journal      = {Journal of Astronomical Telescopes, Instruments, and Systems},
    volume       = 1,
    pages        = {014003},
    doi          = {10.1117/1.JATIS.1.1.014003},
    eid          = {014003},
    adsurl       = {https://ui.adsabs.harvard.edu/abs/2015JATIS...1a4003R}
}

@article{2019ApJ...871L..26F,
    title        = {{A Hot Ultraviolet Flare on the M Dwarf Star GJ 674}},
    author       = {{Froning}, Cynthia S. and {Kowalski}, Adam and {France}, Kevin and {Loyd}, R.~O. Parke and {Schneider}, P. Christian and {Youngblood}, Allison and {Wilson}, David and {Brown}, Alexander and {Berta-Thompson}, Zachory and {Pineda}, J. Sebastian and {Linsky}, Jeffrey and {Rugheimer}, Sarah and {Miguel}, Yamila},
    year         = 2019,
    month        = feb,
    journal      = {\apjl},
    volume       = 871,
    number       = 2,
    pages        = {L26},
    doi          = {10.3847/2041-8213/aaffcd},
    eid          = {L26},
    archiveprefix = {arXiv},
    eprint       = {1901.08647},
    primaryclass = {astro-ph.SR},
    adsurl       = {https://ui.adsabs.harvard.edu/abs/2019ApJ...871L..26F}
}

@article{2024ApJ...971...24P,
    title        = {{A Multiwavelength Survey of Nearby M Dwarfs: Optical and Near-ultraviolet Flares and Activity with Contemporaneous TESS, Kepler/K2, Swift, and HST Observations}},
    author       = {{Paudel}, Rishi R. and {Barclay}, Thomas and {Youngblood}, Allison and {Quintana}, Elisa V. and {Schlieder}, Joshua E. and {Vega}, Laura D. and {Gilbert}, Emily A. and {Osten}, Rachel A. and {Peacock}, Sarah and {Tristan}, Isaiah I. and {Feliz}, Dax L. and {Boyd}, Patricia T. and {Davenport}, James R.~A. and {Huber}, Daniel and {Kowalski}, Adam F. and {Monsue}, Teresa and {Silverstein}, Michele L.},
    year         = 2024,
    month        = aug,
    journal      = {\apj},
    volume       = 971,
    number       = 1,
    pages        = 24,
    doi          = {10.3847/1538-4357/ad487d},
    eid          = 24,
    archiveprefix = {arXiv},
    eprint       = {2404.12310},
    primaryclass = {astro-ph.SR},
    adsurl       = {https://ui.adsabs.harvard.edu/abs/2024ApJ...971...24P}
}

@article{2012ApJ...761..166H,
    title        = {{Photochemistry in Terrestrial Exoplanet Atmospheres. I. Photochemistry Model and Benchmark Cases}},
    author       = {{Hu}, Renyu and {Seager}, Sara and {Bains}, William},
    year         = 2012,
    month        = dec,
    journal      = {\apj},
    volume       = 761,
    number       = 2,
    pages        = 166,
    doi          = {10.1088/0004-637X/761/2/166},
    eid          = 166,
    archiveprefix = {arXiv},
    eprint       = {1210.6885},
    primaryclass = {astro-ph.EP},
    adsurl       = {https://ui.adsabs.harvard.edu/abs/2012ApJ...761..166H}
}

@article{2025ApJ...993...41G,
    title        = {{Simulations of Flare Chemistry in Brown Dwarf Companions to Active M Dwarfs}},
    author       = {{Gibbs}, Aidan and {Fitzgerald}, Michael P.},
    year         = 2025,
    month        = nov,
    journal      = {\apj},
    volume       = 993,
    number       = 1,
    pages        = 41,
    doi          = {10.3847/1538-4357/ae0477},
    eid          = 41,
    archiveprefix = {arXiv},
    eprint       = {2509.07063},
    primaryclass = {astro-ph.SR},
    adsurl       = {https://ui.adsabs.harvard.edu/abs/2025ApJ...993...41G}
}

@article{2022A&A...667A..15K,
    title        = {{Impact of stellar flares on the chemical composition and transmission spectra of gaseous exoplanets orbiting M dwarfs}},
    author       = {{Konings}, T. and {Baeyens}, R. and {Decin}, L.},
    year         = 2022,
    month        = nov,
    journal      = {\aap},
    volume       = 667,
    pages        = {A15},
    doi          = {10.1051/0004-6361/202243436},
    eid          = {A15},
    archiveprefix = {arXiv},
    eprint       = {2209.02483},
    primaryclass = {astro-ph.EP},
    adsurl       = {https://ui.adsabs.harvard.edu/abs/2022A&A...667A..15K}
}

@article{2023MNRAS.523.5681N,
    title        = {{Temperature-chemistry coupling in the evolution of gas giant atmospheres driven by stellar flares}},
    author       = {{Nicholls}, Harrison and {H{\'e}brard}, Eric and {Venot}, Olivia and {Drummond}, Benjamin and {Evans}, Elise},
    year         = 2023,
    month        = aug,
    journal      = {\mnras},
    volume       = 523,
    number       = 4,
    pages        = {5681--5702},
    doi          = {10.1093/mnras/stad1734},
    archiveprefix = {arXiv},
    eprint       = {2306.03673},
    primaryclass = {astro-ph.EP},
    adsurl       = {https://ui.adsabs.harvard.edu/abs/2023MNRAS.523.5681N}
}

@dataset{2020yCat.1350....0G,
    title        = {{VizieR Online Data Catalog: Gaia EDR3 (Gaia Collaboration, 2020)}},
    author       = {{Gaia Collaboration}},
    year         = 2020,
    month        = nov,
    doi          = {10.26093/cds/vizier.1350},
    howpublished = {VizieR On-line Data Catalog: I/350.  Originally published in: 2021A\&A...649A...1G},
    eid          = {I/350},
    adsurl       = {https://ui.adsabs.harvard.edu/abs/2020yCat.1350....0G}
}

@article{2021NatAs...5..298C,
    title        = {{Persistence of flare-driven atmospheric chemistry on rocky habitable zone worlds}},
    author       = {{Chen}, Howard and {Zhan}, Zhuchang and {Youngblood}, Allison and {Wolf}, Eric T. and {Feinstein}, Adina D. and {Horton}, Daniel E.},
    year         = 2021,
    month        = jan,
    journal      = {Nature Astronomy},
    volume       = 5,
    pages        = {298--310},
    doi          = {10.1038/s41550-020-01264-1},
    archiveprefix = {arXiv},
    eprint       = {2101.04507},
    primaryclass = {astro-ph.EP},
    adsurl       = {https://ui.adsabs.harvard.edu/abs/2021NatAs...5..298C}
}

@article{2023MNRAS.518.2472R,
    title        = {{3D modelling of the impact of stellar activity on tidally locked terrestrial exoplanets: atmospheric composition and habitability}},
    author       = {{Ridgway}, R.~J. and {Zamyatina}, M. and {Mayne}, N.~J. and {Manners}, J. and {Lambert}, F.~H. and {Braam}, M. and {Drummond}, B. and {H{\'e}brard}, E. and {Palmer}, P.~I. and {Kohary}, K.},
    year         = 2023,
    month        = jan,
    journal      = {\mnras},
    volume       = 518,
    number       = 2,
    pages        = {2472--2496},
    doi          = {10.1093/mnras/stac3105},
    archiveprefix = {arXiv},
    eprint       = {2210.13257},
    primaryclass = {astro-ph.EP},
    adsurl       = {https://ui.adsabs.harvard.edu/abs/2023MNRAS.518.2472R}
}

@article{2025AJ....170...40C,
    title        = {{Effects of Transient Stellar Emissions on Planetary Climates of Tidally Locked Exo-Earths}},
    author       = {{Chen}, Howard and {De Luca}, Paolo and {Hochman}, Assaf and {Komacek}, Thaddeus D.},
    year         = 2025,
    month        = jul,
    journal      = {\aj},
    volume       = 170,
    number       = 1,
    pages        = 40,
    doi          = {10.3847/1538-3881/add33e},
    eid          = 40,
    archiveprefix = {arXiv},
    eprint       = {2505.03723},
    primaryclass = {astro-ph.EP},
    adsurl       = {https://ui.adsabs.harvard.edu/abs/2025AJ....170...40C}
}

@article{2026A&A...705A.165M,
    title        = {{Young M-dwarfs flare activity model: Towards better exoplanetary atmospheric characterisation}},
    author       = {{Mamonova}, E. and {Kowalski}, A.~F. and {Herbst}, K. and {Wedemeyer}, S. and {Werner}, S.~C.},
    year         = 2026,
    month        = jan,
    journal      = {\aap},
    volume       = 705,
    pages        = {A165},
    doi          = {10.1051/0004-6361/202556844},
    eid          = {A165},
    archiveprefix = {arXiv},
    eprint       = {2511.23129},
    primaryclass = {astro-ph.SR},
    adsurl       = {https://ui.adsabs.harvard.edu/abs/2026A&A...705A.165M}
}

@article{2018AsBio..18..630M,
    title        = {{Exoplanet Biosignatures: Understanding Oxygen as a Biosignature in the Context of Its Environment}},
    author       = {{Meadows}, Victoria S. and {Reinhard}, Christopher T. and {Arney}, Giada N. and {Parenteau}, Mary N. and {Schwieterman}, Edward W. and {Domagal-Goldman}, Shawn D. and {Lincowski}, Andrew P. and {Stapelfeldt}, Karl R. and {Rauer}, Heike and {DasSarma}, Shiladitya and {Hegde}, Siddharth and {Narita}, Norio and {Deitrick}, Russell and {Lustig-Yaeger}, Jacob and {Lyons}, Timothy W. and {Siegler}, Nicholas and {Grenfell}, J. Lee},
    year         = 2018,
    month        = jun,
    journal      = {Astrobiology},
    volume       = 18,
    number       = 6,
    pages        = {630--662},
    doi          = {10.1089/ast.2017.1727},
    archiveprefix = {arXiv},
    eprint       = {1705.07560},
    primaryclass = {astro-ph.EP},
    adsurl       = {https://ui.adsabs.harvard.edu/abs/2018AsBio..18..630M}
}

@article{2006SSRv..123..485G,
    title        = {{The James Webb Space Telescope}},
    author       = {{Gardner}, Jonathan P. and {Mather}, John C. and {Clampin}, Mark and {Doyon}, Rene and {Greenhouse}, Matthew A. and {Hammel}, Heidi B. and {Hutchings}, John B. and {Jakobsen}, Peter and {Lilly}, Simon J. and {Long}, Knox S. and {Lunine}, Jonathan I. and {McCaughrean}, Mark J. and {Mountain}, Matt and {Nella}, John and {Rieke}, George H. and {Rieke}, Marcia J. and {Rix}, Hans-Walter and {Smith}, Eric P. and {Sonneborn}, George and {Stiavelli}, Massimo and {Stockman}, H.~S. and {Windhorst}, Rogier A. and {Wright}, Gillian S.},
    year         = 2006,
    month        = apr,
    journal      = {\ssr},
    volume       = 123,
    number       = 4,
    pages        = {485--606},
    doi          = {10.1007/s11214-006-8315-7},
    archiveprefix = {arXiv},
    eprint       = {astro-ph/0606175},
    primaryclass = {astro-ph},
    adsurl       = {https://ui.adsabs.harvard.edu/abs/2006SSRv..123..485G}
}

@article{2018ExA....46..135T,
    title        = {{A chemical survey of exoplanets with ARIEL}},
    author       = {{Tinetti}, Giovanna and {Drossart}, Pierre and {Eccleston}, Paul and {Hartogh}, Paul and {Heske}, Astrid and {Leconte}, J{\'e}r{\'e}my and {Micela}, Giusi and {Ollivier}, Marc and {Pilbratt}, G{\"o}ran and {Puig}, Ludovic and {Turrini}, Diego and {Vandenbussche}, Bart and {Wolkenberg}, Paulina and {Beaulieu}, Jean-Philippe and {Buchave}, Lars A. and {Ferus}, Martin and {Griffin}, Matt and {Guedel}, Manuel and {Justtanont}, Kay and {Lagage}, Pierre-Olivier and {Machado}, Pedro and {Malaguti}, Giuseppe and {Min}, Michiel and {N{\o}rgaard-Nielsen}, Hans Ulrik and {Rataj}, Mirek and {Ray}, Tom and {Ribas}, Ignasi and {Swain}, Mark and {Szabo}, Robert and {Werner}, Stephanie and {Barstow}, Joanna and {Burleigh}, Matt and {Cho}, James and {Coud{\'e} du Foresto}, Vincent and {Coustenis}, Athena and {Decin}, Leen and {Encrenaz}, Therese and {Galand}, Marina and {Gillon}, Michael and {Helled}, Ravit and {Morales}, Juan Carlos and {Garc{\'\i}a Mu{\~n}oz}, Antonio and {Moneti}, Andrea and {Pagano}, Isabella and {Pascale}, Enzo and {Piccioni}, Giuseppe and {Pinfield}, David and {Sarkar}, Subhajit and {Selsis}, Franck and {Tennyson}, Jonathan and {Triaud}, Amaury and {Venot}, Olivia and {Waldmann}, Ingo and {Waltham}, David and {Wright}, Gillian and {Amiaux}, Jerome and {Augu{\`e}res}, Jean-Louis and {Berth{\'e}}, Michel and {Bezawada}, Naidu and {Bishop}, Georgia and {Bowles}, Neil and {Coffey}, Deirdre and {Colom{\'e}}, Josep and {Crook}, Martin and {Crouzet}, Pierre-Elie and {Da Peppo}, Vania and {Sanz}, Isabel Escudero and {Focardi}, Mauro and {Frericks}, Martin and {Hunt}, Tom and {Kohley}, Ralf and {Middleton}, Kevin and {Morgante}, Gianluca and {Ottensamer}, Roland and {Pace}, Emanuele and {Pearson}, Chris and {Stamper}, Richard and {Symonds}, Kate and {Rengel}, Miriam and {Renotte}, Etienne and {Ade}, Peter and {Affer}, Laura and {Alard}, Christophe and {Allard}, Nicole and {Altieri}, Francesca and {Andr{\'e}}, Yves and {Arena}, Claudio and {Argyriou}, Ioannis and {Aylward}, Alan and {Baccani}, Cristian and {Bakos}, Gaspar and {Banaszkiewicz}, Marek and {Barlow}, Mike and {Batista}, Virginie and {Bellucci}, Giancarlo and {Benatti}, Serena and {Bernardi}, Pernelle and {B{\'e}zard}, Bruno and {Blecka}, Maria and {Bolmont}, Emeline and {Bonfond}, Bertrand and {Bonito}, Rosaria and {Bonomo}, Aldo S. and {Brucato}, John Robert and {Brun}, Allan Sacha and {Bryson}, Ian and {Bujwan}, Waldemar and {Casewell}, Sarah and {Charnay}, Bejamin and {Pestellini}, Cesare Cecchi and {Chen}, Guo and {Ciaravella}, Angela and {Claudi}, Riccardo and {Cl{\'e}dassou}, Rodolphe and {Damasso}, Mario and {Damiano}, Mario and {Danielski}, Camilla and {Deroo}, Pieter and {Di Giorgio}, Anna Maria and {Dominik}, Carsten and {Doublier}, Vanessa and {Doyle}, Simon and {Doyon}, Ren{\'e} and {Drummond}, Benjamin and {Duong}, Bastien and {Eales}, Stephen and {Edwards}, Billy and {Farina}, Maria and {Flaccomio}, Ettore and {Fletcher}, Leigh and {Forget}, Fran{\c{c}}ois and {Fossey}, Steve and {Fr{\"a}nz}, Markus and {Fujii}, Yuka and {Garc{\'\i}a-Piquer}, {\'A}lvaro and {Gear}, Walter and {Geoffray}, Herv{\'e} and {G{\'e}rard}, Jean Claude and {Gesa}, Lluis and {Gomez}, H. and {Graczyk}, Rafa{\l} and {Griffith}, Caitlin and {Grodent}, Denis and {Guarcello}, Mario Giuseppe and {Gustin}, Jacques and {Hamano}, Keiko and {Hargrave}, Peter and {Hello}, Yann and {Heng}, Kevin and {Herrero}, Enrique and {Hornstrup}, Allan and {Hubert}, Benoit and {Ida}, Shigeru and {Ikoma}, Masahiro and {Iro}, Nicolas and {Irwin}, Patrick and {Jarchow}, Christopher and {Jaubert}, Jean and {Jones}, Hugh and {Julien}, Queyrel and {Kameda}, Shingo and {Kerschbaum}, Franz and {Kervella}, Pierre and {Koskinen}, Tommi and {Krijger}, Matthijs and {Krupp}, Norbert and {Lafarga}, Marina and {Landini}, Federico and {Lellouch}, Emanuel and {Leto}, Giuseppe and {Luntzer}, A. and {Rank-L{\"u}ftinger}, Theresa and {Maggio}, Antonio and {Maldonado}, Jesus and {Maillard}, Jean-Pierre and {Mall}, Urs and {Marquette}, Jean-Baptiste and {Mathis}, Stephane and {Maxted}, Pierre and {Matsuo}, Taro and {Medvedev}, Alexander and {Miguel}, Yamila and {Minier}, Vincent and {Morello}, Giuseppe and {Mura}, Alessandro and {Narita}, Norio and {Nascimbeni}, Valerio and {Nguyen Tong}, N. and {Noce}, Vladimiro and {Oliva}, Fabrizio and {Palle}, Enric and {Palmer}, Paul and {Pancrazzi}, Maurizio and {Papageorgiou}, Andreas and {Parmentier}, Vivien and {Perger}, Manuel and {Petralia}, Antonino and {Pezzuto}, Stefano and {Pierrehumbert}, Ray and {Pillitteri}, Ignazio},
    year         = 2018,
    month        = nov,
    journal      = {Experimental Astronomy},
    volume       = 46,
    number       = 1,
    pages        = {135--209},
    doi          = {10.1007/s10686-018-9598-x},
    adsurl       = {https://ui.adsabs.harvard.edu/abs/2018ExA....46..135T}
}

@article{2023AJ....165..195B,
    title        = {{Coronal X-Ray Emission from Nearby, Low-mass, Exoplanet Host Stars Observed by the MUSCLES and Mega-MUSCLES HST Treasury Survey Projects}},
    author       = {{Brown}, Alexander and {Schneider}, P. Christian and {France}, Kevin and {Froning}, Cynthia S. and {Youngblood}, Allison A. and {J. Wilson}, David and {Loyd}, R.~O. Parke and {Pineda}, J. Sebastian and {Duvvuri}, Girish M. and {Kowalski}, Adam F. and {Berta-Thompson}, Zachory K.},
    year         = 2023,
    month        = may,
    journal      = {\aj},
    volume       = 165,
    number       = 5,
    pages        = 195,
    doi          = {10.3847/1538-3881/acc38a},
    eid          = 195,
    archiveprefix = {arXiv},
    eprint       = {2303.12929},
    primaryclass = {astro-ph.SR},
    adsurl       = {https://ui.adsabs.harvard.edu/abs/2023AJ....165..195B}
}

\begin{appendix}

\section{Modelling approach limitations}
\label{sec:appedix2}

\subsection{Helios T-P profiles and 1D atmospheric models}
While HELIOS generates reliable atmospheric profiles under radiative equilibrium and offers flexibility in input chemistry and stellar parameters, the code is subject to several limitations. The 1D framework neglects horizontal inhomogeneities and atmospheric circulation that can affect thermal structures. Our interpretation also excludes cloud formation, the time-dependent stellar variability is not yet realised in HELIOS, and the computations consider equilibrium chemistry, limiting the code application in atmospheres dominated by dynamic or transient stellar phenomena.

Following \citet{2019ApJ...886..142M}, who use ideal gas approximations, \citet{2023ApJ...953...57K} propose an alternative method deriving adiabatic coefficients and specific heat capacities from gas entropy accounting for local chemical composition. Their T-P profiles for planets with Earth-like gravity and 500 K equilibrium temperature align closely with ours near the upper atmosphere, despite differences at higher pressures due to structural assumptions (with or without surface, the former assumption is adopted in this study, see Fig.~\ref{fig:18}, right panel in Sect.~\ref{sec:resultfa}).

\subsection{YMDF and FF models during the largest flares in the dataset}
\label{sec:appendixsf}
\begin{figure}\resizebox{\hsize}{!}{
   \centering
   \includegraphics{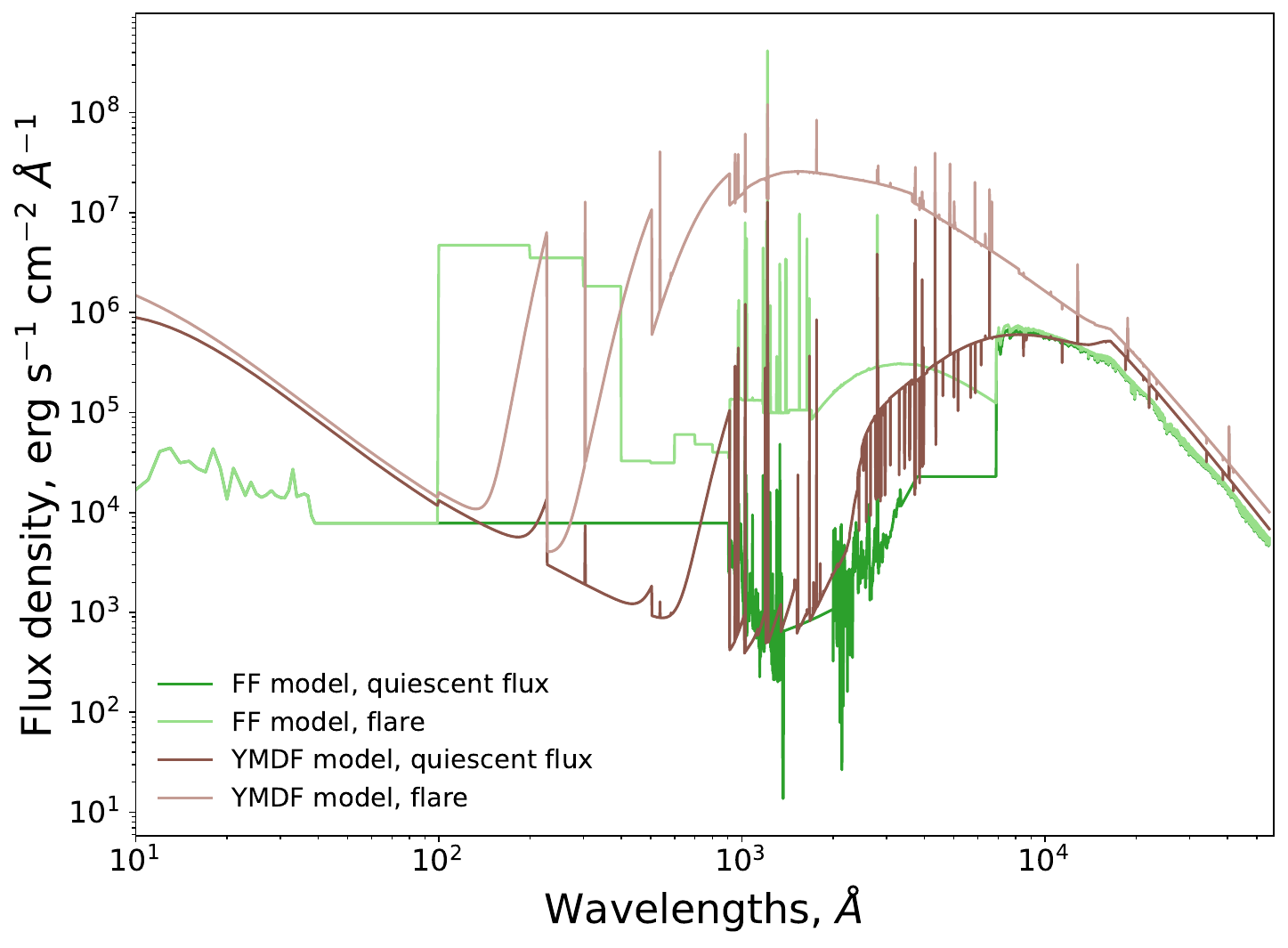}
    }
\caption{Flux densities for the FF and YMDF models: quiescent and averaged over the duration of the largest synthetic flare in the respective datasets. Fluxes are shown for the quiescent state (solid brown and green lines for the YMDF and FF models, respectively) and flare-averaged state (solid light-brown and light-green lines for the YMDF and FF models, respectively). The YMDF model exhibits a clear excess in UV and NUV flux density during its largest recorded flare in the dataset compared to the largest recorded flare in the FF model, whilst at wavelengths beyond $\sim$ 3200 $\AA$ both models show good agreement with each other and only a marginal excess above the respective quiescent stellar continuum.}
\label{fig:98}
\end{figure}

Figure~\ref{fig:98} shows the flux density spectra of the YMDF and FF stellar models in both quiescent and flare states. The spectra of the largest synthetic flares in the YMDF and FF datasets (see the flare data stated in Sect.~\ref{sec:resultsc}) are plotted as the average flux density over the duration of the flare. In the EUV band, the FF flare is represented as a coarse histogram due to missing spectral data in this range in the panchromatic quiescent spectra from \citet{2022AJ....164..110F}, and this effect is described in \citetalias{2026A&A...705A.165M}. The YMDF flare therefore shows smaller flux values compared to the FF model. This spectral region is important for the photodissociation of molecules such as H$_2$O and CO$_2$, and both models suffer from a lack of observational data that could help better characterise this regime. In the FUV, both flare spectra are highly structured owing to Lyman-series absorption and emission features. However, the most energetic flare in the FF model is a factor of $\sim$ 10$^2$ lower than the YMDF model across this region. The Lyman-$\alpha$ line appears as a sharp absorption trough in both models, though the associated line emission flanking it is considerably more pronounced in YMDF. In the NUV, the YMDF flare remains elevated, declining only gradually toward longer wavelengths. The FF flare rises modestly in this window, with the YMDF excess over FF somewhat smaller than in the FUV. This region overlaps strongly with the photolysis cross-sections of photochemically active species, making the flux discrepancy between the two models particularly consequential for photochemical modelling. Beyond $\sim$ 3200 \AA, the two flare spectra converge toward their respective quiescent spectra, both following the photospheric continuum into the infrared.

\subsection{Rate coefficients' evolution induced by flares}
\label{sec:appendixrt}

We introduced variable stellar flux to the atmospheres, where calculates chemical kinetic evolution of species, and these changes directly affect the rate coefficients $J_s(z)$, which govern the photodissociation of each chemical species $s$ at altitude $z$. The rate coefficient is computed as
\begin{equation}
    J_s(z) = \int_{\lambda} \Phi(z, \lambda)\, \sigma_s(\lambda)\, \mathrm{d}\lambda,
\end{equation}
where $\Phi(z, \lambda)$ is the actinic flux, representing the total photon flux integrated over all directions available to a molecule at altitude $z$ and wavelength $\lambda$, and $\sigma_s(\lambda)$ is the photodissociation cross-section of species $s$. The actinic flux is derived from the total radiative intensity via $\Phi = F_\mathrm{tot} / (hc/\lambda)$, where $h$ is the Planck constant and $c$ is the speed of light, converting energy flux in erg s$^{-1}$ cm$^{-2}$ \AA$^{-1}$ into photon number flux in photons s$^{-1}$ cm$^{-2}$ \AA$^{-1}$. The stellar spectrum is interpolated onto a uniform wavelength grid of 240 bins with a resolution of 0.1 nm below 240 nm and 2.0 nm above 240 nm, and the cross-sections are binned onto the same grid.

\begin{figure*}\resizebox{\hsize}{!}{
   \centering
   \includegraphics{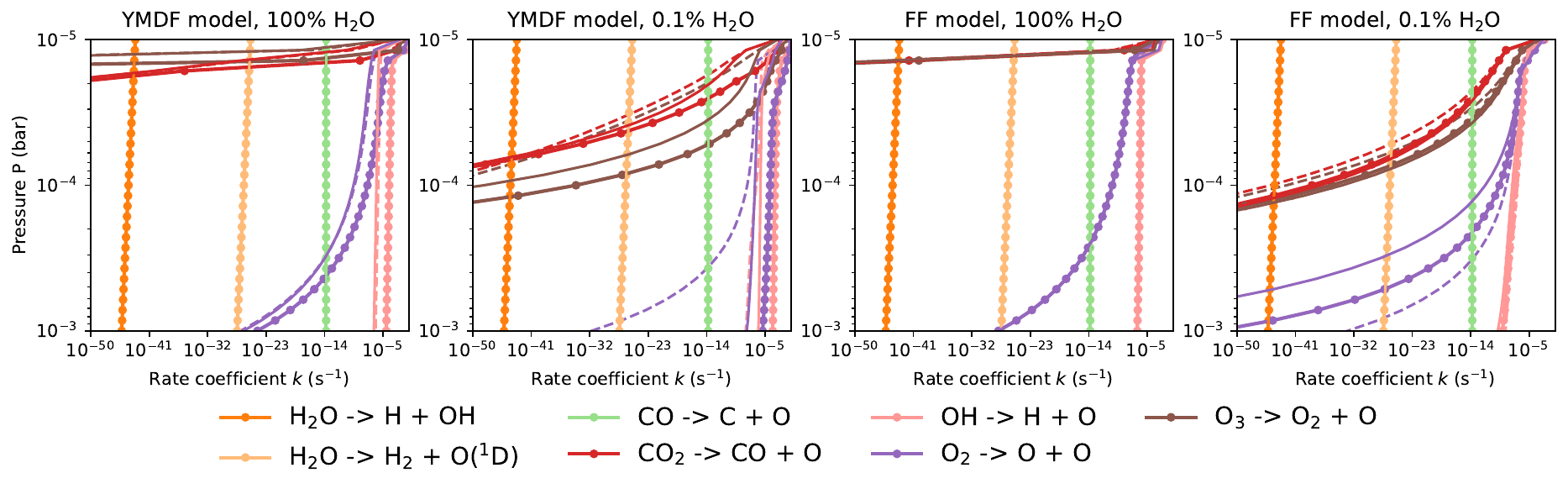}
    }
\caption{Photolysis rate coefficients $J_s(z)$ as a function of altitude for selected chemical reactions, computed in VULCAN for the YMDF and FF stellar models, and shown for the upper layers of the simulated atmospheres. Dashed lines show the atmospheric state at the end of the 10-day integration, solid lines at 360 days. In the left two panels, lines with markers correspond to the YMDF simulation at 260 days after the largest flare at day 253, and, in the two panel to the right, these lines correspond to the FF simulation at 140 days after the largest flare at day 136.}
\label{fig:99}
\end{figure*}

The photolysis rate coefficients were sampled several days after the largest flare in each model rather than at the moment of peak flux, as the atmospheric chemistry requires time to respond to the elevated radiation field and sampling at a later epoch reveals for how long the cumulative photochemical perturbation persists. Figure~\ref{fig:99} shows the resulting changes in selected photolysis rate coefficients across the atmosphere for both stellar models. The dashed lines correspond to the state of the atmosphere at the end of a 10-day integration, whilst the solid lines show the final state at 360 days. Lines with markers indicate the atmospheric state in the YMDF simulation at 260 days, following the largest flare which occurred at day 243, and in the FF simulation at 140 days, following the largest flare which occurred at day 136. The elevated photolysis rates driven by the flare flux persist differently between the two models, reflecting the spectral differences in the FUV and NUV discussed above. We note that, in the FF model, the chemical signature of the largest flare has largely dissipated by day 4, whilst in the YMDF model the flare-driven perturbation remains clearly detectable 17 days after the event, reflecting the substantially larger energy input. However, examining the photolysis rates at the end of the simulations, the FF model shows that the accumulated atmospheric stress either didn't change as for H$_2$O photodissociation reactions or converges to values close to those observed after the largest flare in other cases, whilst in the YMDF model the substantially larger flare-driven perturbations tend to relax without departing severely from the initial values for the similar reactions. It should be emphasised that a large rate coefficient does not necessarily produce significant chemical change if the number density of the target species is low; conversely, in H$_2$O-rich atmospheres photodissociation proceeds efficiently even at modest rates due to the large molecular reservoir.

\subsection{Thermochemical assumptions}
VULCAN, like all 1D models, is inherently limited by its neglect of atmospheric dynamics, spatial heterogeneity, clouds, and surface-atmosphere interactions. Uncertainties in reaction rates and limited reaction networks constrain chemical pathway accuracy, especially under poorly explored conditions. The current version of VULCAN does not incorporate ionisation processes, which may become significant in highly irradiated or strongly ionised environments. Despite these limitations, VULCAN provides a robust and widely validated framework for probing the fundamental processes of photochemistry and thermochemistry in young, volatile-rich planetary atmospheres, as encountered around active early M dwarfs.

\subsection{MLT in eddy diffusion coefficient calculation}
\label{sec:appmlt}

The vertical eddy diffusion coefficient $K_{zz}$ is parameterised using a mixing-length theory (MLT) approach, $K_{zz} = C\cdot l \cdot w_t$, where $C\sim$1, $l$ is a characteristic mixing length, and $w_t$ is a turbulent velocity under the assumption of free convection. This formulation is widely used for H$_2$-dominated atmospheres where convection provides efficient vertical mixing over large pressure ranges \citep{2013A&A...558A..91P,2013ApJ...777...34M,2015MNRAS.446..345M}. We adopt the empirically calibrated pressure scaling from \citet{2022ExA....53..279M} and \citet{2023MNRAS.521.3333L}, where they assumed that the turbulent velocity is govern by stellar effective temperature, as given in Sect.~\ref{sec:methodsvm}, which yields $K_{zz}$ values broadly consistent with global values can be found in the literature (10$^5$--10$^{10}$cm$^2$s$^{-1}$, \citealt{2023MNRAS.521.3333L}, their Figure 4) while allowing exploration across our atmosphere grid. Applied to rocky planets, this MLT-based prescription should be regarded as a physically motivated scaling law rather than a fully self-consistent dynamical model.

\begin{figure}\resizebox{\hsize}{!}{
   \centering
   \includegraphics{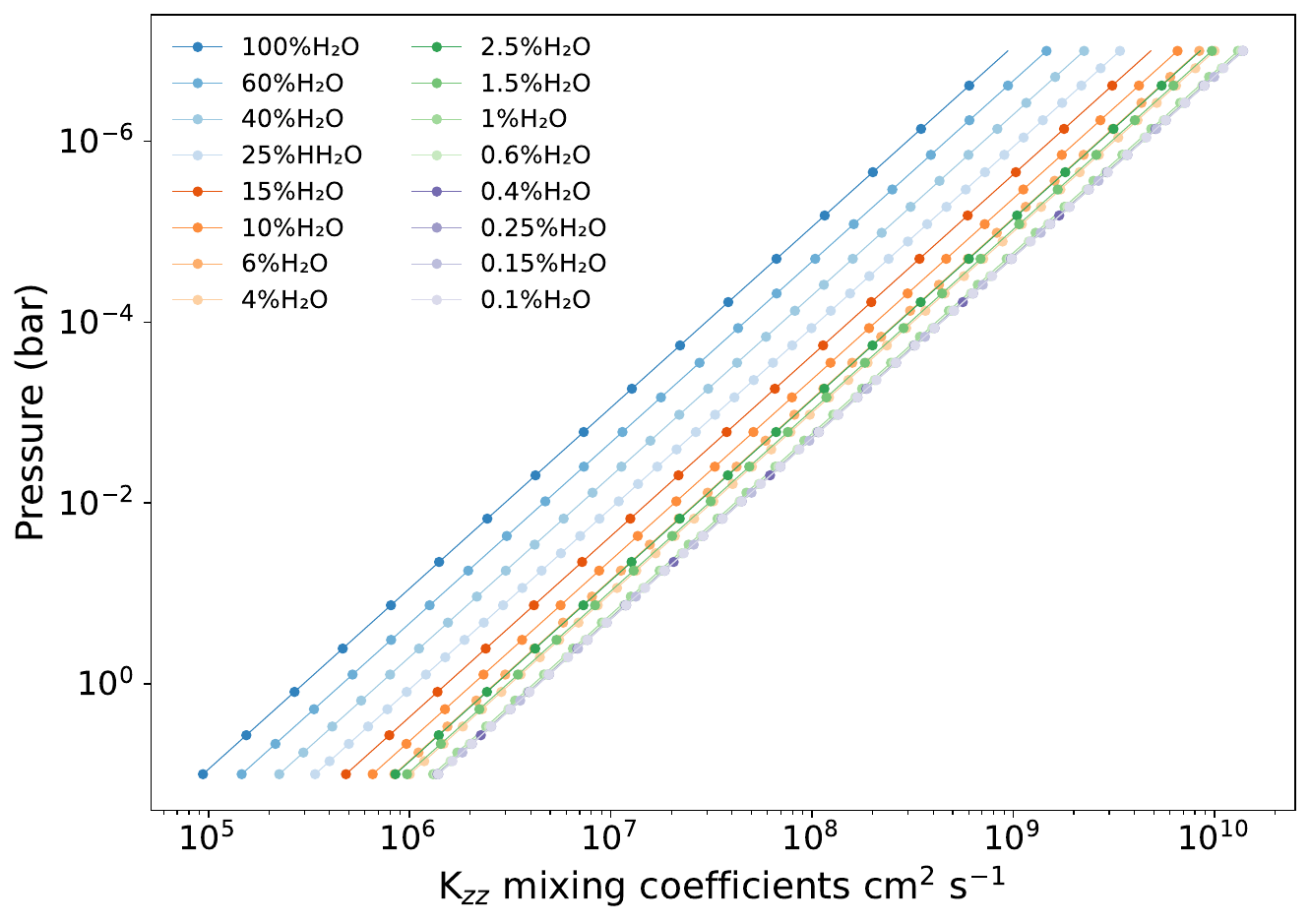}
    }
\caption{Eddy diffusion mixing coefficient ($K_{zz}$) profiles for the grid of the atmospheres with H$_2$O from 100\% to 0.1\%. Solid coloured lines with spheres represent $K_{zz}$ coefficients plotted against pressure, following the colour coding indicated in the plot legend.}
\label{fig:48}
\end{figure}

Figure~\ref{fig:48} shows the resulting $K_{zz}(P)$ profiles across the full atmospheric grid. profiles across the full atmospheric grid. The profiles demonstrate enhanced eddy diffusion towards lower pressures, facilitating vertical mixing of photochemically produced species throughout the atmosphere, whilst at the uppermost layers molecular diffusion takes over as the dominant transport mechanism. The profiles reflect the pressure-dependent atmospheric structure computed by HELIOS for each bulk composition from equilibrium state reached in Fastchem. For atmospheres with 10--100\% H$_2$O, the $K_{zz}$ profiles decrease monotonically with water abundance, as expected from the corresponding increase in mean molecular weight, which reduces the atmospheric scale height and consequently the eddy diffusion coefficient. In the intermediate range (0.6--6\% H$_2$O), where carbon, nitrogen, and hydrogen species computed in chemical equilibrium contribute significantly to the bulk composition, the altitude-dependent pressure structure deviates from a simple water-scaling, resulting in non-monotonic ordering of the profiles. At the lowest water abundances (0.1--0.4\% H$_2$O), where other species dominate the composition, the $K_{zz}$ profiles converge towards a common behaviour that is no longer sensitive to the residual water fraction. While a fully self-consistent $K_{zz}$ derived from 3D GCMs for rocky planets around active M dwarfs would be ideal, our MLT parameterisation provides a physically motivated baseline for the present study.

\subsection{Molecular diffusion parameters}
\label{sec:app3}
In VULCAN, at the atmospheric boundaries, the discretized molecular diffusion coefficients D$_{zz}$ incorporate molecular diffusion, gravitational settling, and thermal diffusion.
The thermal diffusion factor $\alpha$ scales the effect of temperature gradients on species transport, dominating only near the boundaries where vertical fluxes are constrained.

In hydrogen-dominated atmospheres (H$_2$O <40\%), particularly in the extended and low-density upper layers, molecular diffusion processes are significantly influenced by thermal diffusion factors, $\alpha$, that vary according to species. In our setup, for atomic hydrogen, $\alpha$ was set to -0.1. It reflects a slight tendency of this light species to migrate upward along temperature gradients, enhancing its transport and potential escape. For  H$_2$ $\alpha$=0, as this species experiences minimal thermal diffusion effects, consistent with its intermediate mass and relatively well-mixed nature. Heavier species in this regime typically have positive $\alpha$ values around 0.25, which bias their diffusion toward cooler, denser atmospheric layers due to gravitational settling and reduced thermal diffusion, effectively constraining their vertical distribution.

Conversely, in water-rich atmospheres (H$_2$O$\ge$40\%) characterised by higher densities and more compact vertical structures, thermal diffusion factors were adjusted to reflect the stronger collisional coupling and differing transport dynamics. Light species such as H, He, and H$_2$, in our setup, have negative $\alpha$ values near $-0.25$, consistent with enhanced upward thermal diffusion driven by temperature gradients. Heavier molecules in this environment generally have $\alpha \approx 0$, indicating negligible thermal diffusion effects within the tightly coupled atmospheric matrix. The weighting of diffusion-related terms: molecular diffusivity, gravitational settling, and thermal diffusion, varies accordingly. Molecular diffusion coefficients $D_{zz}$ are generally larger in low-density hydrogen atmospheres, amplifying all diffusive fluxes, while in dense water-dominated atmospheres, smaller diffusion coefficients diminish thermal and molecular diffusion contributions, with gravitational and collisional effects dominating heavier species transport.

\section{Simulations in detail and additional figures}
\label{sec:appendix1}

\subsection{Evolution outlook: initial atmospheric states}
\begin{figure*}\resizebox{\hsize}{!}{
   \centering
   \includegraphics[width=0.95\textwidth]{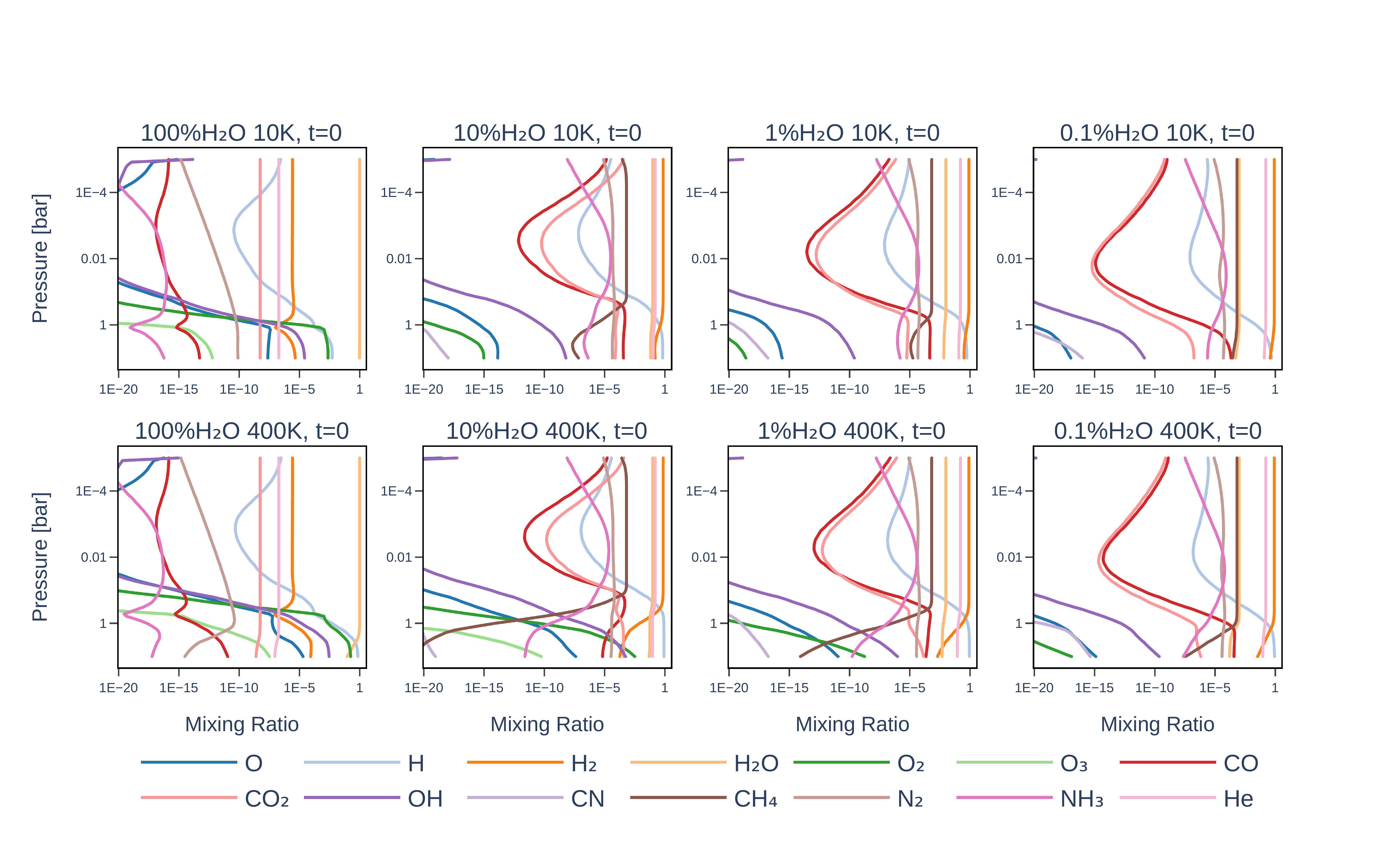}
   }
   \caption{Initial atmospheric mixing ratios for four atmospheres with water fractions of 100\%, 10\%, 1\%, and 0.1\% arranged by columns, and two atmospheric regimes arranged by rows with interior heating of 10K and 400K, respectively. Concentrations are shown for the initial simulation time ($t$=0).}
\label{fig:93}
\end{figure*}

We started all our simulations in the grid by calculating initial concentrations by variation of Fastchem code incorporated in VULCAN. As input, this version of Fastchem use T-P profiles we recalculated in HELIOS as described in Sect.~\ref{sec:metodstp}. In Fig.~\ref{fig:93} we present these concentrations at the beginning of the simulation (t=0)) before applying the external flare forcing. The figure provides a visual framework to compare atmospheric mixing ratios of key species across simulation regimes and initial states before chemical kinetics evolve the system.

\subsection{Number density profiles at the end of simulation}
\label{sec:appendixnd}

\begin{figure*}\resizebox{\hsize}{!}{
   \centering
   \includegraphics[width=0.95\textwidth]{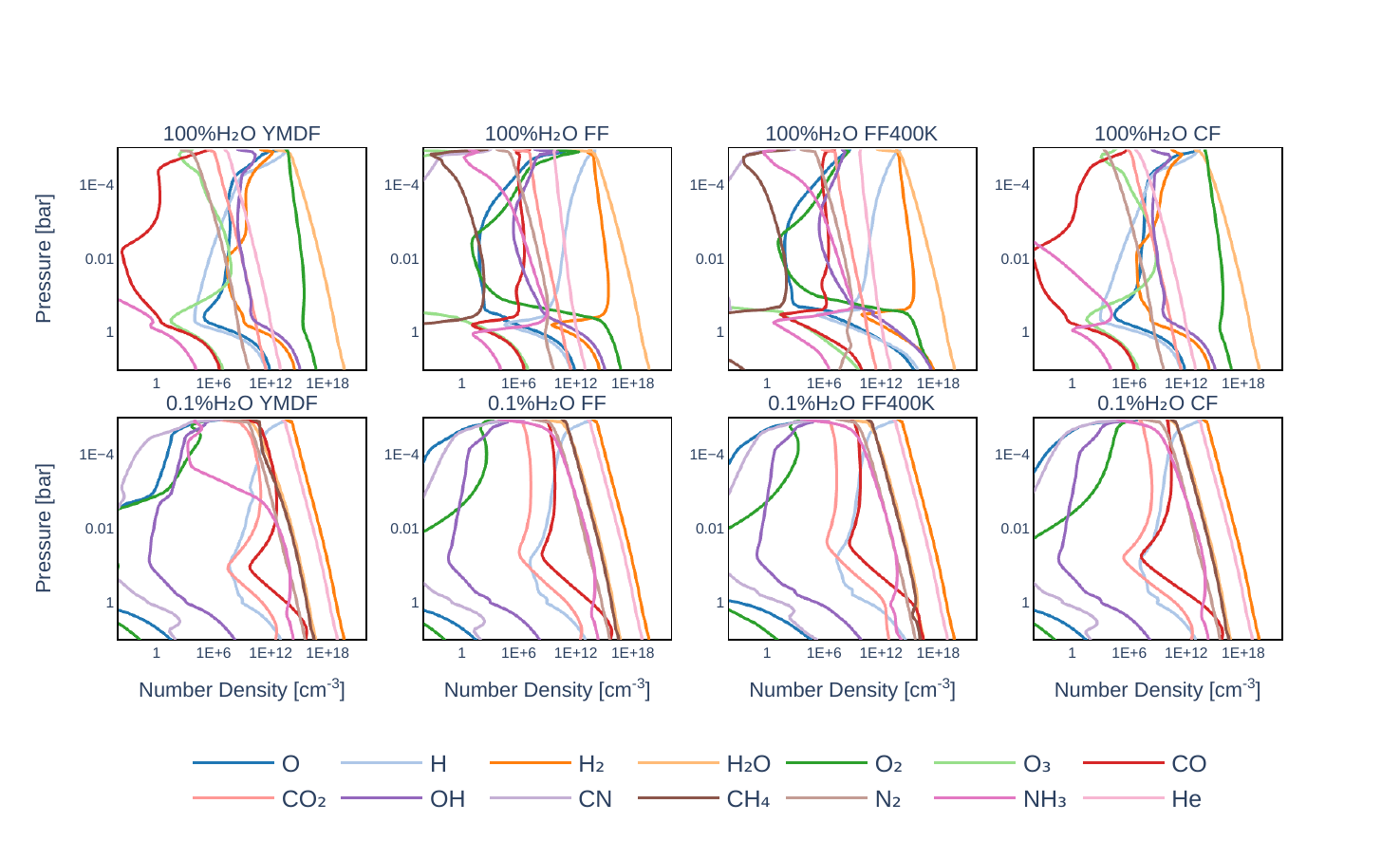}
   }
   \caption{Number density profiles [cm$^{-3}$] as a function of pressure [bar] for key atmospheric species across the four flare models (YMDF, FF, FF400K, and CF), shown for the 100\% H$_2$O (top row) and 0.1\% H$_2$O (bottom row) atmospheric compositions. Species are colour-coded as indicated in the legend.}
\label{fig:37}
\end{figure*}

Figure~\ref{fig:37} presents the number density profiles of the key atmospheric species across the four flare models (YMDF, FF, FF400K, and CF) for the 100\% and 0.1\% H$_2$O boundary compositions. In both cases, the general decline in number densities toward lower pressures is a natural consequence of the decreasing atmospheric density with altitude and does not reflect chemical depletion for some species mentioned before as stable (i.e. H$_2$O, He, and N$_2$). In the 100\% H$_2$O atmosphere, H$_2$O and H$_2$ dominate the column at all pressure levels across all models, with the remaining species occupying a broad but lower number density range. The most notable model-dependent differences appear in the upper atmosphere, where O$_2$ reaches appreciably higher number densities and atomic O is slightly more abundant in the YMDF and CF models compared to FF and FF400K, consistent with enhanced production of oxygen-bearing species driven by more energetic flare events. By contrast, H and CO number densities are elevated in the FF and FF400K models, particularly in the middle atmosphere.

In the 0.1\% H$_2$O atmosphere, H$_2$O remains present throughout the column at number densities consistent with its prescribed mixing ratio, with the exception of the uppermost layers, but falls beneath the dynamical range dominated by H$_2$ and He. The number density profiles of the remaining species are broadly similar across models, with the notable exception of the YMDF case, where O and O$_2$ reach comparable number densities to one another, whereas in the other models atomic oxygen is more strongly depleted relative to O$_2$. This demonstrates that even in water-poor atmospheres, the flare energy distribution governs the relative abundances and vertical distribution of oxygen-bearing species.

\subsection{Simulations' results: additional figures}
\label{sec:appendixb}

\begin{figure*}\resizebox{\hsize}{!}{
   \centering
   \includegraphics[width=0.95\textwidth]{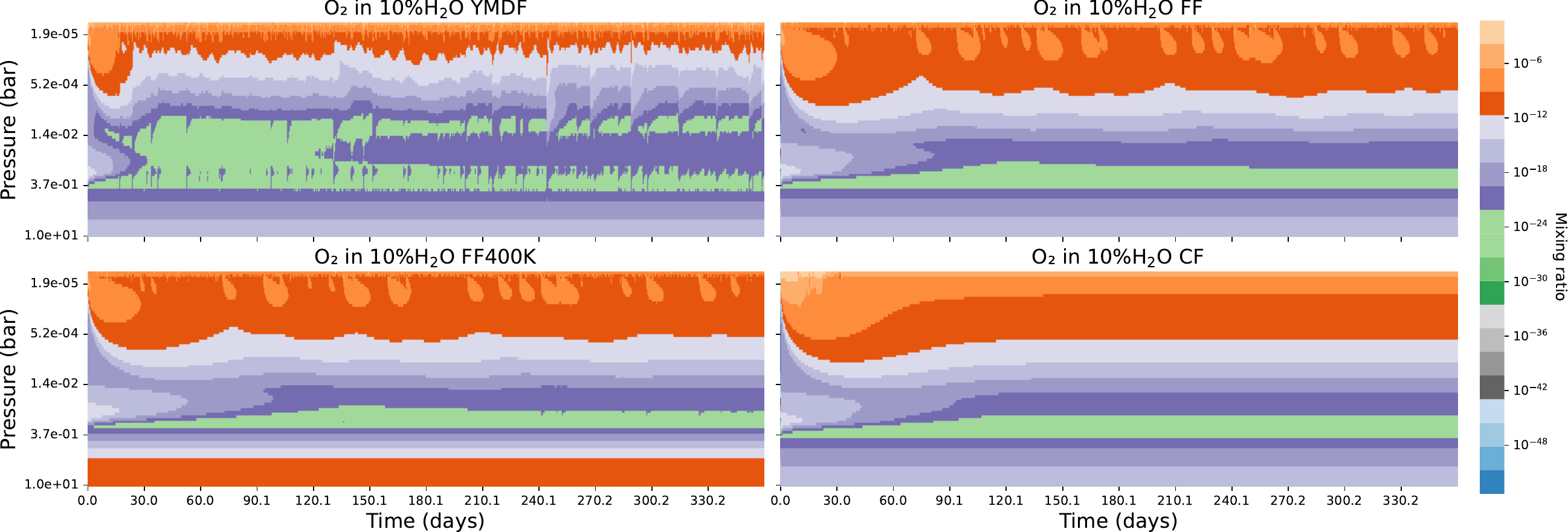}
    }
\caption{Heatmaps of O$_2$ mixing ratio over the 360-day simulation period for a solar abundance atmosphere containing 10\% water vapour. The upper row shows results from the YMDF and FF models, while the bottom row shows the FF model heated to 400 K and the CF model, as indicated on the panels. The vertical axis represents atmospheric pressure in bars, and the horizontal axis represents simulation time in days.}
\label{fig:72}
\end{figure*}

Figure~\ref{fig:72} shows heatmaps of the O$_2$ mixing ratio over the full 360-day simulation for a solar abundance atmosphere containing 10\% water vapour, for all models. A clear atmospheric response to the flare events is visible in the FF model, though it is less pronounced than the largest flare effect in YMDF, consistent with the lower FUV and NUV flux of the FF flares. Unlike the YMDF dataset, which consist one large flare event and numerous mid-size flares, the FF dataset contains multiple flares of energy comparable to its largest event, alongside a higher frequency of smaller flares. This produces a more distributed and cumulative photochemical perturbation in the FF model, where the combined effect of repeated moderate flares contributes to the long-term evolution. Disturbance of the atmospheric composition, forced by stellar activity, is visible in both the FF and YMDF panels, but not in CF model. However, the molecular oxygen mixing ratios in the uppermost layers are similar to those in YMDF model, consistent with values presented in Fig.~\ref{fig:2}.

\begin{figure*}\resizebox{\hsize}{!}{
   \centering
   \includegraphics[width=0.95\textwidth]{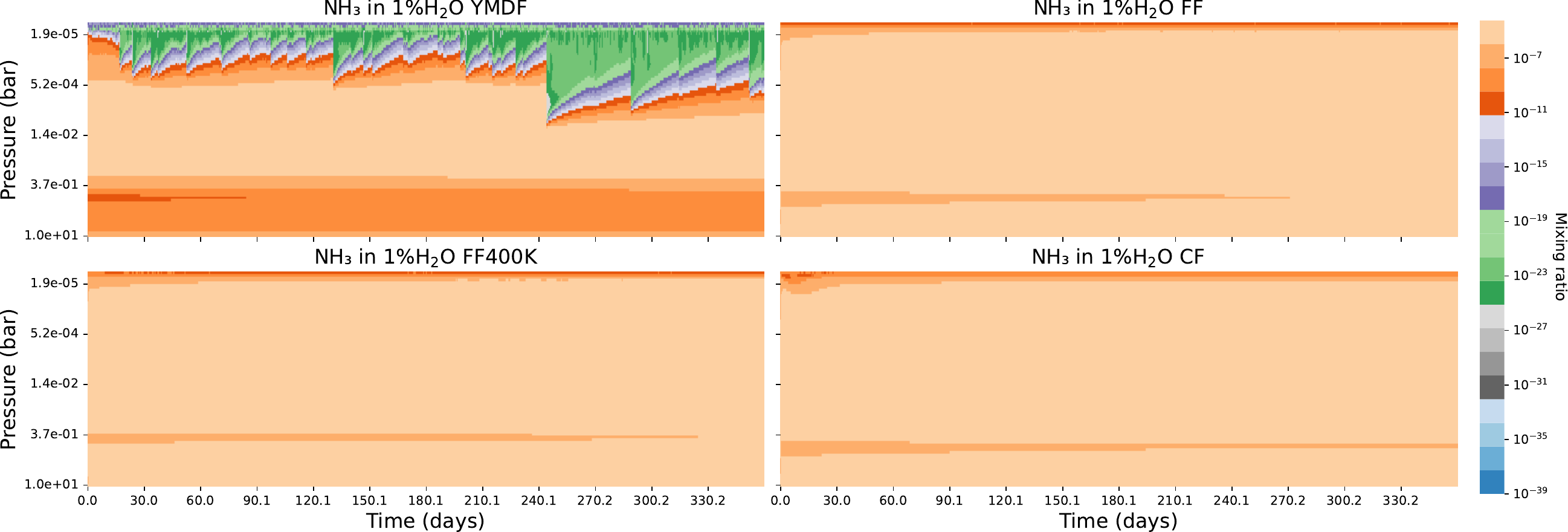}
    }
\caption{Heatmaps of NH$_3$ mixing ratio over the 360-day simulation period for a solar abundance atmosphere containing 1\% water vapour. The upper row shows results from the YMDF and FF models, while the bottom row shows the FF model heated to 400 K and the CF model, as indicated on the panels. The vertical axis represents atmospheric pressure in bars, and the horizontal axis represents simulation time in days.}
\label{fig:27}
\end{figure*}

\begin{figure*}\resizebox{\hsize}{!}{
   \centering
   \includegraphics[width=0.95\textwidth]{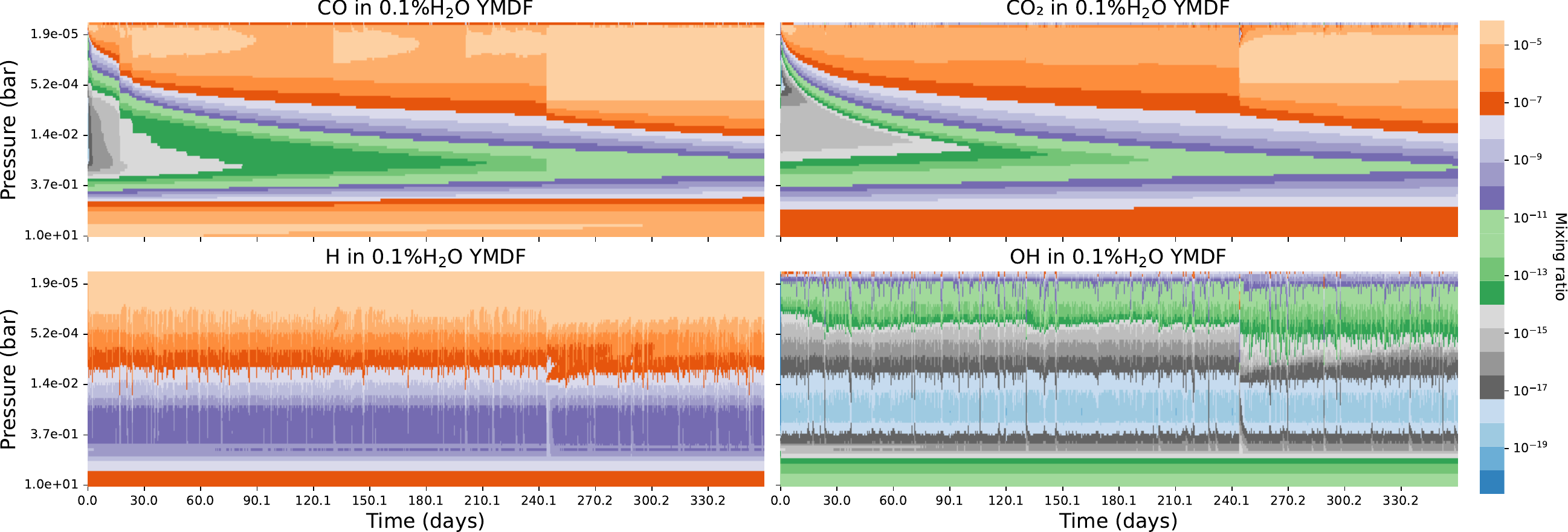}
    }
\caption{Heatmaps of CO, CO$_2$, H and OH mixing ratio over the 360-day simulation period for a solar abundance atmosphere containing 0.1\% water vapour analogous to the previous Fig.~\ref{fig:27}.}
\label{fig:288}
\end{figure*}

\begin{figure*}\resizebox{\hsize}{!}{
   \centering
   \includegraphics[width=0.95\textwidth]{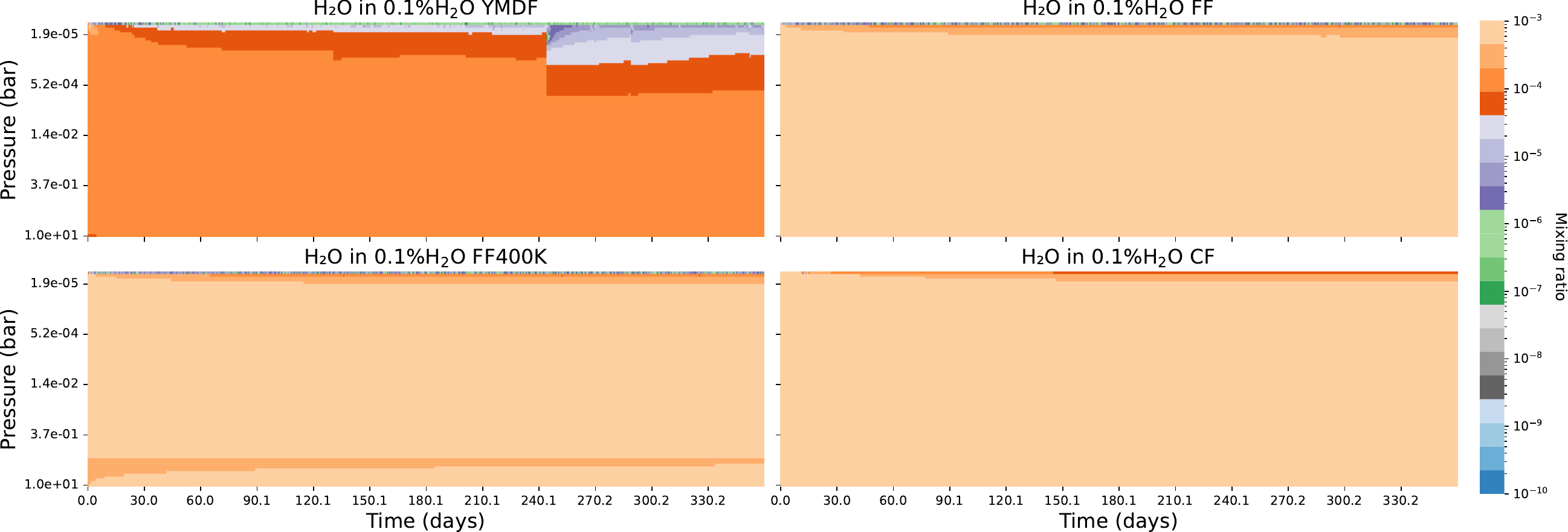}
    }
\caption{Heatmaps of H$_2$O mixing ratio over the 360-day simulation period for a solar abundance atmosphere containing 0.1\% water vapour analogous to the previous Fig.~\ref{fig:27}.}
\label{fig:28}
\end{figure*}

\begin{figure*}
\resizebox{\hsize}{!}{
   \centering
   \includegraphics{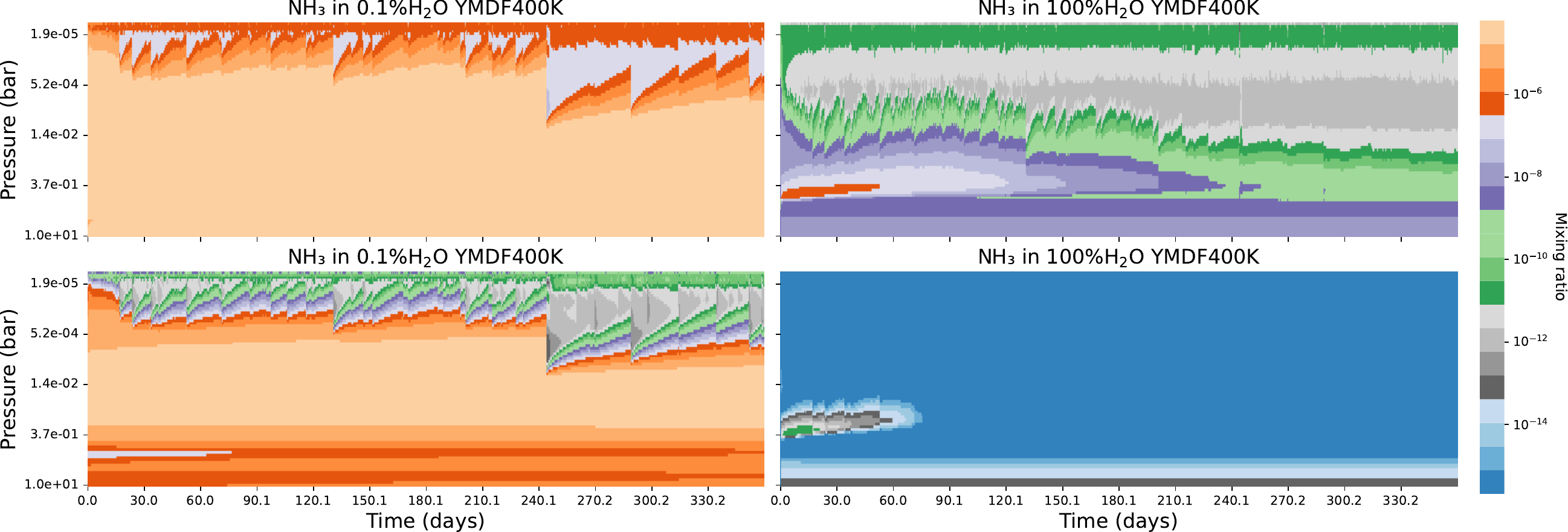}
    }
\caption{Heatmaps showing the evolution of NH$_3$ mixing ratios over the 360-day simulation period. The left and right panels represent atmospheres with 0.1\% and 100\% water vapour fractions, respectively. The top and bottom rows correspond to the YMDF model without and with interior heating (10K and 400K, respectively).}
\label{fig:91}
\end{figure*}

\subsection{Additional interior heating in the YMDF model}
\label{sec:appendixymdf}
\begin{figure*}
\resizebox{\hsize}{!}{
   \centering
   \includegraphics{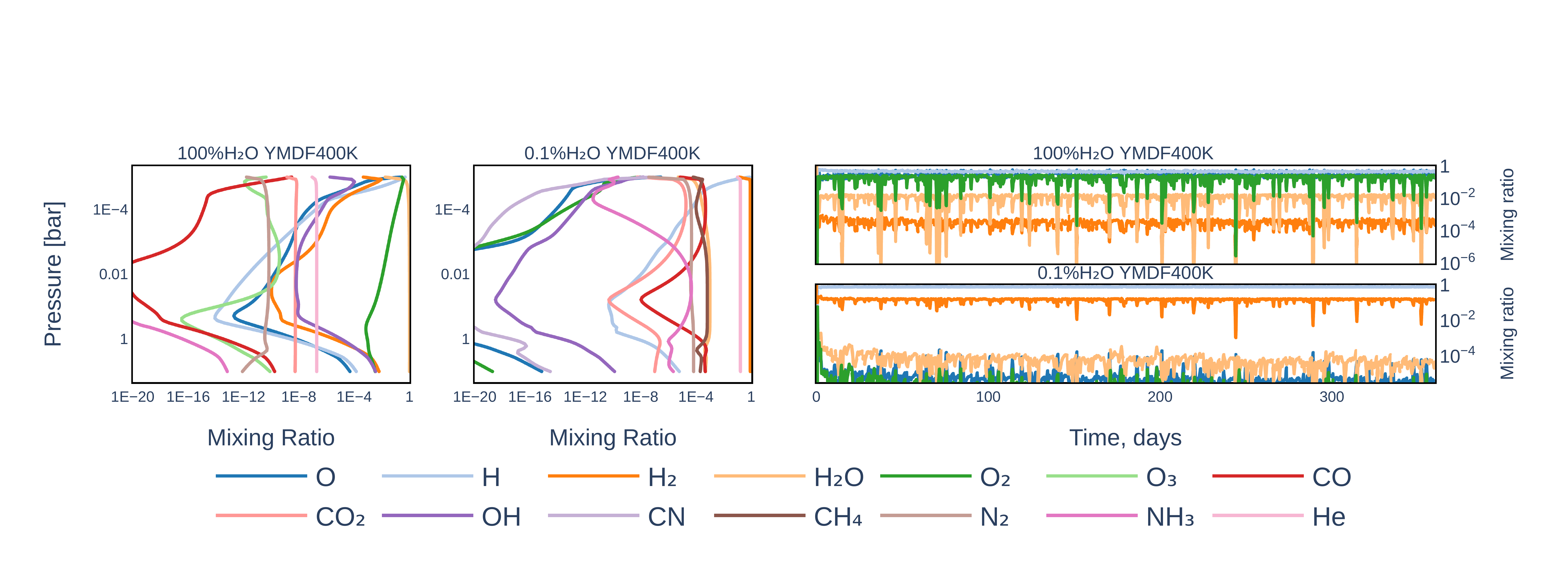}
    }
\caption{Atmospheric mixing ratios for atmospheres with water fractions of 100\%, and 0.1\% under stellar forcing in the YMDF model and with 400K internal hitting at the bottom boundary (two vertical panels on the left), and 360-days evolution of the mixing ratios of photodissociation-driven species with concentrations above 10$^{-4}$ in the uppermost atmospheric layer of these two atmospheric regimes (two horizontal panels on the right).
}
\label{fig:9}
\end{figure*}

We analysed two extreme regimes in our atmospheric grid under the YMDF model flaring, simulating 100\% and 0.1\% H$_2$O with temperature-pressure profiles including T=400K interior heating (Fig.~\ref{fig:9}). This heating increases oxygen abundances in both upper and lower layers, with species such as H and OH showing similar trends. Most species depicted in the two left square panels of Fig.~\ref{fig:9}, representing the mixing ratios at the end of the 360-day simulation, show elevated concentrations in the upper and lower layers, while their overall abundances in the middle of the atmosphere remain comparable to those in the YMDF model without interior heating. Species like CH$_4$ and NH$_3$ display more complex responses to the modified temperature-pressure profile. The two right horizontal panels illustrate the temporal evolution in the upper layers and show compositions similar to the YMDF model without heating, although flare responses are  more smothered and prolonged.

In Fig.~\ref{fig:91} we present the temporal evolution of the ammonia mixing ratio for atmospheres with 100\% and 0.1\% H$_2$O fractions, comparing both the YMDF and YMDF400K models. When the H$_2$O abundance is low, ammonia levels gradually accumulate in the atmosphere, reaching mixing ratios of $10^{-6}$ and higher. These values are highly sensitive to large flares, which deplete NH$_3$ not only in the upper layers but also deeper within the atmosphere. Partial recovery occurs over time, yet full replenishment is not achieved within the simulation period. Introducing interior heating causes further NH$_3$ depletion near the atmospheric base, leading to overall lower concentrations. In water steam-dominated regimes, an initial ammonia reservoir is progressively lost through sustained stellar activity, with complete removal observed when basal heating is included.

\end{appendix}
\end{document}